\documentclass[sigconf]{acmart}
\AtBeginDocument{%
  }

\setcopyright{none}

\renewcommand\footnotetextcopyrightpermission[1]{}
\usepackage{graphicx}
\usepackage{makecell}
\usepackage{multirow}
\usepackage{tabularx}
\usepackage{enumitem}
\usepackage{amsthm}
\usepackage{tikz}
\usepackage{tabularray}
\usepackage{subcaption}
\usepackage{threeparttable}
\usepackage{algorithm}
\usepackage[noend]{algorithmic}
\usepackage{colortbl}
\usepackage{balance}

\DeclareMathOperator*{\argmin}{argmin}
\DeclareMathOperator*{\argmax}{argmax}
\newtheorem{definition}{Definition}

\newcommand*\emptycirc[1][0.8ex]{\tikz\draw (0,0) circle (#1);} 
\newcommand*\halfcirc[1][0.8ex]{%
	\begin{tikzpicture}
	\draw[fill] (0,0)-- (90:#1) arc (90:270:#1) -- cycle ;
	\draw (0,0) circle (#1);
	\end{tikzpicture}}
\newcommand*\fullcirc[1][0.8ex]{\tikz\fill (0,0) circle (#1);} 
\definecolor{darkorange}{RGB}{255, 140, 0}
\definecolor{darkgreen}{RGB}{0, 100, 0}
\definecolor{blueviolet}{RGB}{138, 43, 226}
\definecolor{dodgerblue}{RGB}{30, 144, 255}

\begin{document}

\newcommand\sFORALL[2]{\FORALL{#1}#2\ENDFOR}
\newcommand\sIF[2]{\IF{#1}#2\ENDIF}
\newcommand{\RIGHTCOMMENT}[1]{\hfill

$\triangleright$ {#1}}

\title{Peekaboo, I See Your Queries: Passive Attacks Against DSSE Via Intermittent Observations}

\author{Hao Nie}
\orcid{0009-0009-9961-5817}
\affiliation{%
  \institution{Huazhong University of Science and Technology}
  \city{Wuhan}
  \state{Hubei}
  \country{China}}
\email{nie@hust.edu.cn}

\author{Wei Wang}
\authornote{Corresponding author.}
\orcid{0000-0003-4457-6709}
\affiliation{%
  \institution{Huazhong University of Science and Technology}
  \city{Wuhan}
  \state{Hubei}
  \country{China}}
\email{viviawangwei@hust.edu.cn}

\author{Peng Xu}
\orcid{0000-0003-4268-4976}
\affiliation{%
  \institution{Huazhong University of Science and Technology}
  \city{Wuhan}
  \state{Hubei}
  \country{China}}
\email{xupeng@hust.edu.cn}

\author{Wei Chen}
\orcid{0009-0005-8983-9375}
\affiliation{%
  \institution{Huazhong University of Science and Technology}
  \city{Wuhan}
  \state{Hubei}
  \country{China}}
\email{hust_cw@hust.edu.cn}

\author{Laurence T. Yang}
\orcid{0000-0002-7986-4244}
\affiliation{%
  \institution{St. Francis Xavier University}
  \city{Antigonish}
  \state{Nova Scotia}
  \country{Canada}}
\email{ltyang@gmail.com}

\author{Mauro Conti}
\orcid{0000-0002-3612-1934}
\affiliation{%
  \institution{University of Padua}
  \city{Padua}
  \state{Veneto}
  \country{Italy}}
\affiliation{%
    \institution{Örebro University}
    \city{Örebro}
    \state{Örebro County}
    \country{Sweden}
}
\email{mauro.conti@unipd.it}

\author{Kaitai Liang}
\orcid{0000-0003-0262-7678}
\affiliation{%
  \institution{TU Delft}
  \city{Delft}
  \state{South Holland}
  \country{Netherlands}}
\affiliation{%
  \institution{University of Turku}
  \city{Turku}
  \state{Southwest Finland}
  \country{Finland}}
\email{kaitai.liang@tudelft.nl}





\renewcommand{\shortauthors}{Hao Nie et al.}

\begin{abstract}

Dynamic Searchable Symmetric Encryption (DSSE) allows secure searches over a dynamic encrypted database but suffers from inherent information leakage. Existing passive attacks against DSSE rely on persistent leakage monitoring to infer leakage patterns, whereas this work targets intermittent observation - a more practical threat model. We propose Peekaboo - a new universal attack framework - and the core design relies on inferring the search pattern and further combining it with auxiliary knowledge and other leakage. We instantiate Peekaboo over the SOTA attacks, Sap (USENIX' 21) and Jigsaw (USENIX' 24), to derive their ``+'' variants (Sap+ and Jigsaw+). Extensive experiments demonstrate that our design achieves >0.9 adjusted rand index for search pattern recovery and $\backsim$90\% query accuracy vs. FMA’s $\backsim$30\% (CCS' 23). Peekaboo’s accuracy scales with observation rounds and the number of observed queries but also it resists SOTA countermeasures, with >40\% accuracy against file size padding and >80\% against obfuscation.

\end{abstract}




\maketitle

\begin{center}
\textbf{Full Version.} This is the full version of the paper accepted at ACM CCS 2025. 
The published version is available at 
\href{https://doi.org/10.1145/3719027.3765075}{https://doi.org/10.1145/3719027.3765075}.
\end{center}

\section{Introduction}

\label{sec:intro}

%

Searchable Symmetric Encryption (SSE) \cite{DBLP:conf/sp/SongWP00,DBLP:conf/ccs/KamaraPR12,DBLP:conf/ccs/BostMO17} allows a client to search an encrypted database on a server without disclosing searched keywords and the files, while its variant, dynamic SSE (DSSE) \cite{DBLP:conf/ccs/Bost16,DBLP:conf/ccs/ChamaniPPJ18,DBLP:conf/ndss/Chen0PLS0L23,DBLP:journals/tifs/DouDXWXCJ24}, additionally enables secure updates to the database. 
Existing DSSE leaks certain information, even with forward and backward privacy (FP/BP) \cite{DBLP:conf/ccs/Bost16,DBLP:conf/ccs/ChamaniPPJ18,DBLP:conf/ndss/Chen0PLS0L23,DBLP:journals/tifs/DouDXWXCJ24,DBLP:conf/ccs/BostMO17}. 
Passive attackers\footnote{Passive attackers differ from active attackers \cite{DBLP:conf/uss/ZhangKP16,DBLP:conf/eurosp/PoddarWLP20,DBLP:conf/ndss/BlackstoneKM20,DBLP:conf/uss/Zhang00YL23}, who intentionally inject dummy files into the database, which is orthogonal to this work.} (Table \ref{tab:attack_comparison}) exploit this leakage by observations and prior knowledge (known/similar data) to recover search queries.

These attacks \cite{DBLP:conf/ccs/XuZXYW23,DBLP:conf/codaspy/HaltiwangerH24,DBLP:conf/codaspy/Salmani021} against DSSE typically rely on a ``strong'' assumption that the attacker can perform continuous observations as a so-called \textit{persistent attacker}. 
Such attackers monitor leakage over time and track changes in search queries, enabling them to reconstruct a comprehensive leakage profile across the entire query history. 
For example, the leakage from any two consecutive queries for the same keyword is consistent, as these queries are issued and observed in close succession, and the underlying database is unlikely to update significantly within such a short interval.

\begin{figure}[tp]
    \centering
            
 \subfloat[Intermittent Observation, Case I]
	{
 \label{fig:Intermittent-observation attacker 1}
		\begin{minipage}{0.95\linewidth}
			\centering
			\includegraphics[width=\linewidth]{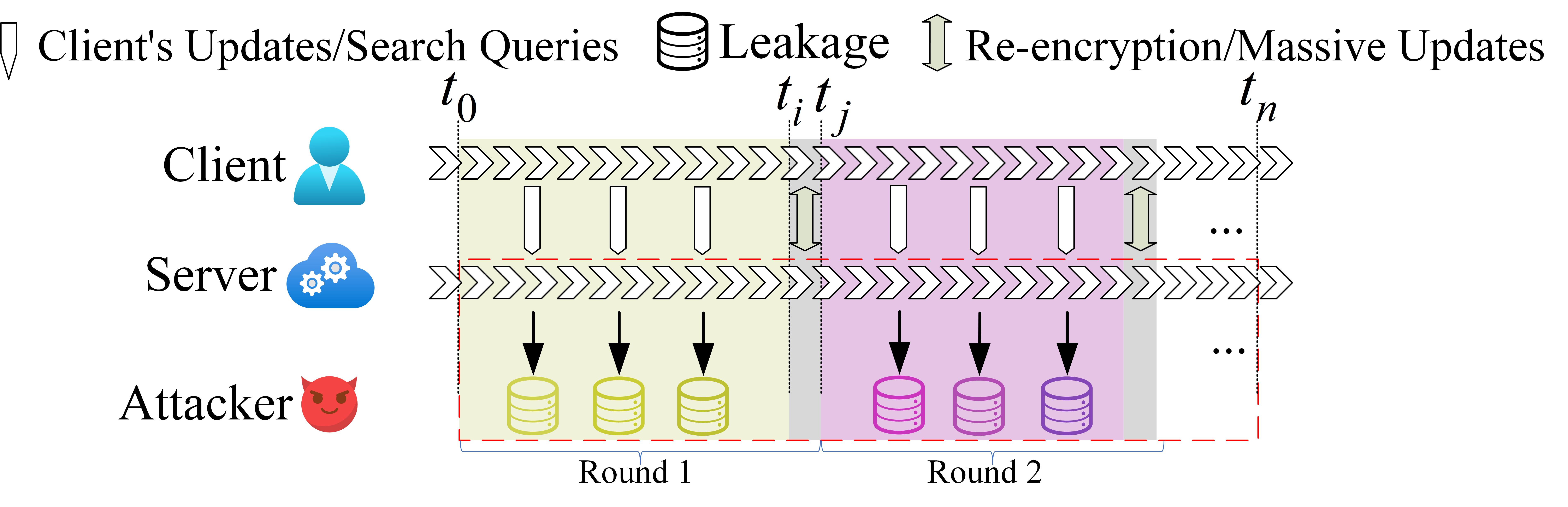}
		\end{minipage}
	}

    \subfloat[Intermittent Observation, Case II]
	{
 \label{fig:Intermittent-observation attacker 2}
		\begin{minipage}{0.7913\linewidth}
			\centering
			\hspace{-0.6cm}\includegraphics[width=\linewidth]{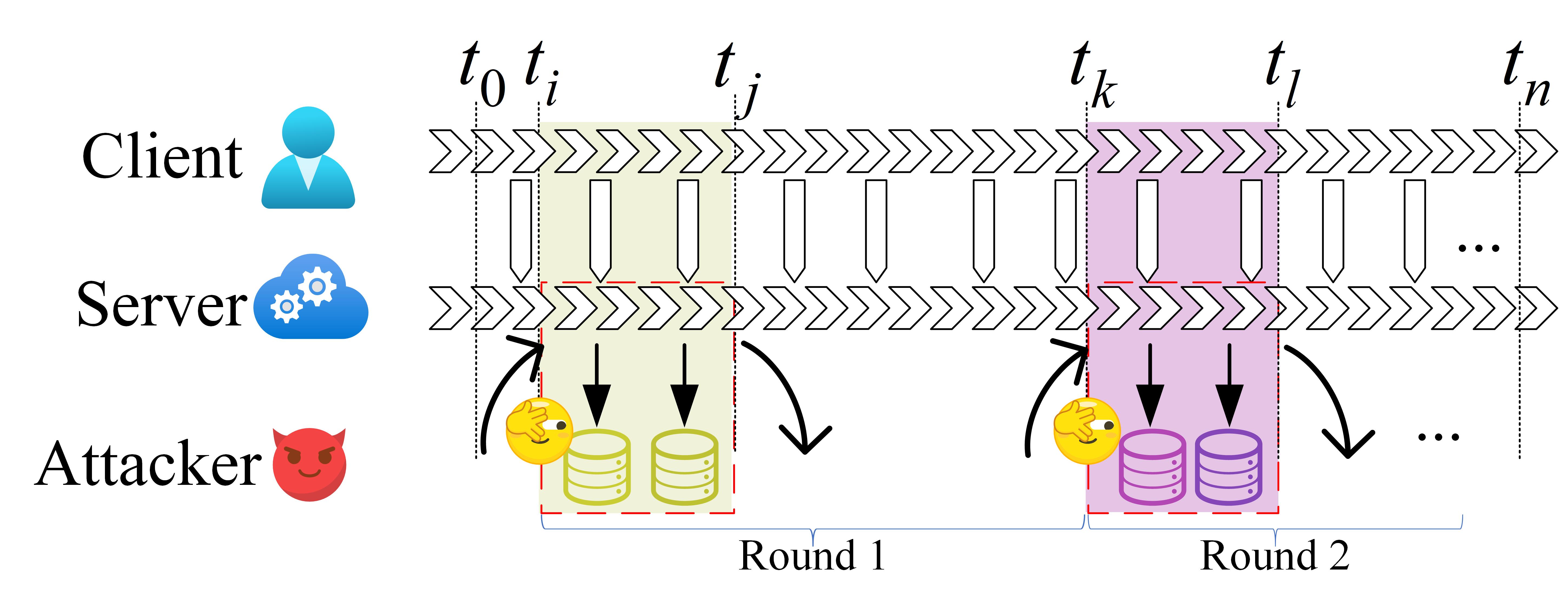}
		\end{minipage}
	}
    
 \caption{Intermittent observations: Case I, the attacker cannot match the queries (and their leakage) between rounds due to database re-encryption or massive updates on most of database files; Case II, the attacker only observes multiple short-term periods and skips the leakage between $t_j$ and $t_k$.} 
    \Description{Figures of intermittent observations. The two cases in this figure are fully described in the text.}
	\label{fig:attackers}
\end{figure}

In real-world scenarios, the assumption of persistent observation may not hold, as attackers are often limited to \textit{intermittent observations} due to practical constraints:
\\
$\bullet$ Case I: 
Database re-encryption following file updates disrupts the attacker’s ability to maintain continuous observations
\footnote{It is possible that the observed leakage is completely shuffled if a DSSE  imports a key-update technology \cite{DBLP:conf/ndss/Chen0PLS0L23} to re-encrypt the database.}.
\\    
$\bullet$ Case II: The attacker's observation must be intermittent, as sustained monitoring increases the risk of detection and potential interruption. 
Modern attackers, such as Trickbot \cite{Trickbot} and certain APT attackers \cite{ambika2020improved}, eavesdropping on the server or the communication channel,
often operate in sleep mode to evade detection, thereby acquiring only intermittent observations. 

\noindent Note that we provide concrete examples in Appendix \ref{app:detailed examples of ioas}.


To adapt to the landscape, we consider more practical attackers, namely \textit{intermittent-observation attackers (IOAs)}.
Like \cite{DBLP:conf/crypto/DuGG22,DBLP:conf/crypto/PersianoY23,DBLP:journals/tkde/VoYSLNW23}, the IOAs, repeating multiple rounds of observing and then staying offline, can only capture the leakage in multiple short-term periods.  
The attackers miss certain leakage of search queries and all related updates during the offline periods.  
As in Figure \ref{fig:Intermittent-observation attacker 1}, an IOA observes until re-encryption or massive updates happen at time $[t_i,t_j]$ changing many files, such that the leakage observed by the attacker before and after can be entirely different and mismatched. 
In Figure \ref{fig:Intermittent-observation attacker 2}, an IOA observes in periods
$[t_i,t_j]$ and $[t_k,t_l]$. 
Due to missing leakage between $t_j$ and $t_k$, it cannot match {the encrypted files} from the former period to those in the latter. 
Please refer to detailed examples in Appendix \ref{app:detailed examples of ioas}.

Re-encryption/massive updates in Figure \ref{fig:Intermittent-observation attacker 1} and updates of files during the offline period in Figure \ref{fig:Intermittent-observation attacker 2} can significantly impact the leakage patterns of the queries. 
The volume pattern (VP, indicating the number of files in response) and the access pattern (AP, indicating the identities of the encrypted files in the response) of search queries can vary entirely between two different rounds\footnote{The volumes of queries naturally change due to additions and deletions, while the AP of the queries for the same keyword is \textit{refreshed}, as required by the forward and backward security \cite{DBLP:conf/ccs/XuZXYW23}.}.
When the attacker re-observes the query for the same keyword in the later period, it loses the search pattern (SP, indicating whether two search queries are for the same keyword) between the two queries {due to the re-encryptions}.
As a result, the observed leakage is ``isolated'' from each other, and further observing for more periods cannot provide extra advantages for the attacker. 
Compared to a persistent attacker, we argue that an IOA is a ``weaker-in-leakage-but-stronger-in-attack'' variant — one that operates with less leakage, without access to ``a full picture'' of leakage profile, yet achieves comparable attack performance. 
We note that some works \cite{DBLP:journals/popets/KamaraKMDPT24,DBLP:conf/ndss/BlackstoneKM20,DBLP:journals/popets/AmjadKM19} also introduce similar ``intermittent'' concepts, but they obtain the copy of encrypted database instead of query leakage; thus, they cannot recover queries \cite{DBLP:journals/popets/KamaraKMDPT24,DBLP:conf/ndss/BlackstoneKM20}.

\begin{table}[!t]
\setlength{\tabcolsep}{2pt}
\centering
\begin{threeparttable}
\caption{Comparison: passive attacks against (D)SSE\tnote{1}}
\label{tab:attack_comparison}

\begin{tabular}{ccccccc}
\hline
{Target} &
{Type} &
  {Attack} &
 {Leakage} &
  Doc &
  Freq &
  {$P$ or $I$\tnote{2}} \\ 
   \hline
\multirow{8}{*}{SSE} &
  \multirow{3}{*}{Known-Data} &
  {$\text{Subgraph}^\text{ID}$} \cite{DBLP:conf/ndss/BlackstoneKM20} &
  AP &
   -&
   -&
  - \\
 &
   &
  Count \cite{DBLP:conf/ccs/CashGPR15} &
  AP,VP &
   -&
   -&
  - \\
 &
   &
  LEAP \cite{DBLP:conf/ccs/NingHPYL0D21} &
  AP &
   -&
   -&
  - \\
 \cline{2-7} 
 &
  \multirow{5}{*}{Similar-Data} &
  GraphM \cite{DBLP:conf/ccs/PouliotW16} &
  AP &
  $\fullcirc$ &
   $\emptycirc$&
  - \\
 &
   &
  SAP \cite{DBLP:conf/uss/OyaK21} &
  SP,VP &
  $\fullcirc$ &
  $\fullcirc$ &
  - \\
   &
   &
  RSA \cite{DBLP:conf/uss/Damie0P21} &
  AP &
  $\fullcirc$ &
   $\emptycirc$&
  - \\
 &
   &
  IHOP \cite{DBLP:conf/uss/OyaK22} &
  SP,AP &
  $\fullcirc$ &
  $\fullcirc$ &
  - \\
 &
   &
  Jigsaw \cite{DBLP:conf/uss/Nie00ZYL24} &
  VP,SP,AP &
  $\fullcirc$ &
  $\halfcirc $ &
  - \\ \hline

\multirow{3}{*}{DSSE} &
  \multirow{3}{*}{Similar-Data} &
 
 FMA \cite{DBLP:conf/ccs/XuZXYW23} &
  FVP/SP &
   $\emptycirc$&
   $\fullcirc$ &
  $P$ \\
  &
   &
  Sap+ &
  FVP/AP &
   $\fullcirc$ &
   $\fullcirc $ &
  $I$ \\
&
   &
  Jigsaw+ &
  FVP/AP &
   $\fullcirc $ &
   $\halfcirc$ &
  $I$ \\
  
  \hline
\end{tabular}

\begin{tablenotes}    
    \footnotesize               
    \item[1] ``AP'' is the access pattern, ``VP'' is the volume pattern, ``FVP'' is the file volume pattern, and ``SP'' is the search pattern. 
    The attacks targeting SSE all require the ``SP'' to get unique queries. We omit the ``SP'' for these attacks and only tag them with ``SP'' when they use the ``SP'' to obtain the frequency of search queries.
    ``Doc'' denotes whether the attacker needs similar documents, and ``Freq'' for similar query frequency. 
    The known-data attacks require a part of the plaintexts of the database rather than similar prior knowledge.
    The ``$\fullcirc$'' indicates heavy dependency on specific prior knowledge, the ``$\halfcirc$'' suggests moderate dependency, and the ``$\emptycirc$'' denotes no prior knowledge is required. 
      
    \item[2] ``$P$'' and ``$I$'' are the persistent and intermittent-observation attackers, respectively.  
    We note that the encrypted database remains ``unchanged'' during observation in the context of SSE (instead of DSSE), so the concepts of persistent and intermittent-observation attackers do not apply.
    
    
\end{tablenotes} 
\end{threeparttable}
\end{table}

\begin{figure}
    \centering
    \includegraphics[width=0.82\linewidth]{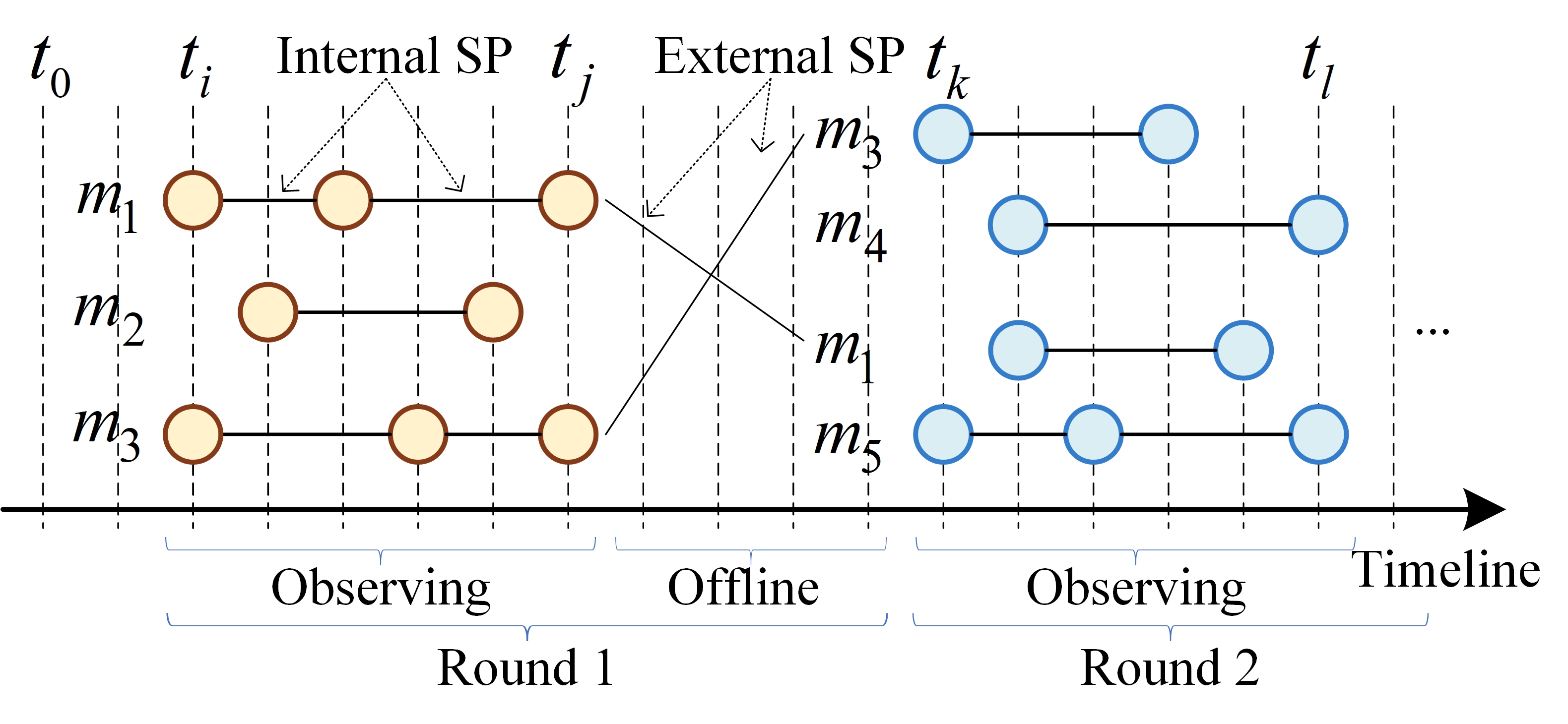}
    \caption{Internal and external SP of search queries. 
    The $m_i$ is the $i$-th keyword. The circles are the search queries observed by the attacker. In each row of an observation round, the queries are for the same keyword $m_i$ in the front.}
    \label{fig:internal and external sp}
    \Description{Figures of internal and external SP of search queries, which are fully described in the text.}
\end{figure}

\noindent \textbf{Challenges: intermittent observations against DSSE.}
In most passive attacks \cite{DBLP:conf/uss/OyaK21,DBLP:conf/uss/OyaK22,DBLP:conf/uss/Damie0P21,DBLP:conf/ccs/PouliotW16,DBLP:journals/isci/LiuZWT14,DBLP:conf/uss/Nie00ZYL24,DBLP:conf/ccs/NingHPYL0D21,DBLP:conf/ndss/IslamKK12,DBLP:conf/ccs/CashGPR15}, the SP is essential.  
The attacker must first categorize all search queries by their underlying keywords to identify unique entries. 
Some attacks \cite{DBLP:conf/uss/OyaK21,DBLP:conf/uss/OyaK22,DBLP:journals/isci/LiuZWT14,DBLP:conf/uss/Nie00ZYL24} also require the search frequency to match the queries with the keywords, while the SP is crucial to deduce the search frequency.
In the context of intermittent observations, SP is not available to the attacker, and deducing it from other leakage is challenging since the observation is intermittent. 
To obtain the ``full picture'' of SP, the attacker must acquire the SP within a single observation period (\textit{internal SP}) and between different observation periods (\textit{external SP}), see Figure \ref{fig:internal and external sp}. 



Previous attacks \cite{DBLP:conf/ccs/XuZXYW23,DBLP:conf/codaspy/Salmani021,DBLP:conf/uss/OyaK22,DBLP:conf/ndss/ShangOPK21} proposed to infer the SP could recover the internal SP. 
In \cite{DBLP:conf/codaspy/Salmani021}, Salmani et al. develop an attack that utilizes the AP to infer the SP. 
The attacker must match the AP precisely between queries, and further, any updates on keywords (e.g., adding or deleting) from files can disable such a matching. 
\cite{DBLP:conf/uss/OyaK22} and \cite{DBLP:conf/ndss/ShangOPK21} propose a method that clusters queries with obfuscated AP using $k$-means, which could also be used to recover the internal SP.
Xu et al. \cite{DBLP:conf/ccs/XuZXYW23} introduce the FMA attack, where the attacker takes the file volume pattern (FVP, indicating the size of each file in the response) to compute the similarity of queries so as to group similar queries to obtain the SP. 
Inferring the external SP is more difficult, as the leakage is periodically hidden from the attacker. 
In this sense, the attacker cannot distinguish whether two encrypted files from different rounds are for the same file (as the inferred identities and the size of files change due to the updates). 
This easily disables Salmani et al.'s attack \cite{DBLP:conf/codaspy/Salmani021} and the $k$-means based clustering \cite{DBLP:conf/uss/OyaK22,DBLP:conf/ndss/ShangOPK21}.  
Furthermore, if multiple updates occur between any two rounds and these updates result in changes to the file size, the similarity of queries calculated in the FVP is significantly distorted, thereby impacting the FMA's performance.  
These attacks cannot perform effectively with intermittent observation.

Besides the SP, current attacks \cite{DBLP:conf/uss/OyaK22,DBLP:conf/uss/Damie0P21,DBLP:conf/ccs/PouliotW16,DBLP:conf/uss/Nie00ZYL24,DBLP:conf/ndss/IslamKK12,DBLP:conf/ndss/BlackstoneKM20,DBLP:conf/ccs/CashGPR15,DBLP:conf/sp/GuiPP23} use other leakages. 
The attacker cannot leverage them from multiple observation rounds.    
From the AP, the inferred identities of the queries for the keyword differ across rounds, meaning the information is ``terminated'' between rounds. 
A similar issue occurs in the VP, as the volume of a keyword can vary significantly in different rounds. 


\noindent \textbf{Inspirations.} Though leakage and updates are not accessible to the IOA during the offline period, the semantic relationships among keywords remain consistent within the database. 
For example, the word ``searchable'' is likely to appear frequently in a file containing the term ``encryption'', both before and after database updates. 
Based on this, we can construct a co-occurrence graph for the database's keywords, incorporating the associated search queries. 
In each round, we group the search queries for the same keyword based on their similarity in leakage patterns, thereby inferring the internal SP. 
Next, we construct the co-occurrence graph for these query groups. 
Finally, we match these graphs and merge  corresponding groups from different rounds, inferring the external SP.  

\noindent \textbf{Contributions.} We investigate the IOA targeting DSSE, marking the first type of attacks with intermittent observation. 
To capture this new attack, we propose Peekaboo, with our core designs focusing on search pattern inference (P1) and query recovery (P2).  
P1 infers both internal and external SP by leveraging either AP or FVP leakage; while P2 takes the output of P1, along with either AP or FVP, and the auxiliary similar knowledge to recover all queries.  
\\
1) We define the IOA, which has not been investigated in prior work, and formalize the leakage under the context. 
The proposed attacker is a practical variant compared to the persistent attacker.  
\\
2) We propose Peekaboo, which first infers the internal SP by grouping search queries. 
In each round, when a new search query comes in, Peekaboo calculates the similarity between the unknown query and the last query in each group using either AP or FVP.
The query then joins the group with the largest similarity that exceeds a defined threshold. 
Next, Peekaboo uses the internal SP to calculate the co-occurrence matrix of the groups in each round. 
Between any two rounds, we match the groups from one round to another by solving a quadratic assignment problem based on the co-occurrence matrix. 
We iterate the process over rounds to infer all of the external SP. 
     We then process the leakage and the auxiliary knowledge based on the recovered SP and provide an interface for query recovery. 
    Existing similar-data attacks (with minor adaptations) can use this interface to recover queries through intermittent observations.  
    We use Sap \cite{DBLP:conf/uss/OyaK21} and Jigsaw \cite{DBLP:conf/uss/Nie00ZYL24} to instantiate Sap+ and Jigsaw+.  
    \\
3) We conduct extensive experiments to demonstrate the attack performance of Peekaboo. 
    P1 demonstrates high accuracy in inferring SP, providing an adjusted rand index above 0.9 in most cases; 
    and P2 achieves strong performance in query recovery. 
    For instance, Peekaboo with Jigsaw+ achieves approximately $90\%$ and $50\%$ accuracy with AP and FVP, respectively. 
    In comparison, the accuracy of FMA \cite{DBLP:conf/ccs/XuZXYW23} is about $30\%$ for both AP and FVP.
    Furthermore, Peekaboo with Jigsaw+ maintains $>40\%$ accuracy under file size padding and $>80\%$ under obfuscation in most cases.

\section{The Framework of Peekaboo}
%
We use upper-case boldface to represent matrices, lower-case boldface for vectors, upper-case italics for collections, and lower-case italics for individual values. 
We use superscripts to denote the identity of the round or time slot. 
For example, $Q^x$ represents the queries in the $x$-th round, and $\mathbf{f}^i$ denotes the frequency in the $i$-th time slot. 
We use $\{\cdot\}$ to denote a set or a list of elements and $[\cdot]$ for a vector of numbers. 
$[a]$ is used as a shorthand for $[1,2,...,a]$.  
$|\cdot|$ represents the size of a set or a list and $\# d$ represents the size of a file $d$.  
$\mathbf{v}[i]$ refers to the $i$-th value of a vector and $\mathbf{M}[i][j]$ is the value in the $i$-th row and $j$-th column of $\mathbf{M}$. 
We also use $L[i]$ to refer to the $i$-th element of a list, with $L[-1]$ specifically denoting the last element.  
$|a|$ is for the absolute value of $a$ and $||\mathbf{v}||$ represents the Euclidean norm of a vector $\mathbf{v}$.
Table \ref{tab:notations} summarizes the frequently used notations.

\begin{table}[htp]
\centering
	    \caption{Summary of notations.}
        \label{tab:notations}
		\centering
        \begin{tabular}{c>{\centering\arraybackslash}p{6.6cm}}
        
			\hline
            Notation & Description  \\
            \hline
            $\eta$ & Total number of rounds.\\
            $\boldsymbol{\sigma}$ & Number of online time slots in each round.\\
            $\boldsymbol{\varsigma}$ & Number of offline time slots in each round.\\
            $\tau$ & $\tau=\sum\limits_{i=1}^{\eta} {\boldsymbol{\sigma}}[i]$ is the total number of time slots of observation during the $\eta$ rounds.\\
            $Q^x$ & $Q^x=\{q^x_1,...,q^x_{l_x}\}$ is the observed query sequence of the $x$-th round.\\
            $AP^x$ & $AP^x=\{DB^x(q^x_1),...\}$ is the access pattern leakage of the $x$-th round.\\
             $FVP^x$ & $FVP^x=\{\{\#d_i|i\in DB^x(q^x_1)\},...\}$ is the file volume pattern leakage of the $x$-th round.\\
         
            $F^x$ & The collection of all observed distinctive files (or distinctive file sizes with the FVP leakage) of the $x$-th round.\\
		  $Gs^x$ & $Gs^x=\{G_1,...\}$ is a partition of the $Q^x$ of the $x$-th round.\\
            {$\mathbf{ID}^x$} & The index matrix of all the queries of groups of the $x$-th round (size $|Gs^x|\times |F|$). \\
            {$\mathbf{IDH}^x$} & The index matrix of the first queries of groups of the $x$-th round (size $|Gs^x|\times |F|$).\\
            {$\mathbf{IDT}^x$} & The index matrix of the last queries of groups of the $x$-th round (size $|Gs^x|\times |F|$).\\
            $\mathbf{F}_r$ & The matrix of the client's search frequency of $\tau$ time slots.\\ 
            $\mathbf{F}_s$ & The matrix of the search frequency of the keywords known by the attacker.\\
            $\mathbf{V}_r$ & The matrix of the max volume of the queries in each group of $\tau$ time slots.\\
            $\mathbf{V}_s$ & The matrix of the volume of keywords of $\tau$ time slots known by the attacker.\\
            $maxlevel$ & The match of the \textsc{InferESP} between two rounds $i$ and $j$ has an interval less than $maxlevel$, i.e. $|i-j|\leq maxlevel$.\\ 
            $p_g$ & The ratio of the removed matches in \textsc{Match}.\\
         
            
            
            
            \hline
		\end{tabular}
\end{table}

\subsection{Scenarios}
\label{sec:scenarios}
A client maintains a database and uses a DSSE scheme to outsource it to a server. 
The DSSE generally includes three protocols: \textit{Setup}, \textit{Update}, and \textit{Query}. Though the plaintexts of the database and search queries are not directly leaked, the attacker can observe patterns and recover search queries. 
Existing DSSE schemes provide forward/backward privacy (FP/BP) \cite{DBLP:conf/ccs/BostMO17,DBLP:conf/ccs/Bost16} and the BP includes Type I, II, and III, each offering progressively weaker security \cite{DBLP:conf/ccs/BostMO17}.
We provide details of DSSE and related attacks in Appendix \ref{app:background}. 

We consider IOAs against DSSE with only passive and intermittent observation of leakage. 
They can observe the leakage from queries over multiple periods while being offline between observations, either intentionally or due to constraints, as illustrated in Figure \ref{fig:attackers}. 
We refer to each online-then-offline cycle as a round and employ multiple rounds to complete an attack. 

\textit{An IOA only needs to} 1) eavesdrop on the server, or 2) eavesdrop on the communication channel, with the same purpose of query recovery. This assumption is weaker than previous attacks assuming the server is the attacker and can observe continuously. 

The IOA can gain VP and FVP (without AP and SP) against DSSE with Type I BP-security with ORAM-like techniques \cite{DBLP:conf/ccs/XuZXYW23};  
for other types of DSSE, the attacker can additionally obtain AP and internal SP\footnote{For certain work \cite{DBLP:conf/codaspy/Salmani021}, the attacker also cannot track the SP from the search tokens.}. 
If the file is re-encrypted before transmission, the IOA eavesdropping on channel cannot acquire AP and SP. {We assume the external SP is always concealed as re-encryptions happen during offline periods.}
We provide a description of these leakages in Section \ref{sec:peekaboo_attacker}.
The server itself may function as a persistent attacker, observing leakage continuously, which differentiates it from an IOA (see Appendix \ref{app:sba}). 




\begin{figure*}[htp]
    \centering
    \includegraphics[width=0.95\linewidth]{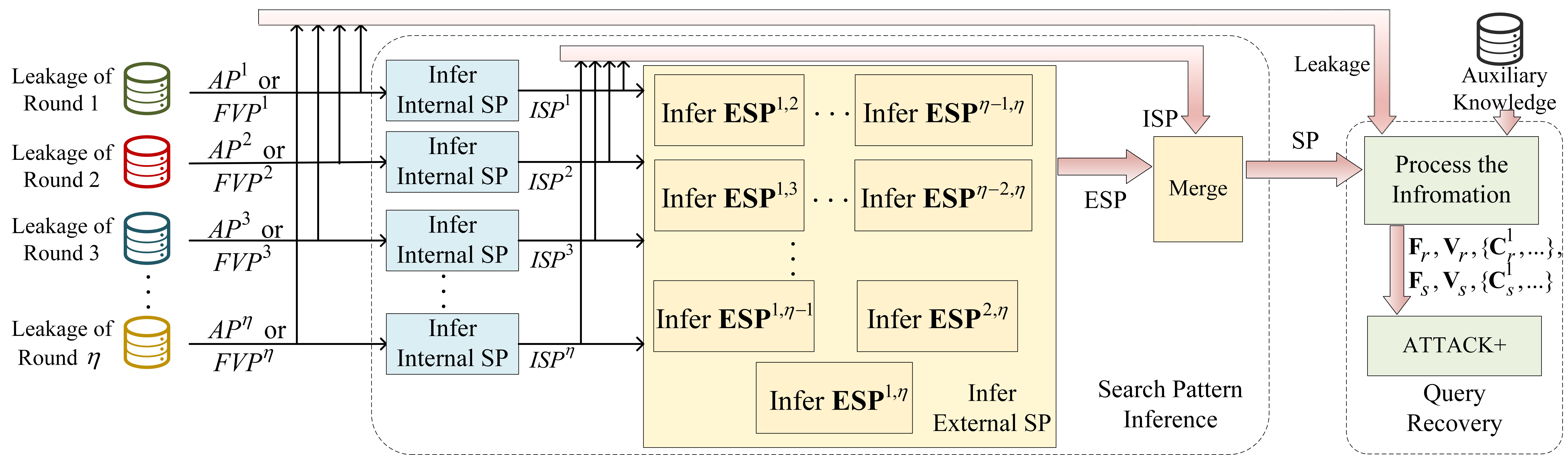}
    \caption{Overview: the Peekaboo attack.}
    \label{fig:Peekaboo Model}
    \Description{The figure of the Peekaboo attack, which are fully described in the text.}
\end{figure*}
{
\subsection{Peekaboo Attacker}
\label{sec:peekaboo_attacker}

\begin{definition}[$(\eta,\boldsymbol{\sigma},\boldsymbol{\varsigma})$ intermittent observation] 
A $(\eta,\boldsymbol{\sigma},\boldsymbol{\varsigma})$ intermittent observation includes $\eta$ rounds of observation. The $i$-th round begins with online observation, obtaining the leakage of search queries $Q^i$ for $\boldsymbol{\sigma}[i]$ time slots. Each online period follows with an offline period of $\boldsymbol{\varsigma}[i]$ time slots without observation.
\end{definition}

The attacker with the above observation ability is denoted as an intermittent-observation attacker (IOA).
The attacker can observe a collection of queries $Q=\{Q^1,...,Q^\eta\}$, where $Q^x=\{q^x_1,...,q^x_{l_x}\}$ is the search queries observed in the $x$-th round.
For the search queries observed from the $\eta$ rounds, we define the leakage patterns, AP, FVP, VP, and SP. 

\begin{definition}[Leakage patterns]
The leakage patterns, AP, FVP, VP, and SP, of $Q=\{Q^1,...,Q^\eta\}$ are:
\\
$\bullet$ The \textbf{AP} of $Q$ is the family of functions $AP:EDB\times Q\rightarrow AP$, where $EDB$ is the encrypted database. The $AP=\{AP^1,...,AP^\eta\}$, where $AP^x=\{DB^x(q^x_1),...,DB^x(q^x_{l_x})\}$ contains all the file identities of the response of each query in $Q^x$. 
\\
$\bullet$ The \textbf{FVP} of $Q$ is the family of functions $FVP:EDB\times Q\rightarrow FVP$. The $FVP=\{FVP^1,...,FVP^\eta\}$, where $FVP^x=$$\{\{\#d_i|i\in DB^x(q^x_1)\}$,...,$\{\#d_i\\|i\in DB^x(q^x_{l_x})\}\}$ contains all the file sizes of the response of the queries in $Q^x$. 
\\
$\bullet$ The \textbf{VP} of $Q$ is the family of functions $VP:EDB\times Q\rightarrow VP$. The $VP=\{VP^1,...,VP^\eta\}$, where $VP^x=\{|DB^x(q_1^x)|,...,|DB^x(q_{l_x}^x)|\}$. 
\\
$\bullet$ The \textbf{SP} of $Q$ is the family of functions $SP:EDB\times Q\rightarrow SP$. The $SP=\{G_1,...,G_n\}$ is a partition of the set $Q^1\cup...\cup Q^\eta$, where the search queries in the $G_i$ are for the same keyword, and the search queries of $G_i$ and $G_j$ ($i\neq j$) are for two different keywords.
\end{definition}

We separate the SP of $Q$ into the internal SP (ISP) and the external SP (ESP). 
The ISP indicates whether two search queries from the same round are for the same keyword, while the ESP focuses on the different rounds, i.e., whether two queries from the different rounds are associated with the same keyword. 

\begin{definition}[Internal and external SP]
The SP of $Q$ consists of two parts:
\\
$\bullet$ The \textbf{Internal SP} of the $i$-th round is the family of functions $ISP: EDB\times Q^i\rightarrow ISP^i$. The $ISP^i$ is 
    a partition $ISP^i=\{G^i_1,...,G^i_n\}$ of $Q^i$, where each group $G^i_j$ contains the queries for the same keyword, and different groups contain queries for different keywords.
    \\
$\bullet$ The \textbf{External SP} between $ISP^i$ and $ISP^j$ is the family of functions $ESP:EDB\times(ISP^i,ISP^j)\rightarrow \mathbf{ESP}^{i,j}$. The $\mathbf{ESP}^{i,j}$ is a binary matrix with the size $|ISP^i|\times|ISP^j|$, where $\mathbf{ESP}^{i,j}[x][y]=1$ indicates the search queries of the $x$-th group of $ISP^i$ and the $y$-th group of $ISP^j$ are for the same keyword, otherwise, $\mathbf{ESP}^{i,j}[x][y]=0$.    
\end{definition}

{\noindent\textbf{The shuffling of database between two successive rounds}}. Between two successive rounds $x$ and $x+1$, the database is shuffled and the files have also been updated during the offline interval between them. 
For the {AP}, suppose in the $x$-th round, the attacker observes that the $i$-th encrypted file is returned to a query; in round $x+1$, the client queries the same keyword again, and the same file is responded. 
But, due to updates, the attacker cannot recognize the file as the same one.
As a result, it re-maps this file in ($x+1$)-th round as the $j$-th encrypted file. 
We here denote the AP leakage as $AP^x=\{h^x(DB(q_1^x),...,h^x(DB(q_{l_x}^x)))\}$ and $AP^{x+1}=\{h^{x+1}(DB(q_1^{x+1}),...,h^{x+1}(DB(q_{l_{x+1}}^{x+1})))\}$, where $h^x$ and $h^{x+1}$ are distinct mappings of file identifies. 
Similarly, with FVP, updates that alter file sizes prevent the attacker from matching files across rounds. 
Moreover, additions and deletions on files make the patterns from two rounds more distorted, making it more difficult to correlate patterns across different observation periods.   

\noindent\textbf{Differences between the persistent attacker and the IOA.}
We note that a persistent attacker can execute $(1,\boldsymbol{\sigma},\boldsymbol{\varsigma})$ intermittent observation, which entails observing a single round for a long duration. 
Since the database undergoes continuous updates, $DB^x(q_i)$ differs from $DB^x(q_j)$ even if $q_i$ and $q_j$ target the same keyword $k$. 
Despite these variations, the persistent attacker can track the updates over time to identify consistent patterns. 
Successive queries for the same keyword typically produce similar leakage so that they are easy to match \cite{DBLP:conf/ccs/XuZXYW23}.

In contrast, when $\eta>1$, an attacker with  $(\eta,\boldsymbol{\sigma},\boldsymbol{\varsigma})$ intermittent observation faces greater difficulty,  
due to fragmented and incomplete leakage. 
As previously mentioned, the response to the same keyword can vary significantly in two rounds, i.e., $DB^x$ and $DB^{x+1}$ for the same keyword $k$ could be entirely different. 

In summary, the persistent attacker requires a continuous and full observation for query recovery (i.e., a sufficiently large $\boldsymbol{\sigma}[1]$) but runs a high risk of being detected. 
On the other hand, the IOA, with intermittent observation, is more practical but gains limited access to leakage. 
Our ultimate goal is to find a way to effectively merge partial leakage from different rounds to enable practical and accurate query recovery under intermittent observation.

\noindent\textbf{The Peekaboo attacker} 
is an IOA.
We assume the Peekaboo attacker knows either AP or FVP, as discussed in Section \ref{sec:scenarios}. 
The attacker can trivially infer the VP. 
We also say the attacker has no prior knowledge of SP.
In existing similar-data attacks, the attacker requires auxiliary knowledge, such as the search frequency of keywords and a similar auxiliary dataset, to recover the search queries. 
We also allow such auxiliary knowledge for the Peekaboo attacker. Formally, we define the Peekaboo attacker as follows.

\begin{definition}[Peekaboo Attacker]
    The Peekaboo attacker is a $(\eta,\boldsymbol{\sigma},\boldsymbol{\varsigma})$ intermittent observation attacker with the $AP$ or $FVP$ of $Q=\{Q^1,...,Q^\eta\}$ as its observed leakage. The Peekaboo attacker also possesses a search frequency $\mathbf{F}_s$ of a keyword set $W_s$ and an auxiliary dataset $DB_s=\{DB_s^1,...,DB_s^\tau\}$, where $\tau=\sum_{i\in [\eta]}\boldsymbol{\sigma}[i]$, and $DB_s^i$ is the auxiliary dataset at the $i$-th time slot.
\end{definition}

\noindent \textbf{Attacker's Target.} The attacker utilizes the observed leakage and attempts to infer an $\widetilde{SP}$ from the $FVP$ or $AP$, making the $\widetilde{SP}$ as similar to the $SP$ as possible.} 
The attacker then utilizes the inferred $\widetilde{SP}$ along with other leakage and the auxiliary knowledge to recover the underlying keywords of search queries.

\subsection{Overview of Peekaboo}

Our proposed attack consists of search pattern inference (P1) and query recovery (P2). 
Note that we illustrate Peekaboo in Figure \ref{fig:Peekaboo Model}.

\noindent\textbf{Search pattern inference.} In P1, Peekaboo first infers the internal SP. 
In each round $i$, the attacker utilizes the $AP^i$ or the $FVP^i$ to group the search queries for the same keyword. For a new search query, it calculates the similarity of the new search query with the last search query in each group by the $AP^i$ or the $FVP^i$.  
As we assume the observing time of a round lasts shortly, the changes of the database are limited. 
Thus, the similarity relying on the $AP^i$ or $FVP^i$ remains accurate.
The search query then joins the group with the largest similarity that exceeds a threshold; otherwise, the search query forms a new group.

The attacker then merges the groups from each round to infer the external SP.
Since the database may change significantly between two rounds, the similarity between queries from different rounds could be heavily distorted by noise.  
Thus, the method of inferring the internal SP is not applicable to the external SP.
To infer the external SP, the attacker calculates the co-occurrence matrix of the groups in each round, relying on the internal SP. 
The problem of inferring the external SP is then converted to 
quadratic assignment problems to match the co-occurrence matrix of different rounds. 
We iterate the matching between groups of all rounds to infer all of the external SP.
If the co-occurrence relation remains across rounds, updates or even re-encryption of the database \cite{DBLP:conf/ndss/Chen0PLS0L23} cannot influence the results. 
Finally, the first part of Peekaboo merges the matched groups and produces the SP. 

\noindent\textbf{Query recovery.} In P2, the attacker utilizes the merged groups of P1, the observed leakage, and the auxiliary knowledge to recover search queries.
For each merged group outputted by P1, the attacker counts the queries in the group in each time slot to get the search frequency. 
It also records the maximum volume of the queries in each time slot in the same group. 
The attacker realigns the FVP or AP leakage of groups in every round according to the groups. 
Based on the FVP or AP leakage, the attacker calculates the co-occurrence matrix of the groups in every round. 

Existing \emph{similar-data} attacks can utilize the above information (with minor adaptations) to recover the keyword for each merged group. 
We show how to instantiate Peekaboo for two recent SOTA attacks, Sap \cite{DBLP:conf/uss/OyaK21} and Jigsaw \cite{DBLP:conf/uss/Nie00ZYL24}, as Sap+ and Jigsaw+. 
Rather than assuming a static database, Sap+ and Jigsaw+ target the DSSE. 
We note that the Sap attack does not rely on co-occurrence information, and the Jigsaw attack has a higher accuracy (similarly in SAP+ and Jigsaw+). 
Therefore, one may implement either of them depending on the attack scenario.
Other \emph{similar-data} attacks can also be instantiated similarly. 
With modifications, it is also possible for one to implement the instantiations of Peekaboo on previous \emph{known-data} attacks.  
We provide discussions in Appendix \ref{app:peekaboo with known-data attack}. 


\section{Peekaboo: Search Pattern Inference}
In most passive attacks \cite{DBLP:conf/uss/OyaK21,DBLP:conf/uss/OyaK22,DBLP:conf/uss/Damie0P21,DBLP:conf/ccs/PouliotW16,DBLP:journals/isci/LiuZWT14,DBLP:conf/uss/Nie00ZYL24,DBLP:conf/ccs/NingHPYL0D21,DBLP:conf/ndss/IslamKK12,DBLP:conf/ccs/CashGPR15,DBLP:conf/sp/GuiPP23}, the attacker relies on the SP (indicating whether two search queries are for the same keyword) to recover search queries. 
As shown in Section \ref{sec:scenarios}, the SP remains concealed in DSSE with Type I BP-security and is not available to the IOAs in the communication channel. 
Under intermittent observation, the external SP always remains concealed from the attacker.
We propose the first part (P1) of Peekaboo, inferring the SP from the FVP ( indicating the size of each file in the response) or AP (indicating the identities of the encrypted files in the response), with intermittent observations. 
In Sections \ref{sec:internal match sec} and \ref{sec:external match sec}, we demonstrate how queries are matched within a single round to infer the internal SP and across multiple rounds to infer the external SP.

\subsection{Inference of Internal SP}
\label{sec:internal match sec}
We here infer the internal SP by the FVP or AP leakage.
To decide whether two queries, the $i$-th and $j$-th query in $Q^x$, are for the same keyword, Xu et al. \cite{DBLP:conf/ccs/XuZXYW23} use the FVP. 
Specifically, they calculate the response similarity between two queries:
\begin{equation}
{ rsp(q^x_{i},q^x_{j}) = |(FVP^x_{i}\Cap FVP^x_{j})|/|(FVP^x_{i}\Cup FVP^x_{j})|}.
\label{eq:rsp_fvp}
\end{equation}
Given two collections $X = \{x_1,...,x_m\}$ and
$Y = \{y_1,...,y_n\}$, they define the intersection $Z =X\Cap Y$ as the collection of elements, including duplication, appearing in both $X$ and $Y$. 
For instance, given $X = \{1, 1, 2, 2\}$ and $Y = \{1, 2, 2\}$, then their intersection is $Z = X \Cap Y = \{1, 2, 2\}$. 
Likewise, their union is defined as $W = X \Cup Y = \{1, 1, 2, 2\}$.
Then, they calculate $qeq$ as follows, where $qeq(q^x_{i},q^x_{j})=1$ represents queries $Q^x_{i}$ and $Q^x_{j}$ are for the same keyword and $qeq(q^x_{i},q^x_{j})=0$ otherwise.
\begin{equation}
\label{eq:qeq}
{
 qeq(q^x_{i},q^x_{j})=\left\{
\begin{aligned}
& 1 &,rsp(q^x_{i},q^x_{j})\geq \delta \\
& 0 &,rsp(q^x_{i},q^x_{j})< \delta
\end{aligned}
\right.
}
\end{equation}

We use the same $rsp$ and $qeq$ in the FVP scenario. 
For the AP scenario, we define the $rsp$ as:
\begin{equation}
{ rsp(q^x_{i},q^x_{j}) = |(AP^x_{i}\cap AP^x_{j})|/|(AP^x_{i}\cup AP^x_{j})|}.
\label{eq:rsp_ap}
\end{equation}
Based on this, we describe the inference of internal SP in Algorithm \ref{alg:dsps}. 
We group the queries for the same keywords, and $Gs^x$ contains all the groups. 
When a new query $q$ is observed, the attacker extracts the last query $q_{end}$ in each group and checks if the two queries are under the same keyword (line \ref{alg1:linecalqeq s}-\ref{alg1:linecalqeq e}). 
If the query $q$ matches other queries, the attacker includes it to the end of the group where the last query $q_{end}$ of this group has the largest $rsp(q,q_{end})$ (line \ref{alg1:add to end s}-\ref{alg1:add to end e}).  
Otherwise, the attacker generates a new group containing only $q$ and puts the new group to $Gs^x$ (line \ref{alg1:new group s}-\ref{alg1:new group e}).

We use the last query $q_{end}$ in each group when calculating the $rsp(q,q_{end})$ to minimize the influence of potential updates by the client. 
Note that if queries are from different rounds, the updates between two rounds could be too substantial, and using the same method for them may result in incorrect matches. 

\begin{figure}[!t]
  \begin{algorithm}[H]
    \caption{Inferring the internal SP.}
    \label{alg:dsps}
    \begin{algorithmic}[1]
        \STATE \textbf{Procedure} \textsc{InferISP}($Q^x$) 
             \STATE $Gs^x\gets\emptyset$; 
             \FORALL{$q \in Q^x$}
                \STATE $Cand\gets \emptyset$;
                \STATE $Q_{end}\gets \{G[-1]|G\in Gs^x\}$;\RIGHTCOMMENT{$G[-1]$ is the last element of $G$;}
                \FORALL{$q_{end} \in Q_{end}$} \label{alg1:linecalqeq s}
                    \STATE Add $(rsp(q,q_{end}),q_{end})$ to $Cand$ if $qeq(q,q_{end})$ is $1$;
                \ENDFOR \label{alg1:linecalqeq e}
                \IF{$Cand$ is empty} \label{alg1:new group s}
                    \STATE Add $G=\{q\}$ to $Gs^x$; \label{alg1:new group e}
                \ELSE
                    \STATE Sort $Cand$ in descending order according to $Cand.rsp$;\label{alg1:add to end s}
                    \STATE Extract $(rsp,q_{end})\gets Cand[1]$ and add $q$ to the end of $G$, where $q_{end}\in G$ and $G\in Gs^x$; \label{alg1:add to end e}
                \ENDIF
             \ENDFOR
             \RETURN $Gs^x$;
        \STATE \textbf{End Procedure}
    \end{algorithmic}
  \end{algorithm}
  \Description{Algorithm of Inferring the internal SP, which are fully described in the text.}
\end{figure}

\begin{figure}[!t]
  \begin{algorithm}[H]
    \caption{Inferring the external SP.}
    \label{alg:dspm}
    \begin{algorithmic}[1]
        \STATE \textbf{Procedure} \textsc{InferESP}($\{Gs^1,...\},\{\mathbf{IDH}^1,...\},$$\{\mathbf{IDT}^1,$$...\}$)
        \STATE Initialize $M$ as a set that contains all groups of $\{Gs^1,...\}$;
        \FOR{$i=1$ to $\min(maxlevel,\eta-1)$} \label{alg2:for s}
            \FOR{$j=1$ to $\eta-i$} \label{alg2:for e}
                \STATE $M_{new}\gets$\textsc{Match}$(Gs^j,\mathbf{IDT}^j,Gs^{j+i},\mathbf{IDH}^{j+i})$;\label{alg2:call_match}
                \FOR{$(G_1,G_2)$ in $M_{new}$}
                    \STATE $GMerge_1\gets \textsc{GetMergedGroup}(G_1,M)$;
                    \STATE $GMerge_2\gets \textsc{GetMergedGroup}(G_2,M)$;
                    \IF{$GMerge_1$ and $GMerge_2$ does not contain queries from the same round}
                        \STATE $M.add(GMerge_1 \cup GMerge_2)$;\label{alg2:merge s}
                        \STATE $M.del(GMerge_1)$;
                        \STATE $M.del(GMerge_2)$;\label{alg2:merge e}
                    \ENDIF
                \ENDFOR
            \ENDFOR
        \ENDFOR
        \RETURN $M$
        \STATE \textbf{End Procedure}
        \STATE
        \STATE \textbf{Procedure} \textsc{Match}($Groups_1,\mathbf{ID}_1,Groups_2,\mathbf{ID}_2$)
            \STATE $\mathbf{C}_1\gets \mathbf{ID}_1\mathbf{ID}^\top_1/|\mathbf{ID}_1[0]|,\mathbf{C}_2\gets \mathbf{ID}_2\mathbf{ID}^\top_2/|\mathbf{ID}_2[0]|$;
            \STATE $P\gets QuadraticAssignment(\mathbf{C}_1,\mathbf{C}_2)$; \RIGHTCOMMENT{Get a match between $Groups_1$ and $Groups_2$ with a quadratic assignment algorithm};
            \STATE Remove some incorrectly matched groups from $\mathbf{P}$ with a ratio of $p_{g}$ according to $\mathbf{C}_1$ and $\mathbf{C}_2$;
            \RETURN Pairs of groups that are matched in $P$;
        \STATE \textbf{End Procedure}
        \STATE
        \STATE \textbf{Procedure} \textsc{GetMergedGroup}($G,M$)
            \FOR {$Group \in M$}
            \IF {$G \subseteq Group$}
                \RETURN $Group$;
            \ENDIF
            \ENDFOR
            \RETURN $\emptyset$;
        \STATE \textbf{End Procedure}
    \end{algorithmic}
  \end{algorithm}
  \Description{Algorithm of Inferring the external SP, which are fully described in the text.}
\end{figure}
\subsection{Inference of External SP and Merge}
\label{sec:external match sec}
We recall that the IOA cannot distinguish whether two encrypted files from two rounds are under the same file with the AP or FVP leakage. 
However, the co-occurrence of queries (i.e., the probability of two queries appearing in the same file) remains across different rounds. 
For instance, the words ``searchable'' and ``encryption'' may frequently show up together in the database both before and after updates. 
We use co-occurrence to match queries across rounds.

First, we use the output of Algorithm \ref{alg:dsps} to get index matrices in one round.
In the $x$-th round, the attack lists all the distinct file identities as $F^x$ (for the FVP scenario, $F^x$ contains the distinct file sizes).
For each $G\in Gs^x$ of the output of Algorithm \ref{alg:dsps}, we can get a binary vector of size $|F^x|$ indicating whether the response of the first query in $G$ contains the files in $F^x$, 1 for yes otherwise 0 (for the FVP scenario, the vector can indicate whether the response of the first query in $G$ contains the file with the size in $F^x$). 
Then, we construct the index matrix $\mathbf{IDH^x}$ of the first query of each group, where each row of $\mathbf{IDH^x}$ corresponds to a $G$ in $Gs^x$ and the row is the binary vector above. 
Similarly, we construct the $\mathbf{IDT^x}$ of the last query of each group. 
We also build the matrix $\mathbf{ID^x}$ for later query recovery, where, in each row $i$, if the response of one of the queries in the group includes the $j$-th file in $F^x$, $\mathbf{ID^x}[i][j]=1$. 

Inferring the external SP (Algorithm \ref{alg:dspm}) takes the groups $\{Gs^1$, ..., $Gs^{\eta}\}$ from the first module and the $\{\mathbf{IDH}^1,...,\mathbf{IDH}^{\eta}\}$ and $\{\mathbf{IDT}^1$, ..., $\mathbf{IDT}^{\eta}\}$ as inputs and outputs the merged groups $M$.
We first initialize the $M$ as a set containing all groups of $Gs^1$,..., and $Gs^\eta$. 
We match groups of any two rounds by iteration (line \ref{alg2:for s}-\ref{alg2:for e}). 
To save running time, we set a $maxlevel$ that limits matching between two rounds when there are many rounds between them. 
When matching groups from round $x$ and $y$ ($x<y$), we extract the $\mathbf{IDT}^x$ and $\mathbf{IDH}^y$ and call the \textsc{Match} (line \ref{alg2:call_match}). 
We note that the files are refreshed between two index matrices from two different rounds.
The \textsc{Match} first constructs the co-occurrence matrix $\mathbf{C}_1=\mathbf{IDT}^x(\mathbf{IDT}^x)^{\top}/|F^x|$ and $\mathbf{C}_2=\mathbf{IDH}^y(\mathbf{IDH}^y)^{\top}/|F^y|$ of $Gs^x$ and $Gs^y$, respectively. 
With $\mathbf{C}_1$ and $\mathbf{C}_2$, we can estimate the mapping $\mathbf{P}$ by
\begin{equation}
\label{eq:mapping} 
{   \mathbf{P}=\argmax\limits_{\mathbf{P}\in \mathcal{P}} \text{Pr}(\mathbf{C}_1|\mathbf{C}_2,\mathbf{P})},
\end{equation}
where $\mathbf{P}$ is a matrix and $\mathbf{P}[i][j]=1$ means the $i$-th group of $Gs^x$ is matched to $j$-th group of $Gs^y$, otherwise $\mathbf{P}[i][j]=0$.
In \cite{DBLP:conf/uss/OyaK22}, 
Oya et al. formalize query recovery using the co-occurrence matrix as a Quadratic Assignment Problem (QAP). 
They aim to match the co-occurrence matrices of keywords and queries,  
and propose an iterative heuristic attack called IHOP.  
Similarly, the problem of finding $\mathbf{P}$ is a QAP and can be solved by existing algorithms \cite{DBLP:conf/uss/OyaK22,DBLP:conf/ndss/IslamKK12,DBLP:conf/ccs/PouliotW16}. 
With some enhancements, we apply IHOP to determine such a mapping. 
Since co-occurrence persists across rounds, database updates do not affect the results. 
As some queries could not appear in both rounds $x$ and $y$ and still participate in the matching, we remove certain incorrectly matched groups from $\mathbf{P}$ by a ratio of $p_g$, according to $\mathbf{C}_1$ and $\mathbf{C}_2$.
The details are in Appendix \ref{app:details about match}.

For any pair of groups $(G_1,G_2)$ in new matches, we call the 
\textsc{GetMergedGroup}
to get the former merged group $GMerge_1$ of $G_1$ from $M$, where  $GMerge_1$ is an element of $M$ that contains $G_1$. 
We also obtain the former merged group $GMerge_2$ of $G_2$ from $M$. 
If both $GMerge_1$ and $GMerge_2$ contain the queries from the same round, indicating a conflict with a previous matching, the current matching will be disregarded.  
For example, in rounds 1, 2, and 3, there is group A in round 1, groups B and C in round 2, and group D in round 3. 
Group A is matched with B, and C is matched with D already. 
If group A now matches D, this conflicts with the previous match. 
Otherwise, we merge the two groups $GMerge_1$ and $GMerge_2$ in $M$ (line \ref{alg2:merge s}-\ref{alg2:merge e}).  
The output $M$ contains all the groups, with each group comprising the queries from all considered rounds corresponding to the same keyword. 

\section{Peekaboo: Query Recovery}
\label{sec:Peekaboo_query_rec}
\begin{figure}
  
  \begin{algorithm}[H]
    \caption{Query recovery.}
    \label{alg:QueryRec}
    \begin{algorithmic}[1]
        \STATE \textbf{Procedure} \textsc{QueryRec}($M$, $\{\mathbf{ID}^1,...\}$, $\{\mathbf{ID}_s^1,...\}$, $\mathbf{F}_s$,$\mathbf{V}_s$) 

        \STATE Initialize $\mathbf{F}_r$ as a $\tau\times |M|$ matrix with all zeros; \label{alg:qr_init_s}
        \STATE Initialize $\mathbf{V}_r$ as a $\tau\times |M|$ matrix with all zeros;
        \FORALL{$k\in [\eta]$}
            \STATE Initialize $\mathbf{ID}_r^k$ as a $|M|\times|\mathbf{ID}^k[0]|$ matrix with all zeros; 
        \ENDFOR \label{alg:qr_init_e}
        \FORALL{$i\in [|M|]$} 
            \STATE $Group\gets$ $M.pop()$;
            \RIGHTCOMMENT{$Group$ contains search queries from all rounds that are matched as the queries for the same keyword};
            \STATE Set the $\tau$ elements in the $i$-th column of $\mathbf{F}_r$ as the numbers of queries of $Group$ in $\tau$ time slots;
            \label{alg:qr_rec_f}
           \STATE Set the $\tau$ elements in the $i$-th column of $\mathbf{V}_r$ as the numbers of queries of $Group$ in $\tau$ time slots;
            \label{alg:qr_rec_v}
            \FORALL{$k\in [\eta]$} \label{alg:qr_realignID_s}
                \STATE Find $j$ so that the $Group$ contains the $j$-th group of $Gs^k$;
                \STATE Set the $i$-th row of $\mathbf{ID}_r^k$ as the $j$-th row of $\mathbf{ID}^k$;
            \ENDFOR \label{alg:qr_realignID_e}
        \ENDFOR
        \STATE Normalize each row of $\mathbf{F}_r$ by dividing the sum of that row;
        \FORALL{$k\in [\eta]$} \label{alg:qr_calC_s}
            \STATE $\mathbf{C}_r^k\gets\mathbf{ID}_r^k(\mathbf{ID}_r^k)^\top/|\mathbf{ID}_r^k[0]|$;
            \STATE $\mathbf{C}_s^k\gets\mathbf{ID}_s^k(\mathbf{ID}_s^k)^\top/|\mathbf{ID}_s^k[0]|$;
        \ENDFOR \label{alg:qr_calC_e}
          
        \STATE {Call $\textsc{Attack+}(\mathbf{F}_r,\mathbf{V}_r,\{\mathbf{C}_r^1,...\},\mathbf{F}_s,\mathbf{V}_s,\{\mathbf{C}_s^1,...\})$ to recover the keyword of each group of $M$;} \label{alg:qr_call_sj}
        \STATE \textbf{End Procedure}
    \end{algorithmic}
  \end{algorithm}
  \Description{Algorithm of Query recovery, which are fully described in the text.}
\end{figure}

We propose the second part (P2) of Peekaboo to use the output of the P1 along with the leakage and the auxiliary knowledge to recover queries, see Figure \ref{fig:Peekaboo Model}.  
The P2 processes the leakage with the $M$ to generate the frequency, volume, and co-occurrence information of groups. 
{
In this part, the attacker can easily call previous similar-data attacks with adaptations to recover the keyword of each group. 
We propose the instantiations, Sap+ and Jigsaw+, based on the SOTA attacks, Sap \cite{DBLP:conf/uss/OyaK21} and Jigsaw \cite{DBLP:conf/uss/Nie00ZYL24}. 
}

\subsection{Attacker's Knowledge}


\noindent \textbf{Attackers’ knowledge derived from auxiliary information.}
From a similar dataset $D_s$, the attacker obtains a similar keyword universe $W_s$, so that it can construct the frequency of each keyword as $\mathbf{F}_s=[\mathbf{f}_s^1,...,\mathbf{f}_s^\tau]$ from public search frequency \cite{DBLP:conf/uss/OyaK21,DBLP:conf/uss/Nie00ZYL24,DBLP:conf/uss/OyaK22,DBLP:conf/ccs/XuZXYW23}, where $\mathbf{f}_s^i$ is a vector of length $|W_s|$ indicating the query frequency of each keyword in $W_s$ in time slot $i$. 
Similarly, the attacker builds the volume of each keyword as $\mathbf{V}_s=[\mathbf{v}_s^1,...,\mathbf{v}_s^\tau]$, where $\mathbf{v}_s^i[j]$ is the max number of files in $D_s$, including the $j$-th keyword in $W_s$ during the time slot $i$. 
The attacker can further construct the index matrix as $\{\mathbf{ID}_s^1,...,\mathbf{ID}_s^\eta\}$ based on $W_s$ and $D_s$ during the observation of each round.

\noindent \textbf{Attackers' knowledge from P1 of Peekaboo.}
The attacker uses the merged groups $M$, the index matrix $\{\mathbf{ID}^1,...,\mathbf{ID}^\eta\}$ corresponding to the groups $\{Gs^1,...,Gs^\eta\}$, and the VP of each search query.

\subsection{Query Recovery with SP}
\label{sec:query recovery}
Recall that in previous similar-data attacks against SSE \cite{DBLP:journals/isci/LiuZWT14,DBLP:conf/ccs/PouliotW16,DBLP:conf/uss/OyaK21,DBLP:conf/uss/Damie0P21,DBLP:conf/uss/OyaK22,DBLP:conf/uss/Nie00ZYL24}, the attacker uses the frequency, volume, and co-occurrence information to recover search queries. 
We process the same knowledge for search queries and keywords, but in a dynamic setting. 
The details are in Algorithm \ref{alg:QueryRec}.

We first initialize the $\mathbf{F}_r$ and $\mathbf{V}_r$ as matrices with size $\tau\times|M|$ to record the search frequency and volume of each group in $M$ in total $\tau$ time slots. 
We set the index matrices of the groups of $M$ in each round as $[\mathbf{ID}_r^1,...,\mathbf{ID}_r^\eta]$, where the matrices are initially zeros (line \ref{alg:qr_init_s}-\ref{alg:qr_init_e}). 
For the $i$-th merged $Group$ in $M$, we count the number of queries in $Group$ of each time slot $k$ and record the number in $\mathbf{F}_r[k][i]$ (line \ref{alg:qr_rec_f}). 
We further mark down the max volume (the number of files in the response) of the queries in $Group$ in each time slot $k$ in $\mathbf{V}_r[k][i]$ (line \ref{alg:qr_rec_v}). 
We realign the rows in $\mathbf{ID}^k$ as $\mathbf{ID}^k_r$ according to the order of merged groups in $M$. 
If the client issues search queries for a keyword in some rounds but not in the $i$-th round, the $\mathbf{ID}^i$ has no records for that keyword. 
We naturally 
keep that row in $\mathbf{ID}_r^i$ with zeros. 
Based on the index matrix of the keywords and the merged groups in $M$, we calculate the co-occurrence matrix. 
For each $k$ of total $\eta$ rounds, we compute $\mathbf{C}_r^k\gets\mathbf{ID}_r^k(\mathbf{ID}_r^k)^\top/|\mathbf{ID}_r^k[0]|$ and
$\mathbf{C}_s^k\gets\mathbf{ID}_s^k(\mathbf{ID}_s^k)^\top/|\mathbf{ID}_s^k[0]|$ (line \ref{alg:qr_calC_s}-\ref{alg:qr_calC_e}).

Peekaboo can instantiate \textsc{Attack+} based on prior similar-data attacks, using the above information, i.e., the frequency, volume, and the co-occurrence of groups and keywords.   
We propose two instantiations in Section\ref{sec:Sap+} and Section \ref{sec:Jigsaw+} and provide a generic idea of instantiations over other attacks in Section \ref{sec:otherattacks}.

\noindent\textbf{About $\eta$ and $\tau$.} 
We use the co-occurrence matrix for $\eta$ rounds, along with the volume and frequency information for $\tau$ time slots.
The volume and frequency vary across slots, and the attacker can observe these changes to acquire additional information. 
While the attacker could use the co-occurrence matrix for $\tau$ time slots, each time slot only involves a subset of the keywords, resulting in 
%
index and the co-occurrence matrices with a high proportion of zeros.

\subsection{Instantiation: Sap+}
\label{sec:Sap+}
We provide a brief review of Sap in Appendix \ref{app:introduction to sap and jigsaw} and refer the reader to \cite{DBLP:conf/uss/OyaK21} for more details. 
%
The Sap attack assumes a static volume of search queries and keywords, using the volume and frequency information to solve a maximum likelihood problem and map search queries to keywords.  
In a DSSE, the database undergoes updates, altering keyword volume.  
Thus the original Sap attack cannot be directly applied to the dynamic volumes. 
In Sap+, we refine the maximum likelihood problem (Appendix \ref{app:introduction to sap and jigsaw}, Equation \ref{eq:sap1}) to
\begin{equation}
\label{eq:sap+ 1}
{\mathbf{P}=\argmax\limits_{\mathbf{P}\in\mathcal{P}} \text{Pr}(\boldsymbol{\rho},\mathbf{F}_r,\mathbf{V}_r,\mathbf{n}_D|\mathbf{F}_s,\mathbf{V}_s,\mathbf{P})},
\end{equation}
where the $\mathbf{V}_r$ records the volume of groups in each time slot instead of the unchanged volume $\mathbf{v}_r$ in Sap and $\mathbf{n}_D$ is the vector of the number of total encrypted files in each time slot. 
Accordingly, we modify the cost matrix in Sap (Appendix \ref{app:introduction to sap and jigsaw}, Equation \ref{eq:cal_Cv}) to 
\begin{equation}
\begin{split}
    \mathbf{C}_v[i][j]= -\sum\limits_{k=1}^{\tau}(\mathbf{n}_D[k]\cdot \mathbf{v}_r^k[j]\cdot\log \mathbf{v}_s^k[i]+ \\
 \mathbf{n}_D[k](1-\mathbf{v}_r^k[j])\cdot \log(1-\mathbf{v}_s^k[i]))
\end{split}
\label{eq:cal_Cv_sap+}
\end{equation}
which summarizes the costs of different time slots.
Like Sap, Sap+ also uses the Hungarian algorithm \cite{kuhn1955hungarian} to find a mapping $\mathbf{P}$ that indicates which keyword matches each group in $M$.

\subsection{Instantiation: Jigsaw+}
\label{sec:Jigsaw+}
We here adapt the Jigsaw attack \cite{DBLP:conf/uss/Nie00ZYL24} to the dynamic scenario, where search frequency, volume, and co-occurrence information change over time. 
Note we review Jigsaw in Appendix \ref{app:introduction to sap and jigsaw}. 

Similar to Sap, Jigsaw uses a static volume of search queries and keywords. 
It also uses static frequency information, i.e., the total search frequency of all time slots and a static co-occurrence relation. 
Thus, we should apply minor adaptation for the instantiation. 
Specifically, we revise the utilization of the total search frequency of queries in Jigsaw to the search frequency of groups in each time slot as $\mathbf{F}_r=\{\mathbf{f}_r^1,...,\mathbf{f}_r^\tau\}$. 
Also, we use the volume of groups in each time slot as $\mathbf{V}_r=\{\mathbf{v}_r^1,...,\mathbf{v}_r^\tau\}$. 
In Jigsaw+, we replace the static co-occurrence in all observed queries with the dynamic co-occurrence of groups as $\{\mathbf{C}_r^1,...,\mathbf{C}_r^\eta\}$. 
At last, the auxiliary information ($\mathbf{f}_s,\mathbf{v}_s,\mathbf{C}_s$) of Jigsaw is extended to the dynamic version ($\{\mathbf{f}_s^1,...,\mathbf{f}_s^\tau\},\{\mathbf{v}_s^1,...,\mathbf{v}_s^\tau\},\{\mathbf{C}_s^1,...,\mathbf{C}_s^\eta\}$). 

To utilize the volume and frequency information of multiple time slots, we modify the differential distance (Appendix \ref{app:introduction to sap and jigsaw}, Equation\ref{eq:differential distance}) of the $i$-th group and the $j$-th group in $M$ as 
\begin{equation}
{
    d_{i}=\min\limits_{j<|M|\land j\ne i}\sum\limits_{k=1}^{\tau}\alpha\cdot|\mathbf{v}_r^k[i]-\mathbf{v}_r^k[j]|+(1-\alpha)|\mathbf{f}_r^k[i]-\mathbf{f}_r^k[j]|},
    \label{eq:differential distance plus}
\end{equation}
which summarizes the distance of all time slots.
Based on this, Jigsaw+ can locate $BaseRec$ distinctive groups instead of distinctive search queries.
Similarly, the distance $s(i,j)$ between the $i$-th group of $M$ and the $j$-th keyword in $W_s$ (Appendix \ref{app:introduction to sap and jigsaw}, Equation\ref{eq:s}) is 
\begin{equation}
{
    s(i,j) = \sum\limits_{k=1}^{\tau}\alpha\cdot|\mathbf{v}_r^k[i]-\mathbf{v}_s^k[j]|+(1-\alpha)|\mathbf{f}_r^k[i]-\mathbf{f}_s^k[j]|}.
\end{equation}

The subsequent confirmation uses the co-occurrence matrix of observations. Instead of using the static co-occurrence matrix, Jigsaw+ extracts the co-occurrence matrix of different rounds as $\{\mathbf{C}_r^1{}',...,\mathbf{C}_r^\eta{}'\}$ of the matched groups and $\{\mathbf{C}_s^1{}',...,\mathbf{C}_s^\eta{}'\}$ of their corresponding keywords based on $\{\mathbf{C}_r^1,...,\mathbf{C}_r^\eta\}$ and $\{\mathbf{C}_s^1,...,\mathbf{C}_s^\eta\}$. 
Then we define the $revconf$ (Appendix \ref{app:introduction to sap and jigsaw}, Equation \ref{eq:revconf}) of the $i$-th recovered distinctive group as $revconf = \sum\limits_{k=1}^{\eta} ||\mathbf{C}_r^{k}{}'[i] - \mathbf{C}_s^{k}{}'[i]||$.

Then, Jigsaw+ extracts the $\{\mathbf{C}_{rs}^1,...,\mathbf{C}_{rs}^\eta\}$ and $\{\mathbf{C}_{ss}^1,...,\mathbf{C}_{ss}^\eta\}$, as in Jigsaw. 
We set the $score$ (Appendix \ref{app:introduction to sap and jigsaw}, Equation \ref{eq:score}) between the $i$-th group of the left unmatched groups and the $j$-th unmatched keyword $j$ as $score =-\ln(\beta\sum\limits_{k=1}^\eta ||\mathbf{C}_{rs}^k[i]-\mathbf{C}_{ss}^k[j]||+(1-\beta)s(i,j))$.

Finally, Jigsaw+ calculates the $certainty$ as the difference between the largest and the second largest $score$, then recovers the top $RefSpeed$ groups with the highest $certainty$. 
Jigsaw+ treats the recovered groups as known matches and repeats the process until all groups are matched to keywords.

\subsection{Instantiations of Other Attacks}
\label{sec:otherattacks}

\noindent\textbf{Similar-data attacks.}
Peekaboo generates the dynamic search frequency, volume, and co-occurrence matrix of merged groups in $M$ and keywords in $W_s$. 
A straightforward way to make instantiations over other similar-data attacks is to average the generated dynamic information to recover search queries.
Another approach is to iterate the calculation in the attacks over multiple rounds and further aggregate the results as in Sap+ and Jigsaw+. 
For example, 
in Jigsaw+, we can compute the differential distances for each round and then summarize them (Equation \ref{eq:differential distance plus}) to replace the differential distance used in Jigsaw.
We also provide an instantiation based on IHOP \cite{DBLP:conf/uss/OyaK22} in Appendix \ref{app:IHOP+}, i.e. IHOP+.
Any future similar-data attacks relying on frequency, volume, or co-occurrence information can adopt this method as an instantiation of Peekaboo.

\noindent\textbf{Known-data attacks.}
We discuss the possibility of instantiating Peekaboo over previous known-data attacks (see Appendix \ref{app:peekaboo with known-data attack}).

\section{Evaluation}
We introduce settings and evaluation metrics, illustrate the performance of Peekaboo, and finally compare the query recovery results with the benchmarks. 
Our code is available in \url{https://github.com/Peekaboo20241115/Peekaboo}.
\subsection{Setup}
\label{sec:setup of exp}
\noindent \textbf{Dataset.} We use the Enron and Lucene datasets in the experiments. 
The Enron email dataset \cite{Enron} has 30,109 emails between 2000-2002, while the Lucene email dataset \cite{Lucene} includes 66,491 emails between 2001-2020, where all these emails are tagged with timestamps indicating the time they were sent. 
We also conduct experiments under another type of dataset of our attacks, i.e., Wikipedia\cite{Wikipedia} (results are in Appendix \ref{app:wiki}).
We use a total of 5,525 keywords extracted from the datasets.  
Among these, we download the daily search trends of 3,000 keywords from PageViews \cite{Pageviews}, covering from July 2019 to July 2024.
The keyword count is consistent with previous works \cite{DBLP:conf/ccs/PouliotW16,DBLP:conf/uss/OyaK21,DBLP:conf/uss/Damie0P21,DBLP:conf/uss/OyaK22}.
This approach is practical and also supported by Zipf's law \cite{zipf2016human,DBLP:conf/uss/Nie00ZYL24}, which suggests that frequently used words are limited in number.  
Practical databases, such as those from  \cite{WHOlist}, often utilize fewer than 3,000 keywords.
We randomly divide the email dataset for each day into two halves, with one half considered as the client's dataset and the other half as a similar dataset known by the attacker. 
In practice, the attacker can obtain such a similar dataset. 
For example, an insider with access to a comparable email database or an industry competitor with a structurally similar database may exploit such resources to facilitate attacks. 
We assume the attacker is given half of the dataset that is a common experimental setting in prior work \cite{DBLP:conf/ccs/PouliotW16,DBLP:conf/uss/OyaK21,DBLP:conf/uss/Damie0P21,DBLP:conf/uss/OyaK22,DBLP:conf/uss/Nie00ZYL24}.

\noindent \textbf{Client.} We use the Enron and Lucene datasets to simulate real-world client behaviors in DSSE. 
We highlight that simulating SSE operations is straightforward: without any updates, there is no distinction between a persistent attacker and an IOA.
The client adds the emails to the encrypted dataset daily according to the timestamps.  
For example, on the $i$-th day of the experiment, the client stores the emails with timestamps from the $i$-th day after the first day of the year 2000 for the Enron dataset. 
After the updates, the client deletes the outdated emails. 
For Enron, we assume the client can delete emails one year prior, while for Lucene, the client deletes the emails three years prior, due to the lower volume of emails per year in Lucene.  
We also assume the client randomly deletes emails from all the stored ones daily to simulate deletion behavior in practice.  
We set the number of randomly deleted emails to be ten percent of the number of newly added emails. 
For each day, the client can issue multiple queries for the extracted keywords according to the search frequency from PageViews.

\noindent \textbf{Attacker.} The attacker updates the dataset and deletes outdated emails to minimize the distributional differences between the attacker's and the client's datasets.
We state that Peekaboo can also apply to the case that there is no update, i.e., the attacker only obtains a static dataset (see Appendix \ref{app:outdated data exp}).
The attacker does not imitate the client's random deletion behavior, as it is unlikely that the attacker would predict which files the client will delete.  
Thus, its deletion cannot contribute to the similarity of the datasets.  
We also assume the attacker can access the extracted keywords and obtain the corresponding search frequency, i.e., the true query frequencies of PageViews in our experiments. 
We note that if the instantiated Attack does not require query frequency, the same holds for Attack+. We provide the results of Jigsaw+ without query frequency in Appendix \ref{app:without_frequency} accordingly.
It thus can generate the $\mathbf{F}_s$, $\mathbf{V}_s$, and $\{\mathbf{ID}^1_s,...,\mathbf{ID}^\eta_s\}$ of the extracted keywords based on the similar dataset and search frequency.
We specifically restrict the attacker to intermittent observations, where it attacks over $\eta$ rounds. 
In the $i$-th round, the attacker observes the leakage for $\boldsymbol{\sigma}[i]$ days, acquiring the search query sequence $Q^i$ and the corresponding leakage, either $AP^i$ or $FVP^i$. 
After that, it goes offline.   
We repeat the above strategy for subsequent rounds. 

\noindent \textbf{Parameter selection.} Peekaboo introduces three hyperparameters, i.e., $\delta$, $maxlevel$, and $p_g$. 
We provide the experimental results and analyses of these hyperparameters in Appendix \ref{app:parameter selection}.
Similar to FMA\cite{DBLP:conf/ccs/XuZXYW23}, $\delta$ is the threshold that reflects the changes in the responded files between successive queries of the same keyword within one round and is recommended to be larger than 0.5 \cite{DBLP:conf/ccs/XuZXYW23}. 
Results (Figure \ref{fig:test_delta}, Appendix \ref{app:parameter selection}) show that it is best to choose a value between 0.6 and 0.95. 
$maxlevel$ is used to reduce the runtime cost. A smaller $maxlevel$ can result in shorter runtime with slightly decreased accuracy. 
From Figure \ref{fig:test_maxlevel}, we see that Jigsaw+ reaches near-maximum accuracy when $maxlevel$ is around 5.
$p_g$ is used to discard the mistakes in matching. 
Experiments in Figures \ref{fig:test_p_g_AP} and \ref{fig:test_p_g_FVP} support selecting a value between 0.05 and 0.2. 
If more mistakes are expected (e.g., using only FVP or against countermeasures), $p_g$ should be set slightly higher. 
Unless otherwise specified, we set $\delta = 0.95$, $maxlevel = 5$, and $p_g = 0.05$. 
Other parameters of Attack+, such as $\alpha$ and $\beta$ in Jigsaw+, are introduced by the original works of the corresponding Attack, and we follow the recommendations provided in the corresponding references.

\noindent \textbf{Evaluation metrics.} We use the \textit{adjusted rand index} (ARI) \cite{hubert1985comparing} for the evaluation of SP inference, which is based on the rand index (RI) \cite{rand1971objective}. 
The RI is a measure of the similarity between two partitions for a set $S$ containing $n$ elements. Concretely, given two partitions $X=\{X_1,...,X_m\}$ and $Y=\{Y_1,...,Y_k\}$, the RI can be calculated as $RI = (a+b)/ \binom{n}{2}$,
where $a$ is the number of times a pair of elements belongs to the same partitions across $X$ and $Y$, and $b$ is the number of times a pair of elements belongs to the different partitions across $X$ and $Y$. 
The ARI is the ``adjusted-for-chance'' version of RI, which is $ARI = (RI - ExpectedRI)/(MaxRI - ExpectedRI)$, 
where $ExpectedRI$ is the expected value of $RI$, and $MaxRI$ is the value of $RI$ in the most ideal partition scenario (always equal to 1).
This value lies between $-1$ and $1$. 
The closer the ARI approaches $1$, the more similar the two partitions are, and vice versa. 
For the evaluation of the SP inference, we group all the queries of the same keyword as partition $X$ and treat the output of the P1 as $Y$.

We use the \textit{accuracy} to evaluate the performance of the query recovery attacks. The accuracy is calculated as the proportion of correctly recovered search queries to the total observed queries. 

\noindent\textbf{Ethical concerns}. All datasets used in our experiments are publicly available and commonly used in previous studies. We provided the discussion of ethical considerations in Appendix \ref{app:ethical consideration}.

\subsection{Evaluation of SP Inferring}

\label{sec:peekaboo_exp}
\begin{figure}
    \centering
    \subfloat[Enron]
	{
 \label{fig:test_peekaboo_1_enron}
		\begin{minipage}{.44\linewidth}
			\centering
                \includegraphics[width=\linewidth]{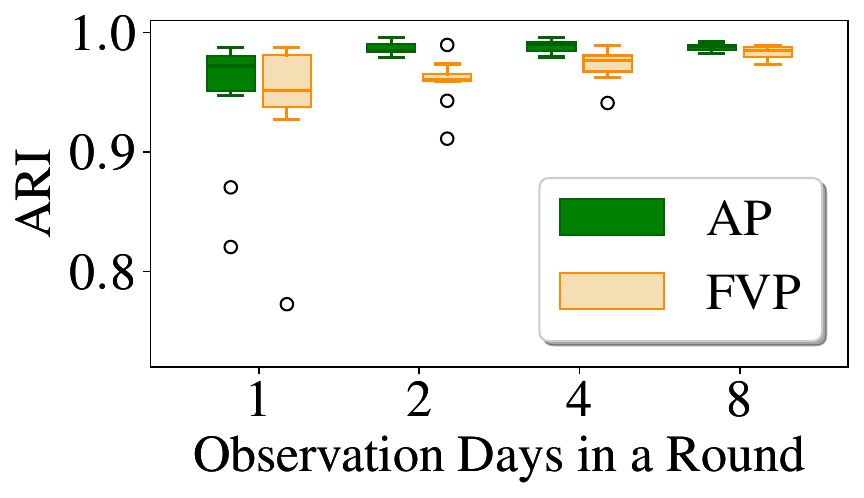}
		\end{minipage}
	}
    \subfloat[Lucene]
	{
 \label{fig:test_peekaboo_1_lucene}
		\begin{minipage}{.44\linewidth}
			\centering
			\includegraphics[width=\linewidth]{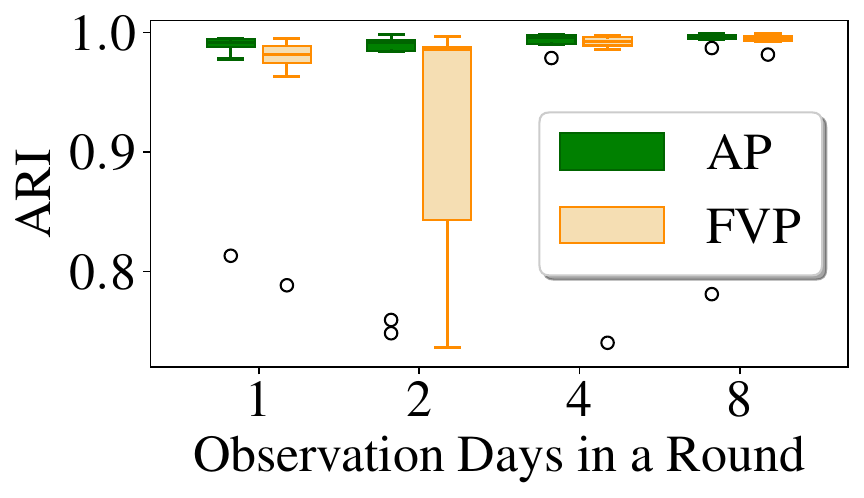}
		\end{minipage}
	}
	\caption{The ARI of the search pattern inferring of Peekaboo with different observation times in each round.}
	\label{fig:test_peekaboo_1}
    \Description{The figures present the ARI results of Peekaboo, which are described in the text.}
\end{figure}

\begin{figure}
    \centering
    \subfloat[Enron]
	{
 \label{fig:test_peekaboo_2_enron}
		\begin{minipage}{.42\linewidth}
			\centering
                \includegraphics[width=\linewidth]{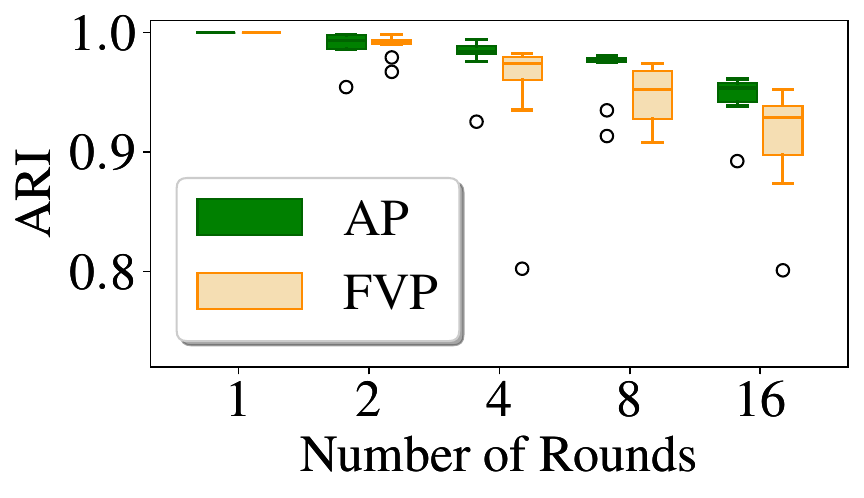}
		\end{minipage}
	}
    \subfloat[Lucene]
	{
 \label{fig:test_peekaboo_2_lucene}
		\begin{minipage}{.42\linewidth}
			\centering
			\includegraphics[width=\linewidth]{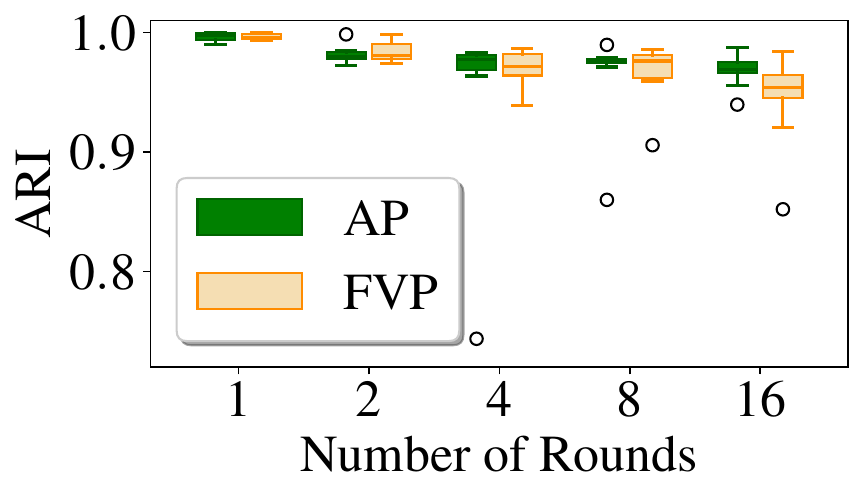}
		\end{minipage}
	}
	\caption{The ARI of the search pattern inferring of Peekaboo with different numbers of rounds.}
	\label{fig:test_peekaboo_2}
    \Description{The ARI results of the search pattern inferring of Peekaboo, which are fully described in the text.}
\end{figure}
\begin{figure*}[htp]
    \centering
            \begin{minipage}{\linewidth}
                \centering
                \begin{tikzpicture}
        \node[draw=gray!50, dashed, rectangle, rounded corners=0pt, thick, inner sep=-1pt] {  
\begin{tabular}{c@{\hskip 4pt}c@{\hskip 4pt}c@{\hskip 4pt}c@{\hskip 4pt}c}
    \begin{tikzpicture}
        \fill[red] (0pt, 0pt) rectangle (6pt, 4pt);
    \end{tikzpicture} 
    \begin{scriptsize}
        FMA
    \end{scriptsize}
    &
    \begin{tikzpicture}
        \fill[darkorange] (0pt, 0pt) rectangle (6pt, 4pt);
    \end{tikzpicture}
    \begin{scriptsize}
        Jigsaw+
    \end{scriptsize}
    &
    \begin{tikzpicture}
        \fill[darkgreen] (0pt, 0pt) rectangle (6pt, 4pt);
    \end{tikzpicture}
    \begin{scriptsize}
        Jigsaw+ with SP
    \end{scriptsize}
    &
    \begin{tikzpicture}
        \fill[blueviolet] (0pt, 0pt) rectangle (6pt, 4pt);
    \end{tikzpicture} 
    \begin{scriptsize}
        Sap+
    \end{scriptsize}
    &
    \begin{tikzpicture}
        \fill[dodgerblue] (0pt, 0pt) rectangle (6pt, 4pt);
    \end{tikzpicture}  
    \begin{scriptsize}
        Sap+ with SP
    \end{scriptsize}

\end{tabular}
};
\end{tikzpicture}
            \end{minipage}
    \subfloat[Enron, AP]
	{
 \label{fig:test_comparison_enron}
		\begin{minipage}{.23\linewidth}
			\centering
                \includegraphics[width=\linewidth]{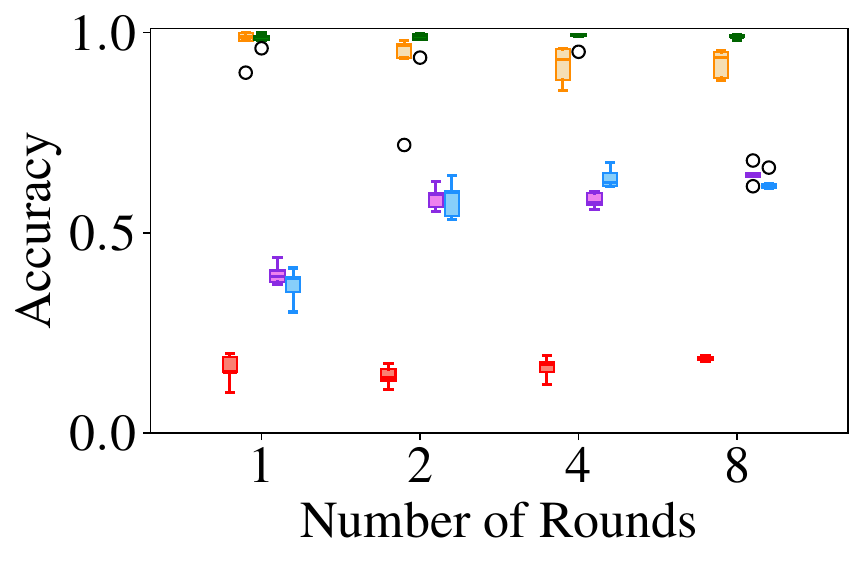}
		\end{minipage}
	}
    \subfloat[Lucene, AP]
	{
 \label{fig:test_comparison_lucene}
		\begin{minipage}{.23\linewidth}
			\centering
			\includegraphics[width=\linewidth]{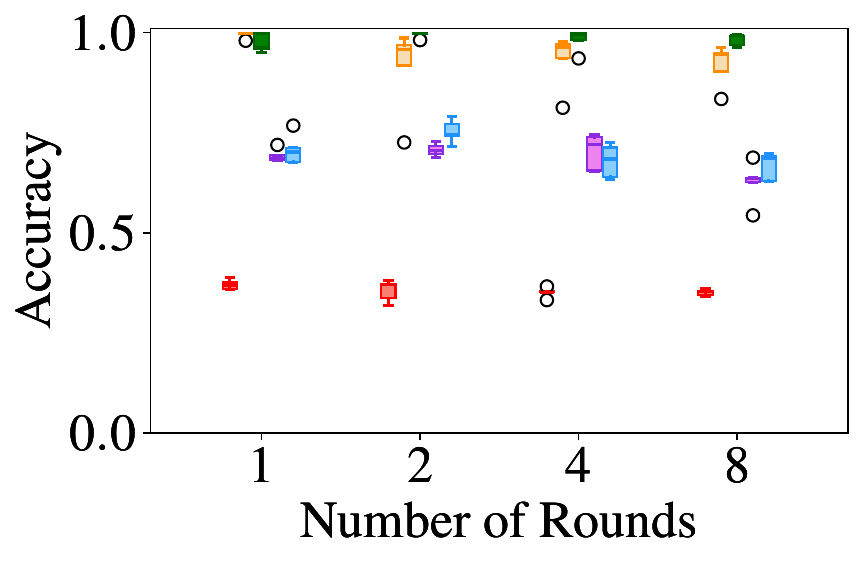}
		\end{minipage}
 	}
    \subfloat[Enron, FVP]
	{
 \label{fig:test_comparison_enron_fvp}
		\begin{minipage}{.23\linewidth}
			\centering
                \includegraphics[width=\linewidth]{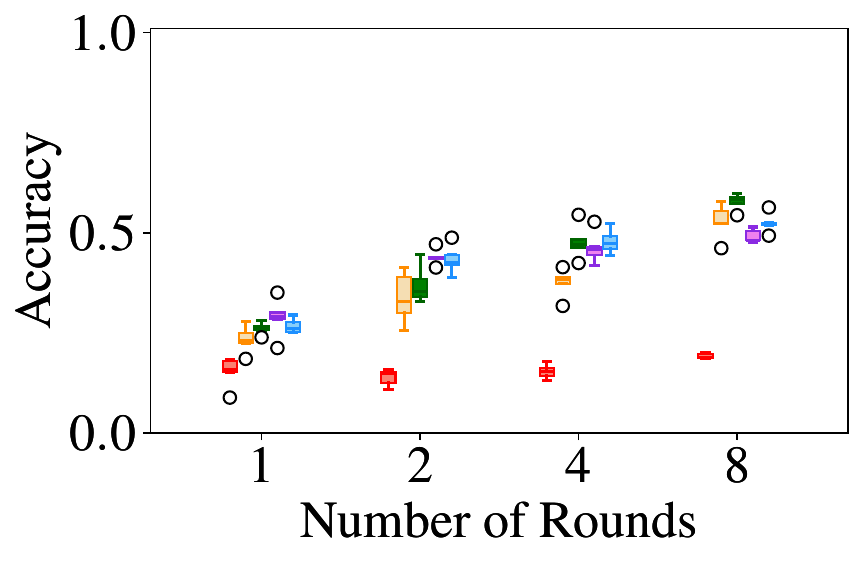}
		\end{minipage}
	}
    \subfloat[Lucene, FVP]
	{
 \label{fig:test_comparison_lucene_fvp}
		\begin{minipage}{.23\linewidth}
			\centering
			\includegraphics[width=\linewidth]{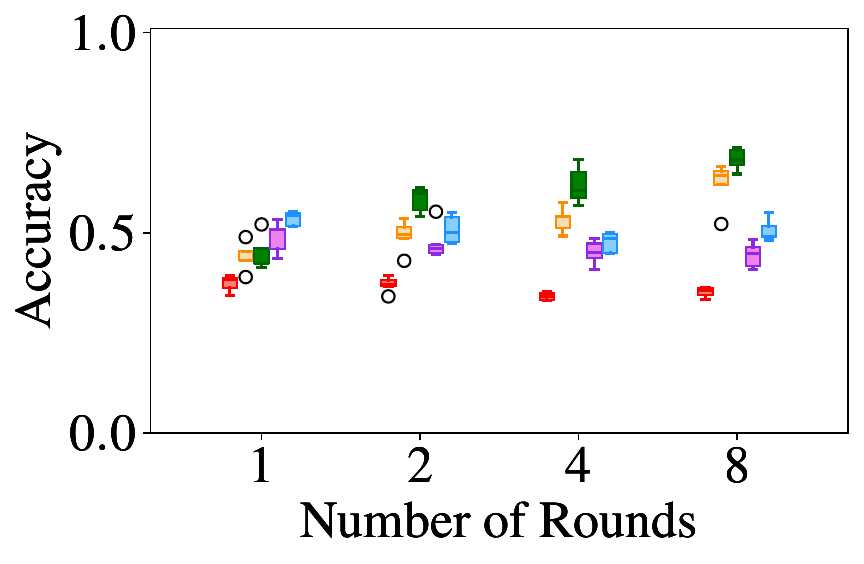}
		\end{minipage}
 	}

	\caption{The accuracy results of Jigsaw+, Sap+, FMA, Jigsaw+ with SP, and Sap+ with SP in Enron and Lucene with different numbers of rounds with the AP or FVP leakage.}
	\label{fig:test_comparison}
    \Description{The figures present the impact of different round numbers for the recovery accuracy, which are fully described in the text.}
\end{figure*}

\begin{figure*}[htp]
    \centering
            \begin{minipage}{\linewidth}
                \centering
                \begin{tikzpicture}
        \node[draw=gray!50, dashed, rectangle, rounded corners=0pt, thick, inner sep=-1pt] {  
\begin{tabular}{c@{\hskip 4pt}c@{\hskip 4pt}c@{\hskip 4pt}c@{\hskip 4pt}c}
    \begin{tikzpicture}
        \fill[red] (0pt, 0pt) rectangle (6pt, 4pt);
    \end{tikzpicture} 
    \begin{scriptsize}
        FMA
    \end{scriptsize}
    &
    \begin{tikzpicture}
        \fill[darkorange] (0pt, 0pt) rectangle (6pt, 4pt);
    \end{tikzpicture}
    \begin{scriptsize}
        Jigsaw+
    \end{scriptsize}
    &
    \begin{tikzpicture}
        \fill[darkgreen] (0pt, 0pt) rectangle (6pt, 4pt);
    \end{tikzpicture}
    \begin{scriptsize}
        Jigsaw+ with SP
    \end{scriptsize}
    &
    \begin{tikzpicture}
        \fill[blueviolet] (0pt, 0pt) rectangle (6pt, 4pt);
    \end{tikzpicture} 
    \begin{scriptsize}
        Sap+
    \end{scriptsize}
    &
    \begin{tikzpicture}
        \fill[dodgerblue] (0pt, 0pt) rectangle (6pt, 4pt);
    \end{tikzpicture}  
    \begin{scriptsize}
        Sap+ with SP
    \end{scriptsize}

\end{tabular}
};
\end{tikzpicture}
            \end{minipage}
    \subfloat[Enron, AP]
	{
 \label{fig:test_random_enron}
		\begin{minipage}{.23\linewidth}
			\centering
                \includegraphics[width=\linewidth]{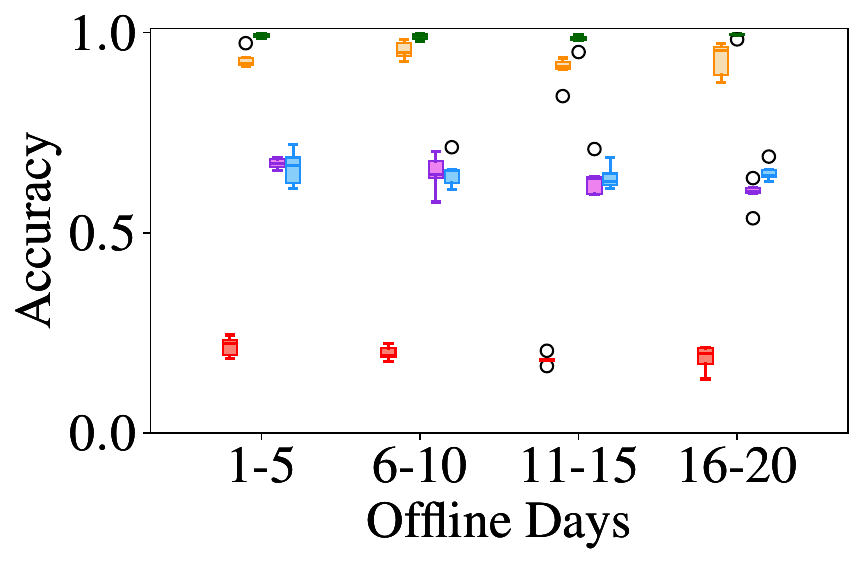}
		\end{minipage}
	}
    \subfloat[Lucene, AP]
	{
 \label{fig:test_random_lucene}
		\begin{minipage}{.23\linewidth}
			\centering
			\includegraphics[width=\linewidth]{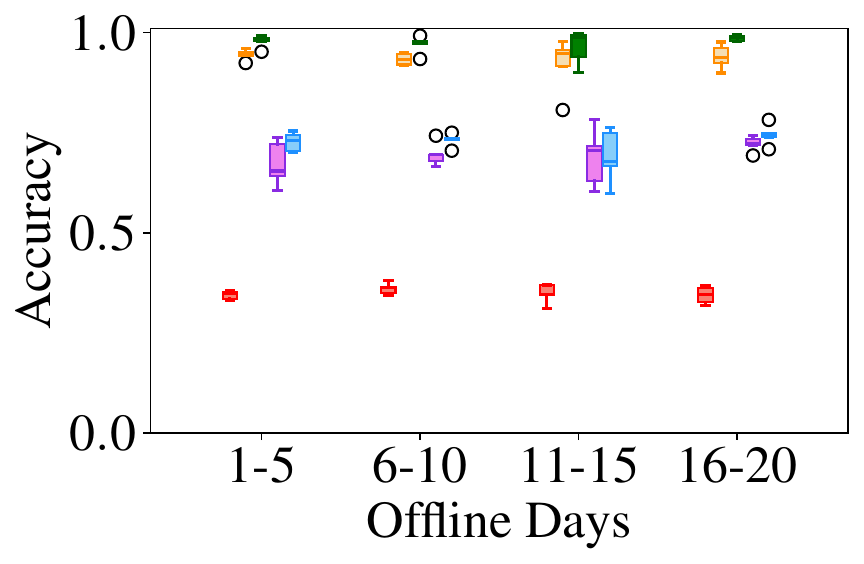}
		\end{minipage}
 	}
    \subfloat[Enron, FVP]
	{
 \label{fig:test_random_enron_fvp}
		\begin{minipage}{.23\linewidth}
			\centering
                \includegraphics[width=\linewidth]{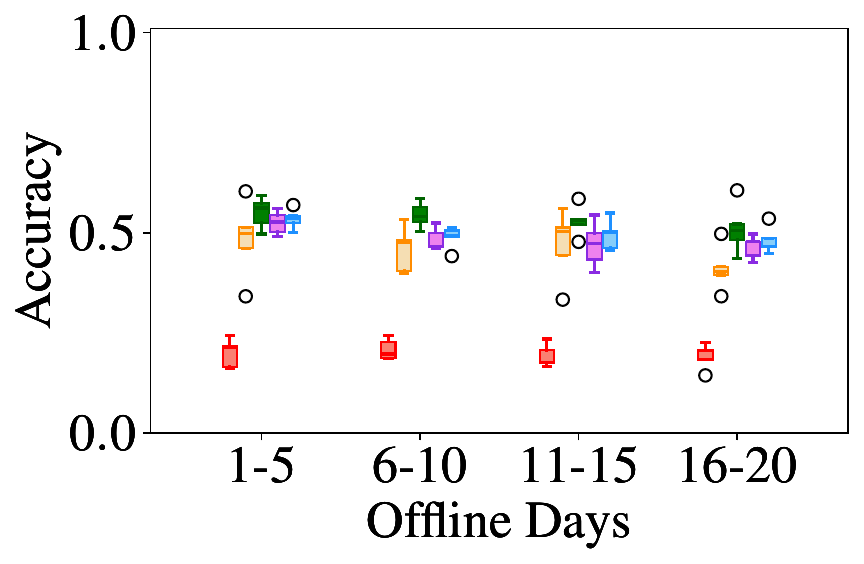}
		\end{minipage}
	}
    \subfloat[Lucene, FVP]
	{
 \label{fig:test_random_lucene_fvp}
		\begin{minipage}{.23\linewidth}
			\centering
			\includegraphics[width=\linewidth]{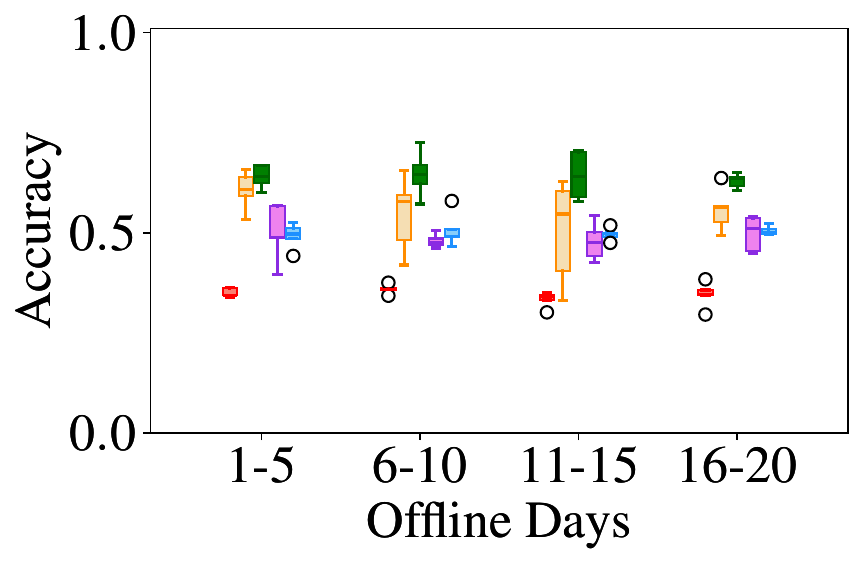}
		\end{minipage}
 	}
	\caption{The accuracy results of Jigsaw+, Sap+, FMA, Jigsaw+ with SP, and Sap+ with SP in Enron and Lucene with different offline days in each round with the AP or FVP leakage.}
	\label{fig:test_random}
    \Description{The figures present the impact of offline days for the recovery accuracy, which are fully described in the text.}
\end{figure*}

\begin{figure*}[htp]
    \centering
    \begin{minipage}{\linewidth}
        \centering
        \begin{tikzpicture}
        \node[draw=gray!50, dashed, rectangle, rounded corners=0pt, thick, inner sep=-1pt] {  
\begin{tabular}{c@{\hskip 4pt}c@{\hskip 4pt}c@{\hskip 4pt}c@{\hskip 4pt}c}
    \begin{tikzpicture}
        \fill[red] (0pt, 0pt) rectangle (6pt, 4pt);
    \end{tikzpicture} 
    \begin{scriptsize}
        FMA
    \end{scriptsize}
    &
    \begin{tikzpicture}
        \fill[darkorange] (0pt, 0pt) rectangle (6pt, 4pt);
    \end{tikzpicture}
    \begin{scriptsize}
        Jigsaw+
    \end{scriptsize}
    &
    \begin{tikzpicture}
        \fill[darkgreen] (0pt, 0pt) rectangle (6pt, 4pt);
    \end{tikzpicture}
    \begin{scriptsize}
        Jigsaw+ with SP
    \end{scriptsize}
    &
    \begin{tikzpicture}
        \fill[blueviolet] (0pt, 0pt) rectangle (6pt, 4pt);
    \end{tikzpicture} 
    \begin{scriptsize}
        Sap+
    \end{scriptsize}
    &
    \begin{tikzpicture}
        \fill[dodgerblue] (0pt, 0pt) rectangle (6pt, 4pt);
    \end{tikzpicture}  
    \begin{scriptsize}
        Sap+ with SP
    \end{scriptsize}

\end{tabular}
};
\end{tikzpicture}
    \end{minipage}
    \subfloat[Enron, AP]
	{
 \label{fig:test_query_number_enron}
		\begin{minipage}{.23\linewidth}
			\centering
                \includegraphics[width=\linewidth]{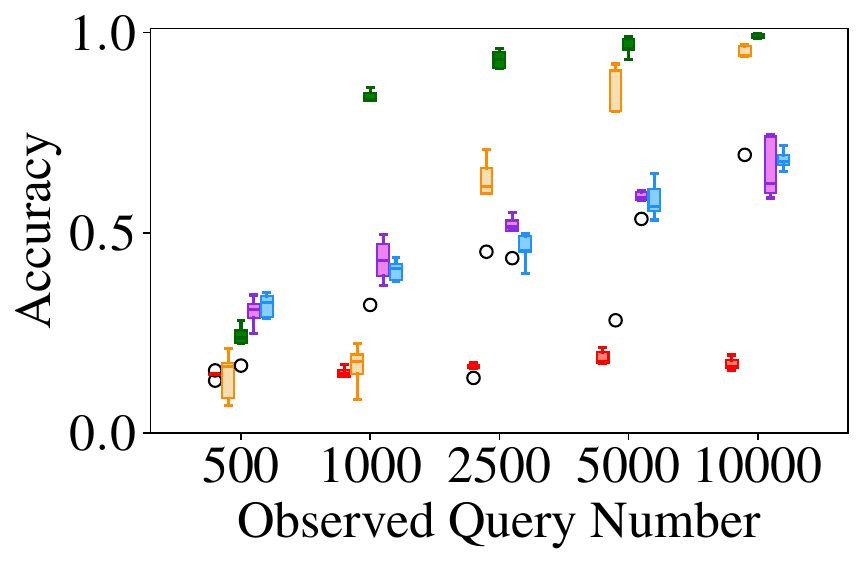}
		\end{minipage}
	}
    \subfloat[Lucene, AP]
	{
 \label{fig:test_query_number_lucene}
		\begin{minipage}{.23\linewidth}
			\centering
			\includegraphics[width=\linewidth]{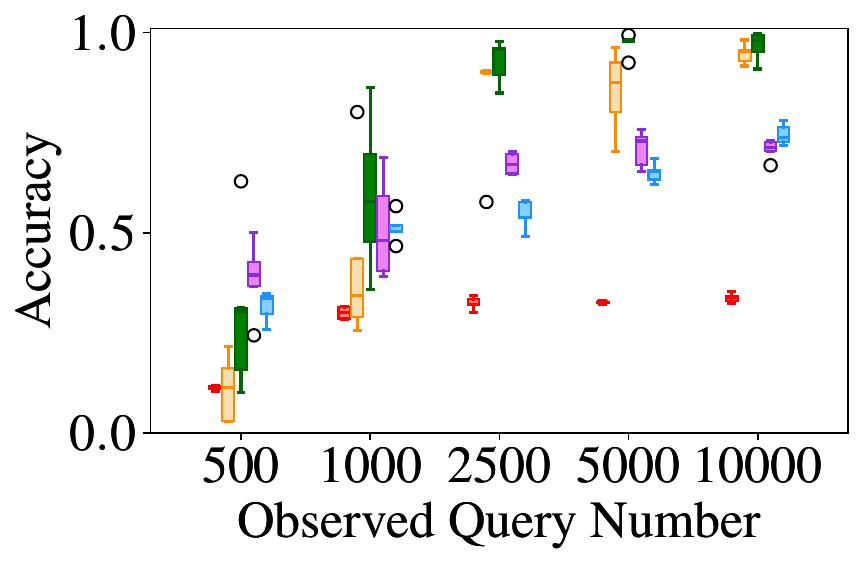}
		\end{minipage}
 	}
    \subfloat[Enron, FVP]
	{
 \label{fig:test_query_number_enron_fvp}
		\begin{minipage}{.23\linewidth}
			\centering
                \includegraphics[width=\linewidth]{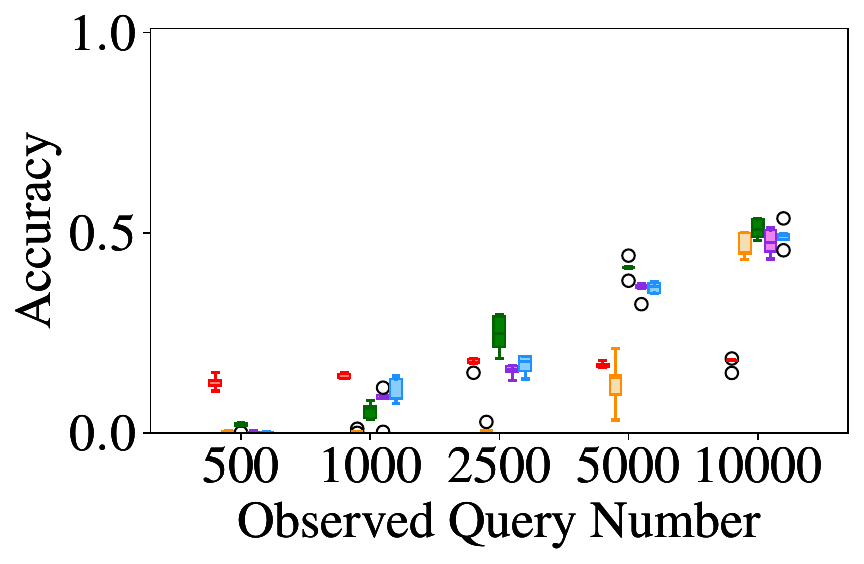}
		\end{minipage}
	}
    \subfloat[Lucene, FVP]
	{
 \label{fig:test_query_number_lucene_fvp}
		\begin{minipage}{.23\linewidth}
			\centering
			\includegraphics[width=\linewidth]{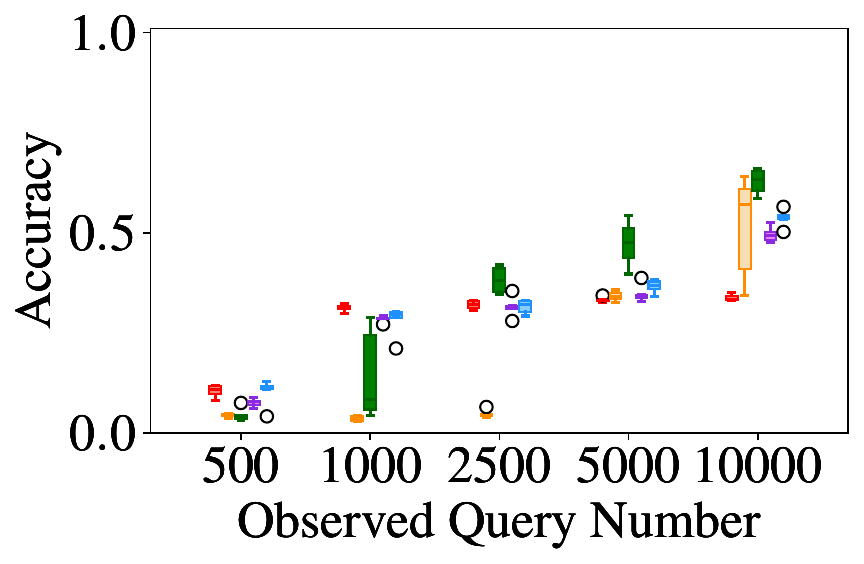}
		\end{minipage}
 	}

	\caption{The accuracy results of Jigsaw+, Sap+, FMA, Jigsaw+ with SP, and Sap+ with SP in Enron and Lucene with different numbers of observed queries in each round with the AP or FVP leakage.}
	\label{fig:test_query_number}
    \Description{The figures present the impact of different numbers of observed search queries, which are fully described in the text.}
\end{figure*}

We use 500 keywords with the biggest volume in each dataset as the keyword universe. 
The client issues 1,000 queries daily. 

\noindent\textbf{The impact of observing days.}
We assume the attack lasts for 5 rounds, with 20 days per round. 
At the beginning of each round, we set that the attacker observes 1, 2, 4, and 8 days, respectively, for the rest of the 20 days, the attacker stays offline.
As shown in Figure \ref{fig:test_peekaboo_1}, the ARI of Peekaboo reaches above 0.9 in most cases, meaning most of the queries are accurately clustered w.r.t. the underlying keywords. 
As the number of days in a round increases, the attacker gains more information in the round to match two groups across rounds, leading to a gradual increase in the ARI. 
The ARI based on AP is higher than that based on FVP. 
Recall that FVP reveals the size of each file. 
But the attacker still cannot distinguish between two files of the same size, which influences the matching process.  

\noindent\textbf{The impact of round number.} 
We set the round number to 1, 2, 4, 8, and 16, and in each round, the attacker observes for 2 days and goes offline for 18 days. 
The results are in Figure \ref{fig:test_peekaboo_2}. 
We see that with an increasing number of rounds, the ARI tends to decrease over time. 
This is so because the errors in earlier matches affect the inference of the external SP in later rounds, causing an accumulation of errors. 
Although the ARI decreases with the increasing number of rounds, it still maintains practical performance, obtaining $>0.9$ even when the round number reaches 16. 
As in Section \ref{sec:peekaboo_exp_2}, the decreasing ARI has minimal impact on query recovery as the round number increases, since the attacker can gain more information about the queries with many extra rounds of observations.

\subsection{Evaluation of Query Recovery}
\label{sec:peekaboo_exp_2}

We provide the comparison among Peekaboo with Jigsaw+ (denoted as Jigsaw+) and Peekaboo with Sap+ (denoted as Sap+). 
Note that since IHOP and Jigsaw exploit the same information and achieve comparable accuracy \cite{DBLP:conf/uss/Nie00ZYL24}, we expect IHOP+ to match Jigsaw+ in performance. 
This is confirmed by the comparison of IHOP+ and Jigsaw+ in Appendix \ref{app:IHOP+}. 
We also present the results of FMA \cite{DBLP:conf/ccs/XuZXYW23}. 
Recall that FMA first recovers the SP of the observed queries using the Equations \ref{eq:rsp_fvp} and \ref{eq:qeq} and then calculates the frequency of queries. 
It treats the keywords with similar frequency of a query as candidates and narrows them down across different time slots. 
If there is only one candidate keyword left, FMA recovers the query.
Besides the FVP leakage, we also test the FMA with AP leakage by replacing Equation \ref{eq:rsp_fvp} of FMA with Equation \ref{eq:rsp_ap}. 
To better understand the inferred SP, we set two benchmark attacks, ``Jigsaw+ with SP'' and ``Sap+ with SP'', where we assume the attacker knows the SP (both the internal and external) instead of inferring the SP, and then utilizes P2 of Peekaboo to recover the queries.

For Jigsaw+ and Jigsaw+ with SP, we set the $\alpha$ to 0.5 and $\beta$ to 0.9. 
For the AP experiments, we set the $BaseRec$ to 25 and $ConfRec$ to 10, while for the FVP experiments, $BaseRec$ is set to 15 and $ConfRec$ to 5. 
For Sap+ and Sap+ with SP, $\alpha$ is set to 0.5. 
For FMA, $\delta$ is 0.95. 
We evaluate the attacks using either AP or FVP leakage. 

\noindent\textbf{The impact of round number.} We set the round number to 1, 2, 4, and 8, with each round consisting of 10 days. 
In each round, the attacker observes 1 day and goes offline for 9 days. 
During the observation, the attacker monitors 10,000 queries each day. 
Other parameters remain the same as those in Section \ref{sec:peekaboo_exp}. 
The results are in Figure \ref{fig:test_comparison}. 
In general, Jigsaw+ and Sap+ outperform FMA.  
With AP, the accuracy of Jigsaw+ and Sap+ is approximately $90\%$ and $60\%$, respectively, while the accuracy of FMA is about $35\%$.  
With FVP, the gap between FMA and Jigsaw+ is roughly $30\%$.
We see that the accuracy of Jigsaw+ and Sap+ increases as the round number increases. 
Observing multiple rounds provides the attacker with more information to recover queries, yielding higher accuracy for Jigsaw+ and Sap+. 
The accuracy of FMA changes little, and we believe this is because FMA does not consider updates between two observations, so the results with multiple rounds are similar to those with just one round. 
We also notice that although the accuracy of Jigsaw+ and Sap+ is lower than their with-SP versions, the gap is small, indicating that the search pattern inferred by Peekaboo is sufficient for query recovery in both Jigsaw+ and Sap+. 

\noindent\textbf{The impact of offline days.} 
We also investigate attack performance with varying numbers of offline days in each round. 
We set the offline days by randomly sampling from 1-5, 6-10, 11-15, and 16-20, respectively. 
The attacker observes 1 day in each round and repeats this for 5 rounds.
Other parameters remain unchanged. 
The results are in Figure \ref{fig:test_random}. 
The accuracy of Jigsaw+ and Sap+ does not decrease as the number of offline days increases, showing the inferred external SP is minimally affected by the offline days. 
For FMA, since it does not consider SP matching for different rounds, its performance remains with different offline days.


\noindent\textbf{The impact of observed search queries.} We also test the situations while the attacker observes a different number of search queries.  
We set the observed queries each day to 500, 1,000, 2,500, 5,000, and 10,000. The round number is set to 5, and the observing and offline days are set to 1 and 19, respectively. 
We keep other parameters the same as in the previous experiments. 
The results are presented in Figure \ref{fig:test_query_number}. 
Increasing the observed queries, we can obtain better accuracy for Jigsaw+, Sap+, and their with-SP versions. 
With fewer observed search queries, the groups from different observations may vary significantly in terms of the underlying keywords, which negatively impacts the matching of groups in P1 of Peekaboo. 
Meanwhile, the attacker has less information to recover queries, resulting in relatively lower accuracy.
But for FMA, as it uses the FVP or AP to group search queries and relies solely on frequency to recover them, the information available to the attacker is limited. 
As a result, the accuracy changes only slightly as the observed query number increases.

\begin{figure}
    \centering
    \begin{minipage}{\linewidth}
        \centering
            \begin{tikzpicture}
        \node[draw=gray!50, dashed, rectangle, rounded corners=0pt, thick, inner sep=-1pt] {  
\begin{tabular}{c@{\hskip 4pt}c@{\hskip 4pt}c@{\hskip 4pt}c@{\hskip 4pt}c}
    \begin{tikzpicture}
        \fill[red] (0pt, 0pt) rectangle (6pt, 4pt);
    \end{tikzpicture} 
    \begin{scriptsize}
        FMA
    \end{scriptsize}
    &
    \begin{tikzpicture}
        \fill[darkorange] (0pt, 0pt) rectangle (6pt, 4pt);
    \end{tikzpicture}
    \begin{scriptsize}
        Jigsaw+
    \end{scriptsize}
    &
    \begin{tikzpicture}
        \fill[darkgreen] (0pt, 0pt) rectangle (6pt, 4pt);
    \end{tikzpicture}
    \begin{scriptsize}
        Jigsaw+ with SP
    \end{scriptsize}
    &
    \begin{tikzpicture}
        \fill[blueviolet] (0pt, 0pt) rectangle (6pt, 4pt);
    \end{tikzpicture} 
    \begin{scriptsize}
        Sap+
    \end{scriptsize}
    &
    \begin{tikzpicture}
        \fill[dodgerblue] (0pt, 0pt) rectangle (6pt, 4pt);
    \end{tikzpicture}  
    \begin{scriptsize}
        Sap+ with SP
    \end{scriptsize}

\end{tabular}
};
\end{tikzpicture}
            \vspace{4pt}
    \end{minipage}
    \subfloat[Enron, AP]
	{
 \label{fig:test_keyword_uni_size_enron}
		\begin{minipage}{.45\linewidth}
			\centering
                \includegraphics[width=\linewidth]{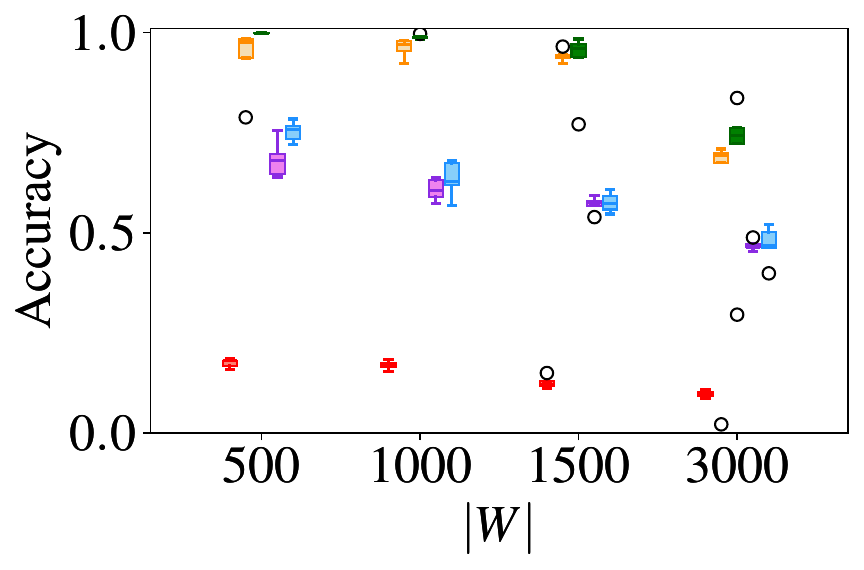}
		\end{minipage}
	}
    \subfloat[Lucene, AP]
	{
 \label{fig:test_keyword_uni_size_lucene}
		\begin{minipage}{.45\linewidth}
			\centering
			\includegraphics[width=\linewidth]{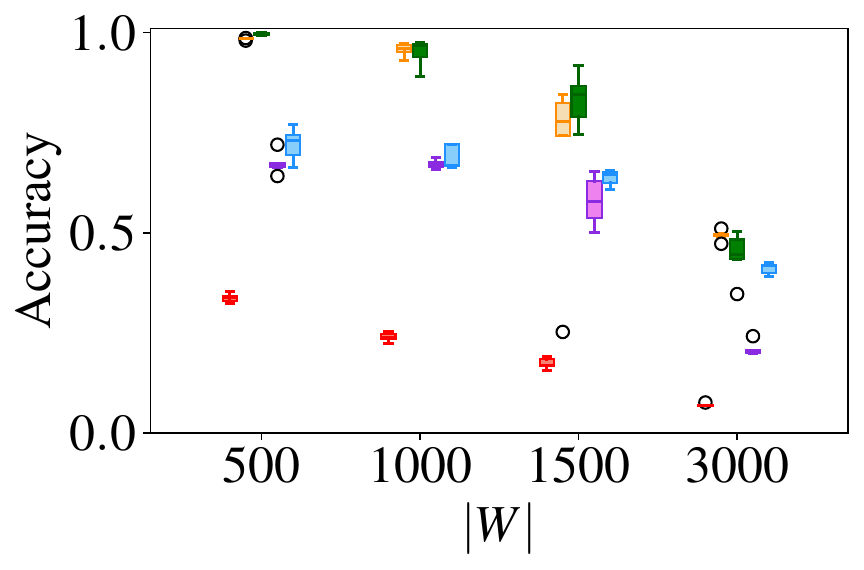}
		\end{minipage}
 	}
	\caption{The accuracy of Jigsaw+, Sap+, FMA, Jigsaw+ with SP, and the Sap+ with SP in Enron and Lucene with different keyword universe sizes $|W|$ with the AP leakage.}
	\label{fig:test_keyword_uni_size}
    \Description{The figures present the impact of different keyword universe sizes with the AP leakage for recovery accuracy, which are fully described in the text.}
\end{figure}

\noindent \textbf{The impact of keyword universe size.} 
We test the attacks with different keyword universe sizes (denoted as $|W|$), using 500, 1000, 1500, and 3000 keywords, with the largest volume in each dataset. 
The attacker observes 20,000 queries per day. 
For Jigsaw+ and Jigsaw+ with SP, we set the $\beta$ to 0.7 as the larger keyword universe contains more low-volume keywords and their co-occurrence information is greatly noised due to the low volume.
The accuracy is captured in Figure \ref{fig:test_keyword_uni_size}. 
The performance of all the attacks declines as the keyword universe size increases.  
In the Enron, the accuracy of Jigsaw+ and Sap+ drops from $>95\%$ to $70\%$ and from $70\%$ to $45\%$, respectively, as $|W|$ jumps from 500 to 3000.
FMA has $<25\%$ accuracy even with only 500 keywords in Enron.
The decline in accuracy when $|W|=3000$ is expected, as a larger keyword universe increases the number of candidate keywords for each search query, introducing more low-volume keywords that are more sensitive to noise given by database updates.  
%
%

\noindent\textbf{Runtime and memory cost}. 
The time complexity of Algorithm \ref{alg:dsps} is $O(|Q^x| \cdot |Gs^x| \cdot \log |Gs^x|)$, while Algorithm \ref{alg:dspm} requires $O(\eta^2 \cdot T_{\textsc{MATCH}})$. 
$T_{\textsc{MATCH}}$ is the runtime of the \textsc{MATCH} function, which depends on the $QuadraticAssignment$ it invokes.
The runtime of Algorithm \ref{alg:QueryRec} mainly depends on the \textsc{Attack+}.
We also present the runtime results in Table \ref{tab:runtimecop},  where we execute \textsc{InferISP} in parallel using 5 threads. 
Peekaboo completes in a few minutes, while FMA requires several hours.  
As the keywords universe size increases, Peekaboo's runtime also increases, as it must perform more computations on the co-occurrence matrix, which grows in size with $|W|$.  
As for FMA, however, its performance improves instead. 
This occurs because, as $|W|$ increases, the volume of new keywords and the average response size for queries decrease, making the calculation of similarities between queries less complex. 
We further state that our attack has modest memory requirements. 
Under the same settings as in Figures \ref{fig:test_comparison_enron}, \ref{fig:test_comparison_lucene}, and \ref{fig:test_wiki_AP}, Jigsaw+ and Sap+ consume at most 6.8, 6.6 and 15.2 GB of memory when evaluated on Enron, Lucene, and Wikipedia, respectively.


\begin{table}[tp]
\centering
    \begin{threeparttable} 
	    \caption{The runtime results of algorithms with different keyword universe sizes $|W|$.
     \label{tab:runtimecop}
     }
		\centering
        {
        
    \begin{tabular}{ccccl}
\hline
\multirow{2}{*}{Algorithm} & \multicolumn{4}{c}{Runtime (s) with different $|W|$} \\ \cline{2-5} 
                           & $500$        & $1000$      & $1500$     & $3000$     \\ \hline
Jigsaw+      & 284.89       & 355.93      & 817.37     & 3089.82    \\
Sap+         & 285.42       & 312.16      & 592.33     & 1304.54    \\
FMA                        & 17490.01     & 10169.8     & 7063.96    & 4006.73    \\ \hline
\end{tabular}
        } 
    \end{threeparttable} 
\end{table}

        

\section{Against Countermeasures}
\label{sec:Against Countermeasures}
We evaluate the attacks in Section \ref{sec:peekaboo_exp_2} under the padding of file size and the obfuscation of AP.

\noindent\textbf{Padding.} 
We assume the client pads each file size to a multiple of $k$, with $k$ set to 100, 200, 500, and 1,000. 
We use this to add noise to the FVP and affect the results of both the SP inference and query recovery.  
For instance, when $k=500$, there are only 57 and 125 distinct file sizes in Enron and Lucene, respectively; when $k=1000$, these numbers decrease to 34 and 80, respectively.
As the FVP is noised, the subsequent calculation of the co-occurrence is also affected by the noise.  
Thus, we use a larger $p_g$ in P1 of Peekaboo to remove more matches between groups. 
We also set a smaller $\beta$ in Jigsaw+ to balance the noise in the co-occurrence matrix. 
We set the $p_g$ to 0.15 and the $\beta$ to $0.7$. 
As previous works \cite{DBLP:conf/uss/OyaK22,DBLP:conf/uss/Nie00ZYL24,DBLP:conf/ndss/ShangOPK21} have shown that adaptations can improve attack accuracy against countermeasures, we apply the same approach.  
Specifically, the attacker adopts a similar adaption as in \cite{DBLP:conf/uss/Nie00ZYL24}, applying the same padding with identical parameters as the client to its own dataset in order to minimize the difference between the two datasets.
All attacks in our experiments incorporate this adaptation.
The attacks are tested over 5 rounds, with 1 day of observation (10,000 search queries per day) and 9 days of offline in each round. 
%
Other parameters are consistent with those in Section \ref{sec:peekaboo_exp_2}. 
As padding does not interfere with the AP, we examine the attacks using only the FVP leakage. 

\begin{figure}
    \centering
    \begin{minipage}{\linewidth}
        \centering
        \begin{tikzpicture}
        \node[draw=gray!50, dashed, rectangle, rounded corners=0pt, thick, inner sep=-1pt] {  
\begin{tabular}{c@{\hskip 4pt}c@{\hskip 4pt}c@{\hskip 4pt}c@{\hskip 4pt}c}
    \begin{tikzpicture}
        \fill[red] (0pt, 0pt) rectangle (6pt, 4pt);
    \end{tikzpicture} 
    \begin{scriptsize}
        FMA
    \end{scriptsize}
    &
    \begin{tikzpicture}
        \fill[darkorange] (0pt, 0pt) rectangle (6pt, 4pt);
    \end{tikzpicture}
    \begin{scriptsize}
        Jigsaw+
    \end{scriptsize}
    &
    \begin{tikzpicture}
        \fill[darkgreen] (0pt, 0pt) rectangle (6pt, 4pt);
    \end{tikzpicture}
    \begin{scriptsize}
        Jigsaw+ with SP
    \end{scriptsize}
    &
    \begin{tikzpicture}
        \fill[blueviolet] (0pt, 0pt) rectangle (6pt, 4pt);
    \end{tikzpicture} 
    \begin{scriptsize}
        Sap+
    \end{scriptsize}
    &
    \begin{tikzpicture}
        \fill[dodgerblue] (0pt, 0pt) rectangle (6pt, 4pt);
    \end{tikzpicture}  
    \begin{scriptsize}
        Sap+ with SP
    \end{scriptsize}

\end{tabular}
};
\end{tikzpicture}
    \end{minipage}
    \subfloat[Enron, FVP]
	{
 \label{fig:test_countermeasure_padding_enron}
		\begin{minipage}{.44\linewidth}
			\centering
                \includegraphics[width=\linewidth]{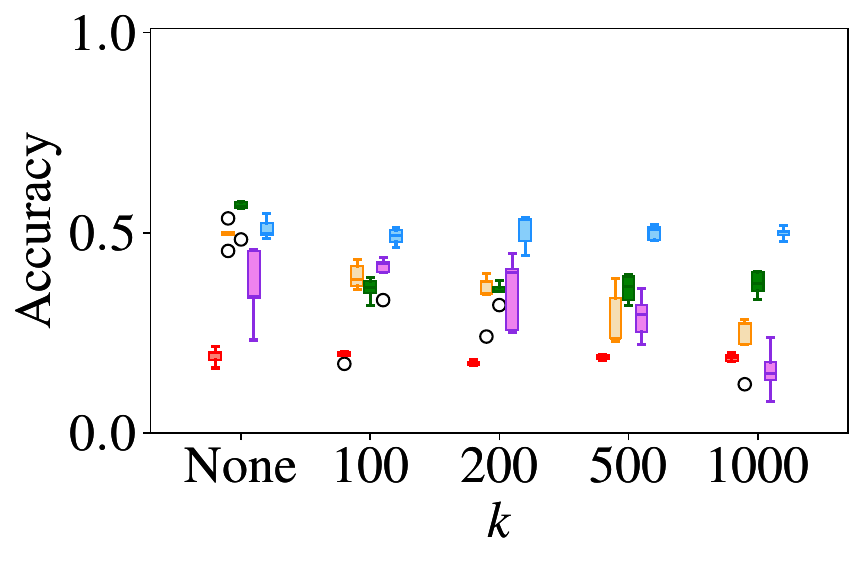}
		\end{minipage}
	}
    \subfloat[Lucene, FVP]
	{
 \label{fig:test_countermeasure_padding_lucene}
		\begin{minipage}{.44\linewidth}
			\centering
			\includegraphics[width=\linewidth]{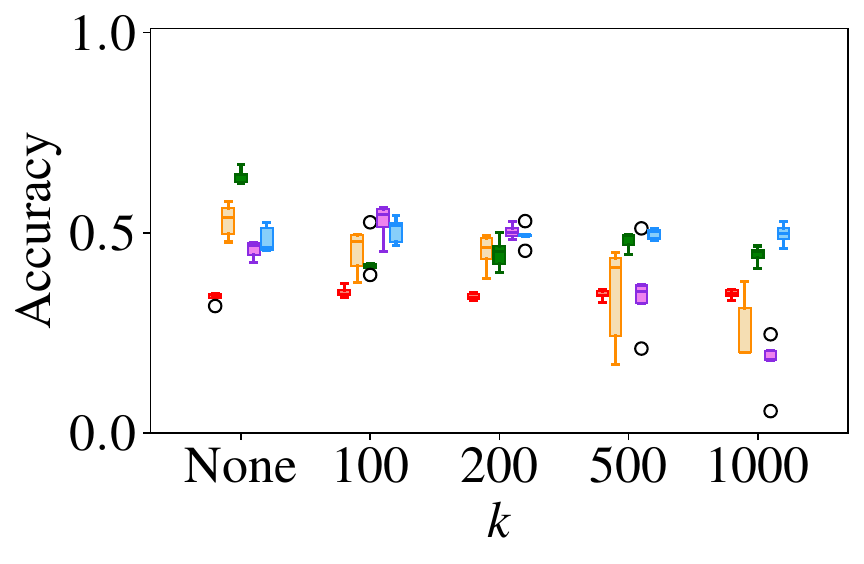}
		\end{minipage}
 	}
	\caption{The accuracy of Jigsaw+, Sap+, FMA, Jigsaw+ with SP, and the Sap+ with SP in Enron and Lucene against the padding of the file size with the FVP leakage.}
	\label{fig:test_countermeasure_padding}
    \Description{The figures present the impact of padding with the FVP leakage for recovery accuracy, which are fully described in the text.}
\end{figure}

We present the results in Figure \ref{fig:test_countermeasure_padding}. 
The accuracy of Sap+ with SP (about $50\%$) remains stable, as the attacker has access to the SP, and Sap+ does not rely on information related to the size of the files. 
In contrast, Jigsaw+ with SP loses $20\%$ performance, as it depends on the co-occurrence information of groups, which is affected by the noise introduced by the padding of file sizes. 
For Jigsaw+ and Sap+, the inference of the SP in Peekaboo is influenced by the padding, resulting in a drop in accuracy from approximately $50\%$ to about $20\%$.  
The performance of FMA remains largely stable, at about $20\%$ in Enron and $35\%$ in Lucene. 
We say that FMA (and P1 of Peekaboo) calculates the similarity between two queries using the intersection and union of file sizes, including duplicates. 
Both the file sizes and the number of files in the leakage contribute to this calculation and the number of files is not affected.
Thus, the padding only delivers minor influence on the similarity calculation. 

\begin{figure}
    \centering
    \begin{minipage}{\linewidth}
        \centering
        \begin{tikzpicture}
        \node[draw=gray!50, dashed, rectangle, rounded corners=0pt, thick, inner sep=-1pt] {  
\begin{tabular}{c@{\hskip 4pt}c@{\hskip 4pt}c@{\hskip 4pt}c@{\hskip 4pt}c}
    \begin{tikzpicture}
        \fill[red] (0pt, 0pt) rectangle (6pt, 4pt);
    \end{tikzpicture} 
    \begin{scriptsize}
        FMA
    \end{scriptsize}
    &
    \begin{tikzpicture}
        \fill[darkorange] (0pt, 0pt) rectangle (6pt, 4pt);
    \end{tikzpicture}
    \begin{scriptsize}
        Jigsaw+
    \end{scriptsize}
    &
    \begin{tikzpicture}
        \fill[darkgreen] (0pt, 0pt) rectangle (6pt, 4pt);
    \end{tikzpicture}
    \begin{scriptsize}
        Jigsaw+ with SP
    \end{scriptsize}
    &
    \begin{tikzpicture}
        \fill[blueviolet] (0pt, 0pt) rectangle (6pt, 4pt);
    \end{tikzpicture} 
    \begin{scriptsize}
        Sap+
    \end{scriptsize}
    &
    \begin{tikzpicture}
        \fill[dodgerblue] (0pt, 0pt) rectangle (6pt, 4pt);
    \end{tikzpicture}  
    \begin{scriptsize}
        Sap+ with SP
    \end{scriptsize}

\end{tabular}
};
\end{tikzpicture}
    \end{minipage}
    \subfloat[Enron, AP]
	{
 \label{fig:test_countermeasure_obfuscation_enron}
		\begin{minipage}{.44\linewidth}
			\centering
                \includegraphics[width=\linewidth]{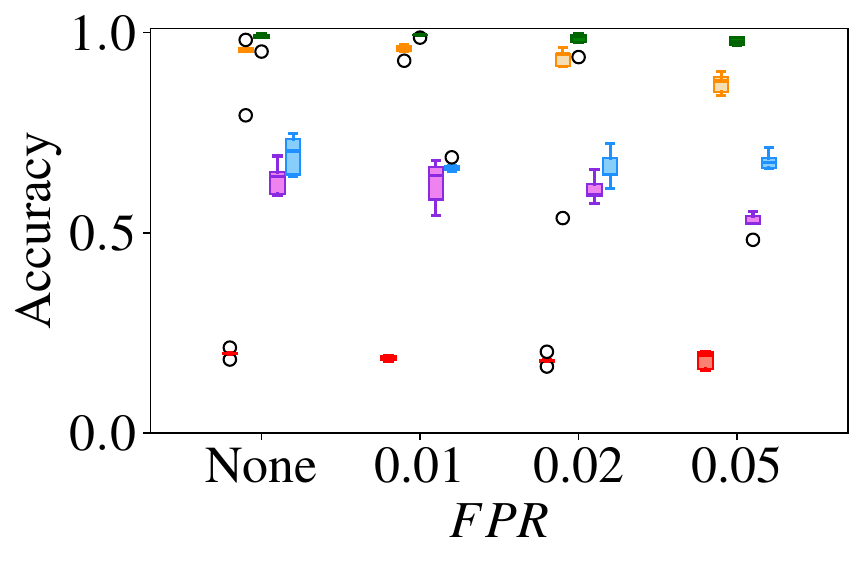}
		\end{minipage}
	}
    \subfloat[Lucene, AP]
	{
 \label{fig:test_countermeasure_obfuscation_lucene}
		\begin{minipage}{.44\linewidth}
			\centering
			\includegraphics[width=\linewidth]{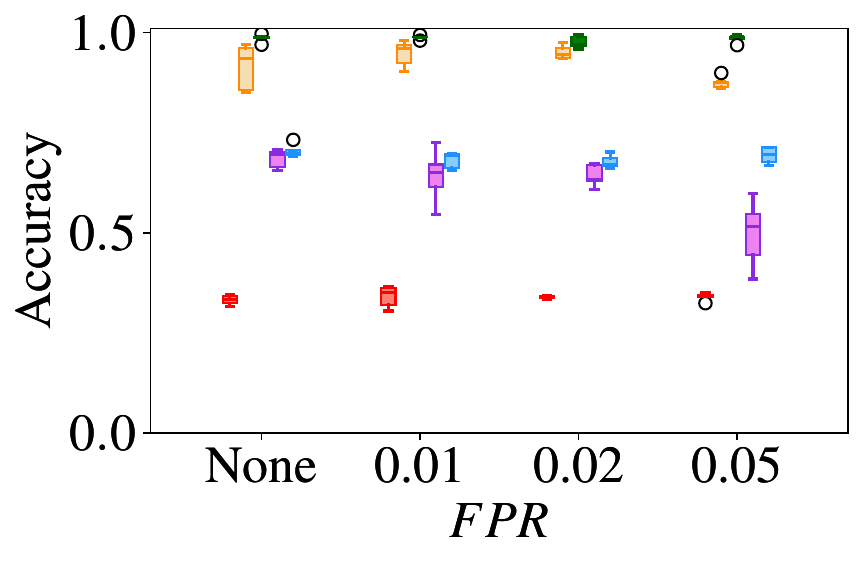}
		\end{minipage}
 	}
	\caption{The accuracy of Jigsaw+, Sap+, FMA, Jigsaw+ with SP, and the Sap+ with SP in Enron and Lucene against obfuscation with the AP leakage.}
	\label{fig:test_countermeasure_obfuscation}
    \Description{The figures present the impact of the obfuscation with the AP leakage for recovery accuracy, which are fully described in the text.}
\end{figure}
 

\noindent\textbf{Obfuscation of the AP.} 
We adopt a similar countermeasure as in \cite{DBLP:conf/infocom/ChenLRZ18} and assume that during the attacker's offline, the client downloads the whole dataset and rebuild the index to simulate dynamic updates under obfuscation.
While reconstructing the index, the client deletes the index of files for each keyword with a probability of $TPR$ and adds files not present in the keyword's response with a probability of $FPR$. 
We set the $TPR$ to 0.999 and $FPR$ to 0.01, 0.02, and 0.05.
The attacker employs a strategy similar to \cite{DBLP:conf/uss/Nie00ZYL24} which applies the same obfuscation to the attacker's dataset using identical parameters as the client, in order to minimize the difference between the two datasets. 
All attacks in our experiments employ the same adaptation.
The observation settings are identical as those in padding and other parameters are consistent with those in Section \ref{sec:peekaboo_exp_2}. 

In Figure \ref{fig:test_countermeasure_obfuscation},  
the results are largely consistent with those from the experiments in the previous section.  
Jigsaw+ and Jigsaw+ with SP lead with about $90\%$ accuracy, followed by Sap+ and Sap+ with SP at around $60\%$.  
FMA achieves only $20\%$ accuracy in Enron and about $35\%$ in Lucene. 
%
The accuracy of Jigsaw+ with SP and Sap+ with SP remains relatively stable across varying $FPR$ values. 
Sap+ with SP relies on frequency and volume. The frequency remains unaffected by obfuscation because it assumes the attacker knows the SP. 
Though the volume is affected by obfuscation, the attacker’s adoption of the same obfuscation method balances out the added volume across keywords, preserving accuracy. 
For Jigsaw+ with SP, the strong performance against obfuscation mirrors that of Jigsaw \cite{DBLP:conf/uss/Nie00ZYL24}, maintaining high accuracy. 
%
The accuracy of Jigsaw+ and Sap+ also remains stable when $FPR\leq 0.02$ but drops by approximately $15\%$ when $FPR$ increases to 0.05. 

\noindent\textbf{Discussions about Countermeasures.}
To counter Peekaboo, a practical approach is to limit the number of search queries the attacker can observe per round by quickly detecting and blocking the attacker’s access. As we assume the attacker eavesdrops on the server or the communication channel, the server can implement stricter intrusion detection systems (Appendix \ref{app:detailed examples of ioas}) to limit the duration and number of rounds the attacker can observe.
This reduces the amount of information available to the attacker, making it harder to perform effective query recovery.
For example, in Figure \ref{fig:test_query_number}, when the number of observed queries is limited, the attack accuracy drops.  
Similarly, as illustrated in Figure \ref{fig:test_comparison_enron_fvp} and \ref{fig:test_comparison_lucene_fvp}, further restricting the round number of observations can also reduce the accuracy when dealing with the FVP leakage. 
However, if the attacker gains access to the AP, even a single observation round can produce high accuracy.

The padding proves effective in mitigating the FVP leakage. 
But with the AP leakage, Peekaboo remains robust even against dynamic obfuscation. 
We believe that Peekaboo can pose severe threats to other DSSE schemes that reveal AP and include padded dummy files in the responses.
A possible approach to counter Peekaboo is to implement ``stronger'' parameters for padding or obfuscation.  
However, developing an efficient dynamic padding or obfuscation remains an open challenge. 
Alternatively, technologies like ORAM or PIR could prevent AP and FVP leakage, providing a possible defense against Peekaboo. 
%
%
%
But they typically involve significant communication or computation overhead.

\section{Conclusion}
In this work, we consider an intermittent-observation attacker who has only intermittent observation ability against DSSE. 
We formalize the leakage and propose a new attack called Peekaboo. 
In Peekaboo, the attacker first infers the SP with the AP or FVP leakage and then combines it with auxiliary knowledge and the leakage to recover search queries.
Peekaboo is a generic interface for similar-data attacks. 
We propose Jigsaw+ and Sap+ as instantiations. 
We conduct extensive experiments to confirm that Peekaboo can achieve a well-inferred SP, and Peekaboo with Jigsaw+ and Sap+ respectively provide $90\%$ and $60\%$ accuracy with AP, and about $50\%$ and $45\%$ with FVP. 
Our design also demonstrates practical efficiency. 
Even against countermeasures, Peekaboo still maintains its threats.


\bibliographystyle{CCS_Template_Camera_Ready/ACM-Reference-Format}
\balance


\begin{thebibliography}{56}


\ifx \showCODEN    \undefined \def \showCODEN     #1{\unskip}     \fi
\ifx \showISBNx    \undefined \def \showISBNx     #1{\unskip}     \fi
\ifx \showISBNxiii \undefined \def \showISBNxiii  #1{\unskip}     \fi
\ifx \showISSN     \undefined \def \showISSN      #1{\unskip}     \fi
\ifx \showLCCN     \undefined \def \showLCCN      #1{\unskip}     \fi
\ifx \shownote     \undefined \def \shownote      #1{#1}          \fi
\ifx \showarticletitle \undefined \def \showarticletitle #1{#1}   \fi
\ifx \showURL      \undefined \def \showURL       {\relax}        \fi
\providecommand\bibfield[2]{#2}
\providecommand\bibinfo[2]{#2}
\providecommand\natexlab[1]{#1}
\providecommand\showeprint[2][]{arXiv:#2}

\bibitem[Tri(2018)]%
        {Trickbot}
 \bibinfo{year}{2018}\natexlab{}.
\newblock \bibinfo{title}{Latest Trickbot Variant has New Tricks Up Its Sleeve.}
\newblock
\urldef\tempurl%
\url{https://www.cyberbit.com/endpoint-security/latest-trickbot-variant-has-new-tricks-up-its-sleeve/}
\showURL{%
\tempurl}


\bibitem[DBI(2022)]%
        {DBIR}
 \bibinfo{year}{2022}\natexlab{}.
\newblock \bibinfo{title}{{DBIR}: Data Breach Investigations Report.}
\newblock
\urldef\tempurl%
\url{https://www.verizon.com/business/en-gb/resources/2022-data-breach-investigations-report-dbir.pdf}
\showURL{%
\tempurl}


\bibitem[WHO(2023)]%
        {WHOlist}
 \bibinfo{year}{2023}\natexlab{}.
\newblock \bibinfo{title}{{WHO} Model List of Essential Medicines.}
\newblock
\urldef\tempurl%
\url{https://www.who.int/publications/i/item/WHO-MHP-HPS-EML-2023.02}
\showURL{%
\tempurl}


\bibitem[DBI(2024)]%
        {DBIR2}
 \bibinfo{year}{2024}\natexlab{}.
\newblock \bibinfo{title}{{DBIR}: 2024 Data Breach Investigations Report.}
\newblock
\urldef\tempurl%
\url{https://www.verizon.com/business/resources/T16f/reports/2024-dbir-data-breach-investigations-report.pdf}
\showURL{%
\tempurl}


\bibitem[spl(2024)]%
        {splunk}
 \bibinfo{year}{2024}\natexlab{}.
\newblock \bibinfo{title}{State of Security 2024.}
\newblock
\urldef\tempurl%
\url{https://www.splunk.com/en_us/pdfs/gated/ebooks/state-of-security-2024.pdf}
\showURL{%
\tempurl}


\bibitem[Alageel et~al\mbox{.}(2025)]%
        {DBLP:journals/corr/abs-2502-08830}
\bibfield{author}{\bibinfo{person}{Almuthanna Alageel}, \bibinfo{person}{Sergio Maffeis}, {and} \bibinfo{person}{Imperial~College London}.} \bibinfo{year}{2025}\natexlab{}.
\newblock \showarticletitle{Investigation of Advanced Persistent Threats Network-based Tactics, Techniques and Procedures}.
\newblock \bibinfo{journal}{\emph{CoRR}}  \bibinfo{volume}{abs/2502.08830} (\bibinfo{year}{2025}).
\newblock
\showeprint[arXiv]{2502.08830}
\href{https://doi.org/10.48550/ARXIV.2502.08830}{doi:\nolinkurl{10.48550/ARXIV.2502.08830}}


\bibitem[Ambika(2020)]%
        {ambika2020improved}
\bibfield{author}{\bibinfo{person}{N Ambika}.} \bibinfo{year}{2020}\natexlab{}.
\newblock \showarticletitle{Improved Methodology to Detect Advanced Persistent Threat Attacks}.
\newblock In \bibinfo{booktitle}{\emph{Quantum Cryptography and the Future of Cyber Security}}.
\newblock


\bibitem[Amjad et~al\mbox{.}(2019)]%
        {DBLP:journals/popets/AmjadKM19}
\bibfield{author}{\bibinfo{person}{Ghous Amjad}, \bibinfo{person}{Seny Kamara}, {and} \bibinfo{person}{Tarik Moataz}.} \bibinfo{year}{2019}\natexlab{}.
\newblock \showarticletitle{Breach-Resistant Structured Encryption}.
\newblock \bibinfo{journal}{\emph{Proc. Priv. Enhancing Technol.}} \bibinfo{volume}{2019}, \bibinfo{number}{1} (\bibinfo{year}{2019}), \bibinfo{pages}{245--265}.
\newblock
\href{https://doi.org/10.2478/POPETS-2019-0014}{doi:\nolinkurl{10.2478/POPETS-2019-0014}}


\bibitem[Blackstone et~al\mbox{.}(2020)]%
        {DBLP:conf/ndss/BlackstoneKM20}
\bibfield{author}{\bibinfo{person}{Laura Blackstone}, \bibinfo{person}{Seny Kamara}, {and} \bibinfo{person}{Tarik Moataz}.} \bibinfo{year}{2020}\natexlab{}.
\newblock \showarticletitle{Revisiting Leakage Abuse Attacks}. In \bibinfo{booktitle}{\emph{27th Annual Network and Distributed System Security Symposium}}.
\newblock
\urldef\tempurl%
\url{https://www.ndss-symposium.org/ndss-paper/revisiting-leakage-abuse-attacks/}
\showURL{%
\tempurl}


\bibitem[Bost(2016)]%
        {DBLP:conf/ccs/Bost16}
\bibfield{author}{\bibinfo{person}{Raphael Bost}.} \bibinfo{year}{2016}\natexlab{}.
\newblock \showarticletitle{{\(\sum\)}o{\(\varphi\)}o{\(\varsigma\)}: Forward Secure Searchable Encryption}. In \bibinfo{booktitle}{\emph{Proceedings of the 2016 ACM SIGSAC Conference on Computer and Communications Security}}. \bibinfo{pages}{1143–1154}.
\newblock
\showISBNx{9781450341394}
\href{https://doi.org/10.1145/2976749.2978303}{doi:\nolinkurl{10.1145/2976749.2978303}}


\bibitem[Bost et~al\mbox{.}(2017)]%
        {DBLP:conf/ccs/BostMO17}
\bibfield{author}{\bibinfo{person}{Rapha\"{e}l Bost}, \bibinfo{person}{Brice Minaud}, {and} \bibinfo{person}{Olga Ohrimenko}.} \bibinfo{year}{2017}\natexlab{}.
\newblock \showarticletitle{Forward and Backward Private Searchable Encryption from Constrained Cryptographic Primitives}. In \bibinfo{booktitle}{\emph{Proceedings of the 2017 ACM SIGSAC Conference on Computer and Communications Security}}. \bibinfo{pages}{1465–1482}.
\newblock
\showISBNx{9781450349468}
\href{https://doi.org/10.1145/3133956.3133980}{doi:\nolinkurl{10.1145/3133956.3133980}}


\bibitem[Cash et~al\mbox{.}(2015)]%
        {DBLP:conf/ccs/CashGPR15}
\bibfield{author}{\bibinfo{person}{David Cash}, \bibinfo{person}{Paul Grubbs}, \bibinfo{person}{Jason Perry}, {and} \bibinfo{person}{Thomas Ristenpart}.} \bibinfo{year}{2015}\natexlab{}.
\newblock \showarticletitle{Leakage-Abuse Attacks Against Searchable Encryption}. In \bibinfo{booktitle}{\emph{Proceedings of the 22nd ACM SIGSAC Conference on Computer and Communications Security}}. \bibinfo{pages}{668–679}.
\newblock
\showISBNx{9781450338325}
\href{https://doi.org/10.1145/2810103.2813700}{doi:\nolinkurl{10.1145/2810103.2813700}}


\bibitem[Chen et~al\mbox{.}(2018)]%
        {DBLP:conf/infocom/ChenLRZ18}
\bibfield{author}{\bibinfo{person}{Guoxing Chen}, \bibinfo{person}{Ten{-}Hwang Lai}, \bibinfo{person}{Michael~K. Reiter}, {and} \bibinfo{person}{Yinqian Zhang}.} \bibinfo{year}{2018}\natexlab{}.
\newblock \showarticletitle{Differentially Private Access Patterns for Searchable Symmetric Encryption}. In \bibinfo{booktitle}{\emph{2018 {IEEE} Conference on Computer Communications}}. \bibinfo{pages}{810--818}.
\newblock
\href{https://doi.org/10.1109/INFOCOM.2018.8486381}{doi:\nolinkurl{10.1109/INFOCOM.2018.8486381}}


\bibitem[Chen et~al\mbox{.}(2023)]%
        {DBLP:conf/ndss/Chen0PLS0L23}
\bibfield{author}{\bibinfo{person}{Tianyang Chen}, \bibinfo{person}{Peng Xu}, \bibinfo{person}{Stjepan Picek}, \bibinfo{person}{Bo Luo}, \bibinfo{person}{Willy Susilo}, \bibinfo{person}{Hai Jin}, {and} \bibinfo{person}{Kaitai Liang}.} \bibinfo{year}{2023}\natexlab{}.
\newblock \showarticletitle{The Power of Bamboo: On the Post-Compromise Security for Searchable Symmetric Encryption}. In \bibinfo{booktitle}{\emph{30th Annual Network and Distributed System Security Symposium}}.
\newblock
\urldef\tempurl%
\url{https://www.ndss-symposium.org/ndss-paper/the-power-of-bamboo-on-the-post-compromise-security-for-searchable-symmetric-encryption/}
\showURL{%
\tempurl}


\bibitem[Chen et~al\mbox{.}(2024)]%
        {DBLP:conf/uss/ChenY0L0XXLW24}
\bibfield{author}{\bibinfo{person}{Yihao Chen}, \bibinfo{person}{Qilei Yin}, \bibinfo{person}{Qi Li}, \bibinfo{person}{Zhuotao Liu}, \bibinfo{person}{Ke Xu}, \bibinfo{person}{Yi Xu}, \bibinfo{person}{Mingwei Xu}, \bibinfo{person}{Ziqian Liu}, {and} \bibinfo{person}{Jianping Wu}.} \bibinfo{year}{2024}\natexlab{}.
\newblock \showarticletitle{Learning with Semantics: Towards a Semantics-Aware Routing Anomaly Detection System}. In \bibinfo{booktitle}{\emph{33rd {USENIX} Security Symposium}}.
\newblock
\urldef\tempurl%
\url{https://www.usenix.org/conference/usenixsecurity24/presentation/chen-yihao}
\showURL{%
\tempurl}


\bibitem[Curtmola et~al\mbox{.}(2006)]%
        {DBLP:conf/ccs/CurtmolaGKO06}
\bibfield{author}{\bibinfo{person}{Reza Curtmola}, \bibinfo{person}{Juan Garay}, \bibinfo{person}{Seny Kamara}, {and} \bibinfo{person}{Rafail Ostrovsky}.} \bibinfo{year}{2006}\natexlab{}.
\newblock \showarticletitle{Searchable symmetric encryption: improved definitions and efficient constructions}. In \bibinfo{booktitle}{\emph{Proceedings of the 13th ACM Conference on Computer and Communications Security}}. \bibinfo{pages}{79–88}.
\newblock
\showISBNx{1595935185}
\href{https://doi.org/10.1145/1180405.1180417}{doi:\nolinkurl{10.1145/1180405.1180417}}


\bibitem[Damie et~al\mbox{.}(2021)]%
        {DBLP:conf/uss/Damie0P21}
\bibfield{author}{\bibinfo{person}{Marc Damie}, \bibinfo{person}{Florian Hahn}, {and} \bibinfo{person}{Andreas Peter}.} \bibinfo{year}{2021}\natexlab{}.
\newblock \showarticletitle{A Highly Accurate Query-Recovery Attack against Searchable Encryption using Non-Indexed Documents}. In \bibinfo{booktitle}{\emph{30th {USENIX} Security Symposium}}. \bibinfo{pages}{143--160}.
\newblock
\urldef\tempurl%
\url{https://www.usenix.org/conference/usenixsecurity21/presentation/damie}
\showURL{%
\tempurl}


\bibitem[Dou et~al\mbox{.}(2024)]%
        {DBLP:journals/tifs/DouDXWXCJ24}
\bibfield{author}{\bibinfo{person}{Haochen Dou}, \bibinfo{person}{Zhenwu Dan}, \bibinfo{person}{Peng Xu}, \bibinfo{person}{Wei Wang}, \bibinfo{person}{Shuning Xu}, \bibinfo{person}{Tianyang Chen}, {and} \bibinfo{person}{Hai Jin}.} \bibinfo{year}{2024}\natexlab{}.
\newblock \showarticletitle{Dynamic Searchable Symmetric Encryption With Strong Security and Robustness}.
\newblock \bibinfo{journal}{\emph{{IEEE} Trans. Inf. Forensics Secur.}}  \bibinfo{volume}{19} (\bibinfo{year}{2024}), \bibinfo{pages}{2370--2384}.
\newblock
\href{https://doi.org/10.1109/TIFS.2024.3350330}{doi:\nolinkurl{10.1109/TIFS.2024.3350330}}


\bibitem[Du et~al\mbox{.}(2022)]%
        {DBLP:conf/crypto/DuGG22}
\bibfield{author}{\bibinfo{person}{Yang Du}, \bibinfo{person}{Daniel Genkin}, {and} \bibinfo{person}{Paul Grubbs}.} \bibinfo{year}{2022}\natexlab{}.
\newblock \showarticletitle{Snapshot-Oblivious RAMs: Sub-logarithmic Efficiency for Short Transcripts}. In \bibinfo{booktitle}{\emph{Advances in Cryptology - {CRYPTO} 2022 - 42nd Annual International Cryptology Conference}}, Vol.~\bibinfo{volume}{13510}. \bibinfo{pages}{152--181}.
\newblock
\href{https://doi.org/10.1007/978-3-031-15985-5\_6}{doi:\nolinkurl{10.1007/978-3-031-15985-5\_6}}


\bibitem[Foundation(1999)]%
        {Lucene}
\bibfield{author}{\bibinfo{person}{Apache Foundation}.} \bibinfo{year}{1999}\natexlab{}.
\newblock \bibinfo{title}{Mail Archieves of Lucene}.
\newblock
\urldef\tempurl%
\url{https://mailarchives.apache.org/mod_mbox/#lucene}
\showURL{%
\tempurl}


\bibitem[Foundation(2020)]%
        {Wikipedia}
\bibfield{author}{\bibinfo{person}{Wikipedia Foundation}.} \bibinfo{year}{2020}\natexlab{}.
\newblock \bibinfo{title}{Wikipedia databases}.
\newblock
\newblock
\shownote{\url{https://www.wikipedia.org}}.


\bibitem[Ghareh~Chamani et~al\mbox{.}(2018)]%
        {DBLP:conf/ccs/ChamaniPPJ18}
\bibfield{author}{\bibinfo{person}{Javad Ghareh~Chamani}, \bibinfo{person}{Dimitrios Papadopoulos}, \bibinfo{person}{Charalampos Papamanthou}, {and} \bibinfo{person}{Rasool Jalili}.} \bibinfo{year}{2018}\natexlab{}.
\newblock \showarticletitle{New Constructions for Forward and Backward Private Symmetric Searchable Encryption}. In \bibinfo{booktitle}{\emph{Proceedings of the 2018 ACM SIGSAC Conference on Computer and Communications Security}}. \bibinfo{pages}{1038–1055}.
\newblock
\showISBNx{9781450356930}
\href{https://doi.org/10.1145/3243734.3243833}{doi:\nolinkurl{10.1145/3243734.3243833}}


\bibitem[Gui et~al\mbox{.}(2023)]%
        {DBLP:conf/sp/GuiPP23}
\bibfield{author}{\bibinfo{person}{Zichen Gui}, \bibinfo{person}{Kenneth~G. Paterson}, {and} \bibinfo{person}{Sikhar Patranabis}.} \bibinfo{year}{2023}\natexlab{}.
\newblock \showarticletitle{Rethinking Searchable Symmetric Encryption}. In \bibinfo{booktitle}{\emph{44th {IEEE} Symposium on Security and Privacy}}. \bibinfo{pages}{1401--1418}.
\newblock
\href{https://doi.org/10.1109/SP46215.2023.10179460}{doi:\nolinkurl{10.1109/SP46215.2023.10179460}}


\bibitem[Haltiwanger and Hoang(2024)]%
        {DBLP:conf/codaspy/HaltiwangerH24}
\bibfield{author}{\bibinfo{person}{Jacob Haltiwanger} {and} \bibinfo{person}{Thang Hoang}.} \bibinfo{year}{2024}\natexlab{}.
\newblock \showarticletitle{Exploiting Update Leakage in Searchable Symmetric Encryption}. In \bibinfo{booktitle}{\emph{Proceedings of the Fourteenth {ACM} Conference on Data and Application Security and Privacy}}. \bibinfo{pages}{115--126}.
\newblock
\href{https://doi.org/10.1145/3626232.3653260}{doi:\nolinkurl{10.1145/3626232.3653260}}


\bibitem[Hubert and Arabie(1985)]%
        {hubert1985comparing}
\bibfield{author}{\bibinfo{person}{Lawrence Hubert} {and} \bibinfo{person}{Phipps Arabie}.} \bibinfo{year}{1985}\natexlab{}.
\newblock \showarticletitle{Comparing partitions}.
\newblock \bibinfo{journal}{\emph{Journal of classification}}  \bibinfo{volume}{2} (\bibinfo{year}{1985}), \bibinfo{pages}{193--218}.
\newblock


\bibitem[Islam et~al\mbox{.}(2012)]%
        {DBLP:conf/ndss/IslamKK12}
\bibfield{author}{\bibinfo{person}{Mohammad~Saiful Islam}, \bibinfo{person}{Mehmet Kuzu}, {and} \bibinfo{person}{Murat Kantarcioglu}.} \bibinfo{year}{2012}\natexlab{}.
\newblock \showarticletitle{Access Pattern disclosure on Searchable Encryption: Ramification, Attack and Mitigation}. In \bibinfo{booktitle}{\emph{19th Annual Network and Distributed System Security Symposium}}.
\newblock
\urldef\tempurl%
\url{https://www.ndss-symposium.org/ndss2012/access-pattern-disclosure-searchable-encryption-ramification-attack-and-mitigation}
\showURL{%
\tempurl}


\bibitem[Kamara et~al\mbox{.}(2024)]%
        {DBLP:journals/popets/KamaraKMDPT24}
\bibfield{author}{\bibinfo{person}{Seny Kamara}, \bibinfo{person}{Abdelkarim Kati}, \bibinfo{person}{Tarik Moataz}, \bibinfo{person}{Jamie DeMaria}, \bibinfo{person}{Andrew Park}, {and} \bibinfo{person}{Amos Treiber}.} \bibinfo{year}{2024}\natexlab{}.
\newblock \showarticletitle{{MAPLE:} MArkov Process Leakage attacks on Encrypted Search}.
\newblock \bibinfo{journal}{\emph{Proc. Priv. Enhancing Technol.}} \bibinfo{volume}{2024}, \bibinfo{number}{1} (\bibinfo{year}{2024}), \bibinfo{pages}{430--446}.
\newblock
\href{https://doi.org/10.56553/POPETS-2024-0025}{doi:\nolinkurl{10.56553/POPETS-2024-0025}}


\bibitem[Kamara et~al\mbox{.}(2012)]%
        {DBLP:conf/ccs/KamaraPR12}
\bibfield{author}{\bibinfo{person}{Seny Kamara}, \bibinfo{person}{Charalampos Papamanthou}, {and} \bibinfo{person}{Tom Roeder}.} \bibinfo{year}{2012}\natexlab{}.
\newblock \showarticletitle{Dynamic searchable symmetric encryption}. In \bibinfo{booktitle}{\emph{Proceedings of the 2012 ACM Conference on Computer and Communications Security}}. \bibinfo{pages}{965–976}.
\newblock
\showISBNx{9781450316514}
\href{https://doi.org/10.1145/2382196.2382298}{doi:\nolinkurl{10.1145/2382196.2382298}}


\bibitem[Kuhn(1955)]%
        {kuhn1955hungarian}
\bibfield{author}{\bibinfo{person}{Harold~W Kuhn}.} \bibinfo{year}{1955}\natexlab{}.
\newblock \showarticletitle{The Hungarian method for the assignment problem}.
\newblock \bibinfo{journal}{\emph{Naval research logistics quarterly}} \bibinfo{volume}{2}, \bibinfo{number}{1-2} (\bibinfo{year}{1955}), \bibinfo{pages}{83--97}.
\newblock


\bibitem[Li et~al\mbox{.}(2024)]%
        {DBLP:conf/ndss/LiDXWSC0C024}
\bibfield{author}{\bibinfo{person}{Shaofei Li}, \bibinfo{person}{Feng Dong}, \bibinfo{person}{Xusheng Xiao}, \bibinfo{person}{Haoyu Wang}, \bibinfo{person}{Fei Shao}, \bibinfo{person}{Jiedong Chen}, \bibinfo{person}{Yao Guo}, \bibinfo{person}{Xiangqun Chen}, {and} \bibinfo{person}{Ding Li}.} \bibinfo{year}{2024}\natexlab{}.
\newblock \showarticletitle{{NODLINK:} An Online System for Fine-Grained {APT} Attack Detection and Investigation}. In \bibinfo{booktitle}{\emph{31st Annual Network and Distributed System Security Symposium}}.
\newblock
\urldef\tempurl%
\url{https://www.ndss-symposium.org/ndss-paper/nodlink-an-online-system-for-fine-grained-apt-attack-detection-and-investigation/}
\showURL{%
\tempurl}


\bibitem[Liu et~al\mbox{.}(2014)]%
        {DBLP:journals/isci/LiuZWT14}
\bibfield{author}{\bibinfo{person}{Chang Liu}, \bibinfo{person}{Liehuang Zhu}, \bibinfo{person}{Mingzhong Wang}, {and} \bibinfo{person}{Yu{-}an Tan}.} \bibinfo{year}{2014}\natexlab{}.
\newblock \showarticletitle{Search pattern leakage in searchable encryption: Attacks and new construction}.
\newblock \bibinfo{journal}{\emph{Inf. Sci.}}  \bibinfo{volume}{265} (\bibinfo{year}{2014}), \bibinfo{pages}{176--188}.
\newblock
\href{https://doi.org/10.1016/J.INS.2013.11.021}{doi:\nolinkurl{10.1016/J.INS.2013.11.021}}


\bibitem[MusikAnimal(2015)]%
        {Pageviews}
\bibfield{author}{\bibinfo{person}{Marcel Ruiz~Forns MusikAnimal, Kaldari}.} \bibinfo{year}{2015}\natexlab{}.
\newblock \bibinfo{title}{Pageviews Toolforge}.
\newblock
\urldef\tempurl%
\url{https://pageviews.toolforge.org/}
\showURL{%
\tempurl}


\bibitem[Naveed et~al\mbox{.}(2014)]%
        {DBLP:conf/sp/NaveedPG14}
\bibfield{author}{\bibinfo{person}{Muhammad Naveed}, \bibinfo{person}{Manoj Prabhakaran}, {and} \bibinfo{person}{Carl~A. Gunter}.} \bibinfo{year}{2014}\natexlab{}.
\newblock \showarticletitle{Dynamic Searchable Encryption via Blind Storage}. In \bibinfo{booktitle}{\emph{2014 {IEEE} Symposium on Security and Privacy}}. \bibinfo{pages}{639--654}.
\newblock
\href{https://doi.org/10.1109/SP.2014.47}{doi:\nolinkurl{10.1109/SP.2014.47}}


\bibitem[Nie et~al\mbox{.}(2024)]%
        {DBLP:conf/uss/Nie00ZYL24}
\bibfield{author}{\bibinfo{person}{Hao Nie}, \bibinfo{person}{Wei Wang}, \bibinfo{person}{Peng Xu}, \bibinfo{person}{Xianglong Zhang}, \bibinfo{person}{Laurence~T. Yang}, {and} \bibinfo{person}{Kaitai Liang}.} \bibinfo{year}{2024}\natexlab{}.
\newblock \showarticletitle{Query Recovery from Easy to Hard: Jigsaw Attack against {SSE}}. In \bibinfo{booktitle}{\emph{33rd {USENIX} Security Symposium}}.
\newblock
\urldef\tempurl%
\url{https://www.usenix.org/conference/usenixsecurity24/presentation/nie}
\showURL{%
\tempurl}


\bibitem[Ning et~al\mbox{.}(2021)]%
        {DBLP:conf/ccs/NingHPYL0D21}
\bibfield{author}{\bibinfo{person}{Jianting Ning}, \bibinfo{person}{Xinyi Huang}, \bibinfo{person}{Geong~Sen Poh}, \bibinfo{person}{Jiaming Yuan}, \bibinfo{person}{Yingjiu Li}, \bibinfo{person}{Jian Weng}, {and} \bibinfo{person}{Robert~H. Deng}.} \bibinfo{year}{2021}\natexlab{}.
\newblock \showarticletitle{LEAP: Leakage-Abuse Attack on Efficiently Deployable, Efficiently Searchable Encryption with Partially Known Dataset}. In \bibinfo{booktitle}{\emph{Proceedings of the 2021 ACM SIGSAC Conference on Computer and Communications Security}}. \bibinfo{pages}{2307–2320}.
\newblock
\showISBNx{9781450384544}
\href{https://doi.org/10.1145/3460120.3484540}{doi:\nolinkurl{10.1145/3460120.3484540}}


\bibitem[Oya and Kerschbaum(2021)]%
        {DBLP:conf/uss/OyaK21}
\bibfield{author}{\bibinfo{person}{Simon Oya} {and} \bibinfo{person}{Florian Kerschbaum}.} \bibinfo{year}{2021}\natexlab{}.
\newblock \showarticletitle{Hiding the Access Pattern is Not Enough: Exploiting Search Pattern Leakage in Searchable Encryption}. In \bibinfo{booktitle}{\emph{30th {USENIX} Security Symposium}}. \bibinfo{pages}{127--142}.
\newblock
\urldef\tempurl%
\url{https://www.usenix.org/conference/usenixsecurity21/presentation/oya}
\showURL{%
\tempurl}


\bibitem[Oya and Kerschbaum(2022)]%
        {DBLP:conf/uss/OyaK22}
\bibfield{author}{\bibinfo{person}{Simon Oya} {and} \bibinfo{person}{Florian Kerschbaum}.} \bibinfo{year}{2022}\natexlab{}.
\newblock \showarticletitle{{IHOP:} Improved Statistical Query Recovery against Searchable Symmetric Encryption through Quadratic Optimization}. In \bibinfo{booktitle}{\emph{31st {USENIX} Security Symposium}}. \bibinfo{pages}{2407--2424}.
\newblock
\urldef\tempurl%
\url{https://www.usenix.org/conference/usenixsecurity22/presentation/oya}
\showURL{%
\tempurl}


\bibitem[Persiano and Yeo(2023)]%
        {DBLP:conf/crypto/PersianoY23}
\bibfield{author}{\bibinfo{person}{Giuseppe Persiano} {and} \bibinfo{person}{Kevin Yeo}.} \bibinfo{year}{2023}\natexlab{}.
\newblock \showarticletitle{Limits of Breach-Resistant and Snapshot-Oblivious RAMs}. In \bibinfo{booktitle}{\emph{Advances in Cryptology - {CRYPTO} 2023 - 43rd Annual International Cryptology Conference}}, Vol.~\bibinfo{volume}{14084}. \bibinfo{pages}{161--196}.
\newblock
\href{https://doi.org/10.1007/978-3-031-38551-3\_6}{doi:\nolinkurl{10.1007/978-3-031-38551-3\_6}}


\bibitem[Poddar et~al\mbox{.}(2020)]%
        {DBLP:conf/eurosp/PoddarWLP20}
\bibfield{author}{\bibinfo{person}{Rishabh Poddar}, \bibinfo{person}{Stephanie Wang}, \bibinfo{person}{Jianan Lu}, {and} \bibinfo{person}{Raluca~Ada Popa}.} \bibinfo{year}{2020}\natexlab{}.
\newblock \showarticletitle{Practical Volume-Based Attacks on Encrypted Databases}. In \bibinfo{booktitle}{\emph{{IEEE} European Symposium on Security and Privacy}}. \bibinfo{pages}{354--369}.
\newblock
\href{https://doi.org/10.1109/EUROSP48549.2020.00030}{doi:\nolinkurl{10.1109/EUROSP48549.2020.00030}}


\bibitem[Pouliot and Wright(2016)]%
        {DBLP:conf/ccs/PouliotW16}
\bibfield{author}{\bibinfo{person}{David Pouliot} {and} \bibinfo{person}{Charles~V. Wright}.} \bibinfo{year}{2016}\natexlab{}.
\newblock \showarticletitle{The Shadow Nemesis: Inference Attacks on Efficiently Deployable, Efficiently Searchable Encryption}. In \bibinfo{booktitle}{\emph{Proceedings of the 2016 ACM SIGSAC Conference on Computer and Communications Security}}. \bibinfo{pages}{1341–1352}.
\newblock
\showISBNx{9781450341394}
\href{https://doi.org/10.1145/2976749.2978401}{doi:\nolinkurl{10.1145/2976749.2978401}}


\bibitem[Rand(1971)]%
        {rand1971objective}
\bibfield{author}{\bibinfo{person}{William~M Rand}.} \bibinfo{year}{1971}\natexlab{}.
\newblock \showarticletitle{Objective criteria for the evaluation of clustering methods}.
\newblock \bibinfo{journal}{\emph{Journal of the American Statistical association}} \bibinfo{volume}{66}, \bibinfo{number}{336} (\bibinfo{year}{1971}), \bibinfo{pages}{846--850}.
\newblock


\bibitem[Rehman et~al\mbox{.}(2024)]%
        {DBLP:conf/sp/RehmanAH24}
\bibfield{author}{\bibinfo{person}{Mati~Ur Rehman}, \bibinfo{person}{Hadi Ahmadi}, {and} \bibinfo{person}{Wajih~Ul Hassan}.} \bibinfo{year}{2024}\natexlab{}.
\newblock \showarticletitle{Flash: {A} Comprehensive Approach to Intrusion Detection via Provenance Graph Representation Learning}. In \bibinfo{booktitle}{\emph{{IEEE} Symposium on Security and Privacy}}. \bibinfo{pages}{3552--3570}.
\newblock
\href{https://doi.org/10.1109/SP54263.2024.00139}{doi:\nolinkurl{10.1109/SP54263.2024.00139}}


\bibitem[Salmani and Barker(2021)]%
        {DBLP:conf/codaspy/Salmani021}
\bibfield{author}{\bibinfo{person}{Khosro Salmani} {and} \bibinfo{person}{Ken Barker}.} \bibinfo{year}{2021}\natexlab{}.
\newblock \showarticletitle{Don't fool yourself with Forward Privacy, Your queries {STILL} belong to us!}. In \bibinfo{booktitle}{\emph{{CODASPY} '21: Eleventh {ACM} Conference on Data and Application Security and Privacy}}. \bibinfo{pages}{131--142}.
\newblock
\href{https://doi.org/10.1145/3422337.3447838}{doi:\nolinkurl{10.1145/3422337.3447838}}


\bibitem[Shang et~al\mbox{.}(2021)]%
        {DBLP:conf/ndss/ShangOPK21}
\bibfield{author}{\bibinfo{person}{Zhiwei Shang}, \bibinfo{person}{Simon Oya}, \bibinfo{person}{Andreas Peter}, {and} \bibinfo{person}{Florian Kerschbaum}.} \bibinfo{year}{2021}\natexlab{}.
\newblock \showarticletitle{Obfuscated Access and Search Patterns in Searchable Encryption}. In \bibinfo{booktitle}{\emph{28th Annual Network and Distributed System Security Symposium, {NDSS} 2021, virtually, February 21-25, 2021}}. \bibinfo{publisher}{The Internet Society}.
\newblock
\urldef\tempurl%
\url{https://www.ndss-symposium.org/ndss-paper/obfuscated-access-and-search-patterns-in-searchable-encryption/}
\showURL{%
\tempurl}


\bibitem[Song et~al\mbox{.}(2000)]%
        {DBLP:conf/sp/SongWP00}
\bibfield{author}{\bibinfo{person}{Dawn~Xiaodong Song}, \bibinfo{person}{David~A. Wagner}, {and} \bibinfo{person}{Adrian Perrig}.} \bibinfo{year}{2000}\natexlab{}.
\newblock \showarticletitle{Practical Techniques for Searches on Encrypted Data}. In \bibinfo{booktitle}{\emph{2000 {IEEE} Symposium on Security and Privacy}}. \bibinfo{pages}{44--55}.
\newblock
\href{https://doi.org/10.1109/SECPRI.2000.848445}{doi:\nolinkurl{10.1109/SECPRI.2000.848445}}


\bibitem[Stefanov et~al\mbox{.}(2014)]%
        {DBLP:conf/ndss/StefanovPS14}
\bibfield{author}{\bibinfo{person}{Emil Stefanov}, \bibinfo{person}{Charalampos Papamanthou}, {and} \bibinfo{person}{Elaine Shi}.} \bibinfo{year}{2014}\natexlab{}.
\newblock \showarticletitle{Practical Dynamic Searchable Encryption with Small Leakage}. In \bibinfo{booktitle}{\emph{21st Annual Network and Distributed System Security Symposium}}.
\newblock
\urldef\tempurl%
\url{https://www.ndss-symposium.org/ndss2014/practical-dynamic-searchable-encryption-small-leakage}
\showURL{%
\tempurl}


\bibitem[Sun et~al\mbox{.}(2018)]%
        {DBLP:conf/ccs/SunYLSSVN18}
\bibfield{author}{\bibinfo{person}{Shi-Feng Sun}, \bibinfo{person}{Xingliang Yuan}, \bibinfo{person}{Joseph~K. Liu}, \bibinfo{person}{Ron Steinfeld}, \bibinfo{person}{Amin Sakzad}, \bibinfo{person}{Viet Vo}, {and} \bibinfo{person}{Surya Nepal}.} \bibinfo{year}{2018}\natexlab{}.
\newblock \showarticletitle{Practical Backward-Secure Searchable Encryption from Symmetric Puncturable Encryption}. In \bibinfo{booktitle}{\emph{Proceedings of the 2018 ACM SIGSAC Conference on Computer and Communications Security}}. \bibinfo{pages}{763–780}.
\newblock
\showISBNx{9781450356930}
\href{https://doi.org/10.1145/3243734.3243782}{doi:\nolinkurl{10.1145/3243734.3243782}}


\bibitem[Vo et~al\mbox{.}(2023)]%
        {DBLP:journals/tkde/VoYSLNW23}
\bibfield{author}{\bibinfo{person}{Viet Vo}, \bibinfo{person}{Xingliang Yuan}, \bibinfo{person}{Shi{-}Feng Sun}, \bibinfo{person}{Joseph~K. Liu}, \bibinfo{person}{Surya Nepal}, {and} \bibinfo{person}{Cong Wang}.} \bibinfo{year}{2023}\natexlab{}.
\newblock \showarticletitle{ShieldDB: An Encrypted Document Database With Padding Countermeasures}.
\newblock \bibinfo{journal}{\emph{{IEEE} Trans. Knowl. Data Eng.}} \bibinfo{volume}{35}, \bibinfo{number}{4} (\bibinfo{year}{2023}), \bibinfo{pages}{4236--4252}.
\newblock
\href{https://doi.org/10.1109/TKDE.2021.3126607}{doi:\nolinkurl{10.1109/TKDE.2021.3126607}}


\bibitem[William W.~Cohen(2015)]%
        {Enron}
\bibfield{author}{\bibinfo{person}{CMU William W.~Cohen, MLD}.} \bibinfo{year}{2015}\natexlab{}.
\newblock \bibinfo{title}{Enron Email Datasets}.
\newblock
\urldef\tempurl%
\url{https://www.cs.cmu.edu/~./enron/}
\showURL{%
\tempurl}


\bibitem[Xing et~al\mbox{.}(2022)]%
        {DBLP:journals/vcomm/XingSQYC22}
\bibfield{author}{\bibinfo{person}{Xiaoshuang Xing}, \bibinfo{person}{Gaofei Sun}, \bibinfo{person}{Jin Qian}, \bibinfo{person}{Dongxiao Yu}, {and} \bibinfo{person}{Xiuzhen Cheng}.} \bibinfo{year}{2022}\natexlab{}.
\newblock \showarticletitle{Intermittent jamming for eavesdropping defense in {WAVE} based vehicular networks}.
\newblock \bibinfo{journal}{\emph{Veh. Commun.}}  \bibinfo{volume}{38} (\bibinfo{year}{2022}), \bibinfo{pages}{100542}.
\newblock
\href{https://doi.org/10.1016/J.VEHCOM.2022.100542}{doi:\nolinkurl{10.1016/J.VEHCOM.2022.100542}}


\bibitem[Xu et~al\mbox{.}(2023)]%
        {DBLP:conf/ccs/XuZXYW23}
\bibfield{author}{\bibinfo{person}{Lei Xu}, \bibinfo{person}{Leqian Zheng}, \bibinfo{person}{Chengzhi Xu}, \bibinfo{person}{Xingliang Yuan}, {and} \bibinfo{person}{Cong Wang}.} \bibinfo{year}{2023}\natexlab{}.
\newblock \showarticletitle{Leakage-Abuse Attacks Against Forward and Backward Private Searchable Symmetric Encryption}. In \bibinfo{booktitle}{\emph{Proceedings of the 2023 ACM SIGSAC Conference on Computer and Communications Security}}. \bibinfo{pages}{3003–3017}.
\newblock
\showISBNx{9798400700507}
\href{https://doi.org/10.1145/3576915.3623085}{doi:\nolinkurl{10.1145/3576915.3623085}}


\bibitem[Xu et~al\mbox{.}(2024)]%
        {DBLP:journals/tdsc/XuZDWWJ24}
\bibfield{author}{\bibinfo{person}{Lei Xu}, \bibinfo{person}{Anxin Zhou}, \bibinfo{person}{Huayi Duan}, \bibinfo{person}{Cong Wang}, \bibinfo{person}{Qian Wang}, {and} \bibinfo{person}{Xiaohua Jia}.} \bibinfo{year}{2024}\natexlab{}.
\newblock \showarticletitle{Toward Full Accounting for Leakage Exploitation and Mitigation in Dynamic Encrypted Databases}.
\newblock \bibinfo{journal}{\emph{{IEEE} Trans. Dependable Secur. Comput.}} \bibinfo{volume}{21}, \bibinfo{number}{4} (\bibinfo{year}{2024}), \bibinfo{pages}{1918--1934}.
\newblock
\href{https://doi.org/10.1109/TDSC.2023.3296189}{doi:\nolinkurl{10.1109/TDSC.2023.3296189}}


\bibitem[Xu et~al\mbox{.}(2022)]%
        {DBLP:journals/tifs/XuSWCWLJ22}
\bibfield{author}{\bibinfo{person}{Peng Xu}, \bibinfo{person}{Willy Susilo}, \bibinfo{person}{Wei Wang}, \bibinfo{person}{Tianyang Chen}, \bibinfo{person}{Qianhong Wu}, \bibinfo{person}{Kaitai Liang}, {and} \bibinfo{person}{Hai Jin}.} \bibinfo{year}{2022}\natexlab{}.
\newblock \showarticletitle{{ROSE:} Robust Searchable Encryption With Forward and Backward Security}.
\newblock \bibinfo{journal}{\emph{{IEEE} Trans. Inf. Forensics Secur.}}  \bibinfo{volume}{17} (\bibinfo{year}{2022}), \bibinfo{pages}{1115--1130}.
\newblock
\href{https://doi.org/10.1109/TIFS.2022.3155977}{doi:\nolinkurl{10.1109/TIFS.2022.3155977}}


\bibitem[Zhang et~al\mbox{.}(2023)]%
        {DBLP:conf/uss/Zhang00YL23}
\bibfield{author}{\bibinfo{person}{Xianglong Zhang}, \bibinfo{person}{Wei Wang}, \bibinfo{person}{Peng Xu}, \bibinfo{person}{Laurence~T. Yang}, {and} \bibinfo{person}{Kaitai Liang}.} \bibinfo{year}{2023}\natexlab{}.
\newblock \showarticletitle{High Recovery with Fewer Injections: Practical Binary Volumetric Injection Attacks against Dynamic Searchable Encryption}. In \bibinfo{booktitle}{\emph{32nd {USENIX} Security Symposium}}. \bibinfo{pages}{5953--5970}.
\newblock
\urldef\tempurl%
\url{https://www.usenix.org/conference/usenixsecurity23/presentation/zhang-xianglong}
\showURL{%
\tempurl}


\bibitem[Zhang et~al\mbox{.}(2016)]%
        {DBLP:conf/uss/ZhangKP16}
\bibfield{author}{\bibinfo{person}{Yupeng Zhang}, \bibinfo{person}{Jonathan Katz}, {and} \bibinfo{person}{Charalampos Papamanthou}.} \bibinfo{year}{2016}\natexlab{}.
\newblock \showarticletitle{All Your Queries Are Belong to Us: The Power of File-Injection Attacks on Searchable Encryption}. In \bibinfo{booktitle}{\emph{25th {USENIX} Security Symposium}}. \bibinfo{pages}{707--720}.
\newblock
\urldef\tempurl%
\url{https://www.usenix.org/conference/usenixsecurity16/technical-sessions/presentation/zhang}
\showURL{%
\tempurl}


\bibitem[Zipf(2016)]%
        {zipf2016human}
\bibfield{author}{\bibinfo{person}{George~Kingsley Zipf}.} \bibinfo{year}{2016}\natexlab{}.
\newblock \bibinfo{booktitle}{\emph{Human behavior and the principle of least effort: An introduction to human ecology}}.
\newblock \bibinfo{publisher}{Ravenio books}.
\newblock


\end{thebibliography}

\appendix



\section{Background}
\label{app:background}

\subsection{SSE and DSSE}
An SSE scheme \cite{DBLP:conf/ccs/Bost16,DBLP:conf/ccs/BostMO17,DBLP:conf/ccs/ChamaniPPJ18,DBLP:conf/ccs/CurtmolaGKO06,DBLP:conf/ccs/KamaraPR12,DBLP:conf/ccs/SunYLSSVN18,DBLP:conf/ndss/Chen0PLS0L23,DBLP:conf/ndss/StefanovPS14,DBLP:conf/sp/NaveedPG14,DBLP:journals/tifs/DouDXWXCJ24,DBLP:conf/sp/SongWP00,DBLP:journals/tifs/XuSWCWLJ22} allows a client to perform keyword search over an encrypted dataset $ED$ stored on a server.
On top of SSE, the DSSE \cite{DBLP:conf/ccs/Bost16,DBLP:conf/ccs/BostMO17,DBLP:conf/ccs/ChamaniPPJ18,DBLP:conf/ccs/KamaraPR12,DBLP:conf/ccs/SunYLSSVN18,DBLP:conf/ndss/Chen0PLS0L23,DBLP:conf/ndss/StefanovPS14,DBLP:conf/sp/NaveedPG14,DBLP:journals/tifs/DouDXWXCJ24,DBLP:journals/tifs/XuSWCWLJ22} allows the client to update the dataset. 
Typically, an SSE scheme consists of setup and search protocols. 
At the setup, the client holds a dataset $DB=\{d_1,...d_n\}$, extracts an index from $DB$, and eventually encrypts the index and $DB$, which will be stored on the server. 
To search a keyword $k$, the client first generates a query token $tk(k)$ for the server and then receives a response. 
The response contains the identity $id(d)$ for all the files $d$ containing $k$. 
The client can then utilize the identities to retrieve the corresponding files. 
We use $DB(k)$ and $DB(tk(k))$ to denote all the identities of files containing the keyword $k$.

Many DSSE schemes provide the forward/backward privacy (FP/BP) \cite{DBLP:conf/ccs/BostMO17,DBLP:conf/ccs/Bost16}. 
Generally, forward privacy requires that newly updated entries of the index cannot relate to previous search results, while backward privacy guarantees that search queries cannot link to entries that have already been deleted.   
Bost et al. \cite{DBLP:conf/ccs/BostMO17} define backward privacy as three ``flavors'', Type I, II, and III, each offering progressively lower levels of security. 
Even with the forward and Type I BP security, there are still leakage patterns that attackers could observe.
We summarize the common leakage patterns of search queries used in previous attacks as follows.
\\
$\bullet$ \textbf{Access pattern (AP)} of a search query refers to the set of identities of the encrypted files in the response under the database. 
\\
$\bullet$ \textbf{File volume pattern (FVP)} of a search query is the set of encrypted file sizes in the response under the database.
\\
$\bullet$ \textbf{Volume pattern (VP)} of a search query is the number of encrypted files in the response under the database. 
The VP can also be inferred from the AP or the FVP, as the attacker can count the number of elements in either.
\\
$\bullet$ \textbf{Search pattern (SP)} of two queries indicates whether the two queries correspond to the same keyword.
One can infer the SP from the AP or the FVP by comparing the leakage patterns of two queries. 
However, in the context of DSSE, updates can introduce errors in this inference. 
Furthermore, the SP is unavailable to an IOA.


\subsection{Attacks Against SSE.} 
Attacks against SSE include known-data and similar-data attacks depending on the prior knowledge. 
The known-data attacks acquire a part of the plaintexts of the client's dataset, while the similar-data attacks rely on a similar dataset or a distribution of the client's dataset.

\emph{Known-data attacks}. Islam et al. \cite{DBLP:conf/ndss/IslamKK12} proposed the first attack, requiring the known queries. 
Cash et al. \cite{DBLP:conf/ccs/CashGPR15} introduced the Count attack, which can recover most search queries without known queries. 
In \cite{DBLP:conf/ndss/BlackstoneKM20}, Blackstone et al. proposed an attack that achieves near-perfect accuracy with fully known files. 
Ning et al. \cite{DBLP:conf/ccs/NingHPYL0D21} proposed LEAP, which, using only 1\% of the plaintexts of encrypted files, is able to recover half of all search queries with perfect accuracy. 

\emph{Similar-data attacks.} Liu et al. \cite{DBLP:journals/isci/LiuZWT14} use the search pattern to obtain the search frequency of queries and match the frequency with the auxiliary frequency of the keywords for query recovery.  
Oya et al. \cite{DBLP:conf/uss/OyaK21} apply the search pattern and volume pattern and adopt the maximum likelihood estimation to find a match between search queries and keywords.  
Pouliot et al. \cite{DBLP:conf/ccs/PouliotW16} exploit the access pattern to construct the co-occurrence matrix of queries and keywords. 
They use the weighted graph matching to design an effective attack called GraphM. 
Damie et al. \cite{DBLP:conf/uss/Damie0P21} propose the refined score attack, which starts with some known queries, iteratively recovers search queries, and treats the recovered queries in each iteration as known queries.
Oya et al. \cite{DBLP:conf/uss/OyaK22} develop the IHOP. 
Utilizing the search frequency and co-occurrence matrix, IHOP models query recovery as a quadratic assignment problem and solves it by proposing an iterative heuristic algorithm. 
Nie et al. \cite{DBLP:conf/uss/Nie00ZYL24} introduce the Jigsaw attack, which first recovers the most distinctive queries. 
Based on the recovered queries, the attacker can gradually recover other search queries. 
However, the aforementioned attacks assume a static database and are not designed to target DSSE. 


\subsection{Attacks Against DSSE.} The most commonly studied attacks targeting DSSE are injection attacks \cite{DBLP:conf/uss/ZhangKP16,DBLP:conf/eurosp/PoddarWLP20,DBLP:conf/ndss/BlackstoneKM20,DBLP:conf/uss/Zhang00YL23}, which actively inject crafted files into the client's database rather than passively observing the leakage.
The passive attacks against DSSE have also been investigated. 
Xu et al. \cite{DBLP:conf/ccs/XuZXYW23} exploit the file volume pattern and proposed the FMA.  
The attacker first calculates the similarity between search queries to deduce the search pattern and query frequency and then uses both the query frequency and the auxiliary frequency to match the queries with keywords. 
Salmani et al. \cite{DBLP:conf/codaspy/Salmani021} also propose an attack against DSSE to infer the SP by the AP leakage.

Other passive attacks exploit the information from updates. 
Haltiwanger \cite{DBLP:conf/codaspy/HaltiwangerH24} designs UF and UFID, where the attacker additionally requires the leakage of updates, such as update frequency and the file identities associated with each update, to recover queries. 
Xu et al. \cite{DBLP:journals/tdsc/XuZDWWJ24} introduce an attack using the plaintexts of the encrypted database. 
Based on the ``known-data'' setting, the attacker can gain more information during database updates. 
Xu et al. \cite{DBLP:conf/ccs/XuZXYW23} develop the VIA attack with the plaintexts of updated files. 
We note that attacks in which the attacker knows the plaintexts of updates are similar to injection attacks \cite{DBLP:conf/uss/ZhangKP16,DBLP:conf/eurosp/PoddarWLP20,DBLP:conf/ndss/BlackstoneKM20,DBLP:conf/uss/Zhang00YL23}, except that, in this case, the attacker does not design the injected files. 


\section{Examples: IOAs and Persistent Attackers}
\label{app:detailed examples of ioas}

\begin{table*}[tp]
\centering
    \begin{threeparttable} 
	    \caption{An Example: leakage observed by an IOA.
     }
     \label{tab:exampleIOAs}
		\centering
        {
    \begin{tabular}{p{7.6cm}|p{4.5cm}|p{4.5cm}}
 \hline
\multirow{2}{*}{Interactions between Client and Server}     &\multicolumn{2}{c}{Attacker's Observations on Queries}\\
\cline{2-3}
&Access Pattern (AP) Leakage &File Volume Pattern (FVP) Leakage  \\ 
\hline
\multicolumn{3}{c}{Setup}\\
\hline
Add: encrypted files ``ab3'' and ``786cd'' to keyword ``apple''&\cellcolor{gray!20}&\cellcolor{gray!20}\\
Add: encrypted files ``ef23fa'' and ``2ac7'' to keyword ``banana''&\cellcolor{gray!20}&\cellcolor{gray!20}\\
\hline
\multicolumn{3}{c}{Attacker Stays Online in Round One}\\
\hline
Query: keyword ``apple''&\cellcolor{blue!20}Encrypted files ``ab3'' and ``786cd'' are responded for token A&\cellcolor{blue!20}Encrypted files with sizes 3 and 5 are responded for token A\\
Query: keyword ``banana''&\cellcolor{blue!20}Encrypted files ``ef23fa'' and ``2ac7'' are responded for token B&\cellcolor{blue!20}Encrypted files with sizes 6 and 4 are responded for token B\\
\hline
\multicolumn{3}{c}{Attacker Goes Offline}
\\
\hline
Update: encrypted files ``ab3'', ``786cd'', ``ef23fa'', and ``2ac7'' to ``ef34'', ``2ae4'', ``1234a'', and ``342ae'', respectively&\cellcolor{gray!20}&\cellcolor{gray!20}\\
\hline
\multicolumn{3}{c}{Attacker Stays Online in Round Two}\\
\hline
Query: keyword ``apple''&\cellcolor{blue!20}Encrypted files ``ef34'' and ``2ae4'' are responded for token C&\cellcolor{blue!20}Encrypted files with sizes 4 and 4 are responded for token C\\
Query: keyword ``banana''&\cellcolor{blue!20}Encrypted files ``1234a'' and ``342ae'' are responded for token D&\cellcolor{blue!20}Encrypted files with sizes 5 and 5 are responded for token D\\
\cline{2-3}
&\multicolumn{2}{p{9.365cm}}
{\cellcolor{blue!20}\makecell{(The connections of encrypted files under ``apple'' and ``banana'' are\\broken and thus the attacker fails to directly track whether \\A or B are for the same keyword with C)}}\\
\hline
\end{tabular}
        } 
    \end{threeparttable} 
\end{table*}



\begin{table*}[tp]
\centering
    \begin{threeparttable} 
	    \caption{An Example: leakage observed by a persistent attacker.
     }
     \label{tab:examplepersistent}
		\centering
        {
    \begin{tabular}{p{7.6cm}|p{4.5cm}|p{4.5cm}}
 \hline
\multirow{2}{*}{Interactions between Client and Server}     &\multicolumn{2}{c}{Attacker's Observation on Queries}\\
\cline{2-3}
&Access Pattern (AP) Leakage &File Volume Pattern (FVP) Leakage  \\ 
\hline
Add: encrypted files ``ab3'' and ``786cd'' to keyword ``apple''&\cellcolor{blue!20}&\cellcolor{blue!20}\\
\hline
Query: keyword ``apple''&\cellcolor{blue!20}Encrypted files ``ab3'' and ``786cd'' are responded for token A&\cellcolor{blue!20}Encrypted files with sizes 3 and 5 are responded for token A
\\
\hline
Update: encrypted file ``ab3'' to ``ef34''&\cellcolor{blue!20}&\cellcolor{blue!20}\\
\hline
Query: keyword ``apple''&\cellcolor{blue!20}Encrypted files ``ef34'' and ``786cd'' are responded for token B&\cellcolor{blue!20}Encrypted files with sizes 4 and 5 are responded for token B\\
\cline{2-3}
&\multicolumn{2}{p{9.365cm}}{\cellcolor{blue!20}\makecell{(For AP, the attacker infers that ``ef34'' and ``ab3'' are of the same file and\\
for FVP, the attacker infers that a file with size 3 is updated to size 4)}}\\
\hline
Update: encrypted file ``786cd'' to ``2ae4''&\cellcolor{blue!20}&\cellcolor{blue!20}\\
\hline
Query: keyword ``apple''&\cellcolor{blue!20}Encrypted files ``ef34'' and ``2ae4'' are responded for token C&\cellcolor{blue!20}Encrypted files with sizes 4 and 4 are responded for token C\\
\cline{2-3}
&\multicolumn{2}{p{9.365cm}}{\cellcolor{blue!20}\makecell{(For AP, the attacker infers that ``2ae4'' and ``786cd'' are of the same file, \\and for FVP, the attacker infers that a file with size 5 is updated to size 4)\\
(The attacker guesses A, B, and C are for the same keyword via inference)}}\\
\hline
\end{tabular}
        } 
    \end{threeparttable} 
\end{table*}

Attackers' observations may be detected by modern intrusion detection systems \cite{DBLP:conf/sp/RehmanAH24,DBLP:conf/ndss/LiDXWSC0C024,DBLP:conf/uss/ChenY0L0XXLW24}, which are deployed both on servers and in communication channels. 
Besides, active or intermittent jamming \cite{DBLP:journals/vcomm/XingSQYC22} may disrupt observations, forcing attackers to frequently restart observing. 
Studies \cite{DBIR,splunk} also report that $>50$ breaches are detected within days or less, and detect-and-respond countermeasures have become increasingly rapid. 
To evade such detection, attackers may actively go offline. 
This strategy is seen in modern malware like Trickbot, which implements sleep-wake cycles to bypass behavioral analysis \cite{Trickbot}, and in APT attacks that include a sleep mode to reduce traceability \cite{ambika2020improved}. 
Such intermittent observations only provide ``fragmented'' leakage.
We illustrate a leakage example for intermittent observation in Table \ref{tab:exampleIOAs} (covering the two cases in Section \ref{sec:intro}). We assume the attacker only knows the access pattern or file volume pattern as discussed in Section \ref{sec:scenarios}. The blue and gray colors indicate the attacker is online and offline, respectively.

\noindent $\bullet$ In setup, the client adds encrypted files ``ab3'' and ``786cd'' for the keyword ``apple'', and ``ef23fa'' and ``2ac7'' for ``banana''.

\noindent $\bullet$ In the first round of observation, the client queries ``apple'' and ``banana''. 
If the AP is available, the attacker observes encrypted file sets \{``ab3'', ``786cd''\} and \{``ef23fa'', ``2ac7''\}, corresponding to queries A and B.

\noindent $\bullet$ The attacker then temporarily goes offline to avoid detection. 
Updates to the files triggering re-encryptions during this period are invisible to the attacker. 

\noindent $\bullet$ After rejoining, in the second round, the client queries the same keywords ``apple'' and ``banana'', but the attacker now sees different encrypted file sets \{``ef34'', ``2ae4''\} and \{``ef23fa'', ``2ac7''\} are responded to queries C and D, making it difficult to correlate these with earlier observations. And the attacker fails to directly track the search pattern.  

\noindent If the FVP is leaked, updates can obscure matching due to changes in file sizes. 
A persistent attacker is assumed to maintain continuous observations, always correlating the encrypted files. 
As shown in Table \ref{tab:examplepersistent}, when first observing the query for keyword ``apple'', the attacker observes encrypted files ``ab3'' and ``786cd'' are responded for query A if AP is available. Then, the encrypted file ``ab3'' is updated to ``ef34''. 
The persistent attacker knows the relation of ``ab3'' and ``ef34'' via the leakage of the update (which can be also obtained through later queries). 
When the client queries ``apple'' again, the attacker sees ``ef34'' and ``786cd'' are responded for query B. 
The attacker can guess the two queries are for the same keyword and ``ab3'' and ``ef34'' are the encryptions of the same file. 
Since the database changes gradually, only some responses differ, making the guess more accurate. 
The same holds in the FVP-available scenario and for other keywords like ``banana'', allowing the persistent attacker to consistently correlate encrypted files and track search patterns - at the cost of continuous observation and a higher risk of detection.


By comparison, an IOA 
can operate more stealthily using the following strategy:
1) Estimate the duration and detection probability of each observation window. 
2) Choose observation periods with acceptable detection risk. 
3) Enter sleep mode post-observation to avoid detection. 
4) Repeat the cycle. 

Such strategies enable the IOA to safely collect sufficient leakage for query recovery.
For example, an APT attacker using njRAT (a remote access trojan) can exfiltrate data at 3 MB/s for about 75 seconds without raising alarms \cite{DBLP:journals/corr/abs-2502-08830}. 
This is enough to transmit leakage from about 5,000 queries (assuming the representations of each encrypted file in access or file volume pattern are 30 bits and each query leakage involves 10,000 files). 
Our evaluation demonstrates that Peekaboo can reach $>50\%$ query recovery accuracy with only 2,500 queries observed in a round, see Figure \ref{fig:test_query_number}.
While real-world APT is not directly implemented (which is out of the scope of this paper), our evaluation demonstrates that even intermittent observation can remain highly effective. 
%




\section{Discussions about Server Being the Attacker}
\label{app:sba}
Some prior works \cite{DBLP:conf/uss/Damie0P21,DBLP:conf/uss/Nie00ZYL24} model the server as the attacker, capable of continuously observing and exploiting query leakage - i.e., a persistent attacker. 
Similarly, an external attacker who fully compromises the server gains the same persistent visibility, such as continuous access to query leakage logs. 
While such attackers can easily launch attacks, they could not be practical in real-world applications. 
In practice, few data breaches originate from servers or cloud providers \cite{DBIR2}, and full persistent control by an external attacker could be unrealistic.
As demonstrated in Appendix \ref{app:detailed examples of ioas}, attack activities are detectable, often forcing attackers to withdraw or jump to dormancy.

\section{Sap and Jigsaw}
\label{app:introduction to sap and jigsaw}
\noindent \textbf{Sap \cite{DBLP:conf/uss/OyaK21}} targets static SSE and makes good use of the frequency and volume pattern to recover queries. 
It solves the maximum likelihood problem
\begin{equation}
\label{eq:sap1}
    \mathbf{P}=\argmax\limits_{\mathbf{P}\in\mathcal{P}} \text{Pr}(\boldsymbol{\rho},\mathbf{F}_r,\mathbf{v}_r,n_D|\mathbf{F}_s,\mathbf{v}_s,\mathbf{P}),
\end{equation}
where the $\mathbf{v}_r$ and $\mathbf{v}_s$ are vectors of the volume for all observed unique queries and keywords. 
The $\mathbf{v}_r$ and $\mathbf{v}_s$ remain unchanged during the attack. 
$n_D$ is the number of files in the client's encrypted database, and $\boldsymbol{\rho}=[\rho^1,...,\rho^\tau]$ is a vector of numbers of observed queries in each time slot. 
Oya et al. assume that the querying behavior and the volume of queries are independent, i.e.,
\begin{equation}
\begin{split}
    \text{Pr}(\mathbf{F}_r,\boldsymbol{\rho},\mathbf{v}_r,n_D|\mathbf{F}_s,\mathbf{v}_s,\mathbf{P})=\\
    \text{Pr}(\boldsymbol{\rho},\mathbf{F}_r|\mathbf{F}_s,\mathbf{P})\text{Pr}(\mathbf{v}_r,n_D|\mathbf{v}_s,\mathbf{P}).
\end{split}
\end{equation}
For the frequency part, assuming the client queries each keyword independently, the query number of each keyword follows a multinomial distribution. 
For the volume part, they model the volume of a query for the $i$-th keyword as a binomial random variable with $n_D$ trials and probability $\mathbf{v}_s[i]$. 
For simplicity in calculation, Oya et al. choose to maximize the logarithm of Equation \ref{eq:sap1}. 
They define two cost matrices $\mathbf{C}_f$ and $\mathbf{C}_v$ whose $(i,j)$-th entries are
\begin{equation}
    \mathbf{C}_f[i][j]=-\sum\limits_{k=1}^\tau (\rho^k\cdot \mathbf{f}_r^k[j]\cdot \log(\mathbf{f}_s^k[i]),
\end{equation}
\begin{equation}
\begin{split}
        \mathbf{C}_v[i][j]= -(n_D\cdot \mathbf{v}_r[j]\cdot\log \mathbf{v}_s[i] \\ +n_D(1-\mathbf{v}_r[j])\cdot \log(1-\mathbf{v}_s[i])).
    \label{eq:cal_Cv}
\end{split}
\end{equation}
Then, they formulate the Equation \ref{eq:sap1} as 
\begin{equation}
    \mathbf{P}=\argmin\limits_{\mathbf{P}\in \mathcal{P}}(\mathbf{P}^\top((1-\alpha)\mathbf{C}_v+\alpha \mathbf{C}_f)).
\end{equation}
They eventually solve this problem with the Hungarian algorithm \cite{kuhn1955hungarian} and get a mapping from the queries to keywords.

\noindent \textbf{Jigsaw \cite{DBLP:conf/uss/Nie00ZYL24}} leverages the search, volume, and access pattern to recover queries. 
In \cite{DBLP:conf/uss/Nie00ZYL24}, Nie et al. show that the frequency and volume of queries follow Zipf's law, from which they conclude that a part of queries with high frequency and volume are much easier to recover than others.
Based on this, they propose Jigsaw. %
Jigsaw receives the query sequence $Td_r$, the volume $\mathbf{v}_r$ of queries in $Td_r$, their total search frequency $\mathbf{f}_r$, and the responded file identities of each query. 
It also takes the auxiliary knowledge as input, which may include the keyword set $W$, the volume $\mathbf{v}_s$ of the keywords in $W$, their total search frequency $\mathbf{f}_s$, and the corresponding file identities. 
In Jigsaw, the attacker first calculates the differential distance $d_{td_i}$ of any query $td_i$ as 
\begin{equation}
    d_{td_i}=\min\limits_{td_j\in Td_r\land j\ne i}\alpha\cdot|\mathbf{v}_{td_i}-\mathbf{v}_{td_j}|+(1-\alpha)|\mathbf{f}_{td_i}-\mathbf{f}_{td_j}|,
    \label{eq:differential distance}
\end{equation}
which measures the distance between the query $td_i$ and its nearest neighbor.

The attacker treats the top $BaseRec$ queries with the largest differential distance as distinctive queries and then calculates the distance $s(td_i,w_j)$ between the distinctive query $td_i$ and every keyword $w_j$ to find and match the nearest keyword for $td_i$. 
The $s(td_i,w_j)$ is calculated as
\begin{equation}
    s(td_i,w_j) = \alpha\cdot|\mathbf{v}_{td_i}-\mathbf{v}_{w_j}|+(1-\alpha)|\mathbf{f}_{td_i}-\mathbf{f}_{w_j}|.
    \label{eq:s}
\end{equation}

For subsequent recovery, the attacker constructs the co-occurrence matrix $\mathbf{C}_r$ and $\mathbf{C}_s$ of queries and keywords, respectively.
It then builds the co-occurrence matrix for the recovered distinctive queries and their matched keywords, extracting the corresponding rows and columns from $\mathbf{C}_r$ and $\mathbf{C}_s$ to form $\mathbf{C}_r'$ and $\mathbf{C}_s'$.
Then, it uses the extracted co-occurrence matrix to confirm the previously recovered queries. 
For the $i$-th distinctive query, the attacker computes the $revconf$ as
\begin{equation}
    revconf = ||\mathbf{C}_r'[i] - \mathbf{C}_s'[i]||.
    \label{eq:revconf}
\end{equation}
The attacker keeps only $ConfRec$ distinctive queries with the smallest $revconf$ and treats them as known queries in later steps.

To recover all the queries, Jigsaw uses a similar method as in \cite{DBLP:conf/uss/Damie0P21} by extracting the $\mathbf{C}_{rs}$ and $\mathbf{C}_{ss}$ from $\mathbf{C}_r$ and $\mathbf{C}_s$, where $\mathbf{C}_{rs}$ is the co-occurrence matrix between unmatched queries and known queries and $\mathbf{C}_{ss}$ is the co-occurrence matrix between the matched keywords and those unmatched.
For each of the unmatched queries $td$, Jigsaw computes the $score$ of each keyword $w$ as
\begin{equation}
    score =-\ln( \beta||\mathbf{C}_{rs}[td]-\mathbf{C}_{ss}[w]||+(1-\beta)s(td,w)).
    \label{eq:score}
\end{equation}
It sets the $certainty$ of each unmatched query as the subtraction between the largest and the second largest $score$. 
Then, the top $RefSpeed$ queries with the largest $certainty$ are matched to the keywords with the largest $score$. 
These matched queries are the known queries in subsequent attacks.  
The attacker can repeat the above process until all queries match the keywords.

\section{Details about \textsc{Match}}
\label{app:details about match}


The \textsc{Match} (Algorithm \ref{alg:match_detailed}) takes two collections of groups $Groups_1$ and $Groups_2$ and their corresponding index matrix $\mathbf{ID}_1$ and $\mathbf{ID}_2$ as input, and outputs the matched pairs of groups.
It first calculates the co-occurrence matrix $\mathbf{C}_1$ and $\mathbf{C}_2$ from the index matrix (line \ref{alg:match_ids}-\ref{alg:match_ide}). For clarity, we define the one with fewer groups as $\mathbf{C}_1$. 
It then calls the IHOP \cite{DBLP:conf/uss/OyaK22} to map all the groups corresponding to $\mathbf{C}_1$ to those corresponding to $\mathbf{C}_2$ (line \ref{alg:match_ihops}-\ref{alg:match_ihope}). 
Afterwards, it removes certain incorrectly matched groups from the output of IHOP (line \ref{alg:match_rs}-\ref{alg:match_re}), as some queries may not appear in both $Groups_1$ and $Groups_2$.
It further extracts the columns and rows from $\mathbf{C}_1$ and $\mathbf{C}_2$ to form $\mathbf{C}_1'$ and $\mathbf{C}_2'$, where the columns and rows correspond to the matched groups in $P$. 
%
The group of each column and row of $\mathbf{C}_1'$ must match the group of the corresponding column and row in $\mathbf{C}_2'$ at the same position. 
For the $i$-th row of $\mathbf{C}_1'$, we compute the distance $dis_i=||\mathbf{C}_1'[i]-\mathbf{C}_2'[i]||$ as the error of the corresponding pair of groups. 
We then sort the errors for each pair and remove $p_g\times|\mathbf{C}_1|$ pairs with the largest errors. 

\noindent\textbf{IHOP \cite{DBLP:conf/uss/OyaK22}} formulates the query recovery problem as a quadratic assignment problem (QAP) and solves it by iterating over linear assignment problems (LAP). 
In each iteration, the attacker begins by assigning search queries to keywords. It then fixes a random part of the assignment and frees the remaining search queries and keywords. Treating the fixed assignment as recovered search queries, the attacker re-computes an assignment of the free search queries to the free keywords by LAP and updates the assignment.
If some assignments are fixed, the attacker can use the quadratic terms while keeping the problem linear.
With more iterations, the results are closer to the optimum.
By modifying the coefficients, IHOP provides a flexible framework for quadratic query recovery, allowing it to adapt to different objective functions. 
\begin{figure}[!t]
  \begin{algorithm}[H]
    \caption{Match the groups from two observations.}
    \label{alg:match_detailed}
    \begin{algorithmic}[1]
        \STATE \textbf{Procedure} \textsc{Match}($Groups_1,\mathbf{ID}_1,Groups_2,\mathbf{ID}_2$)
            \IF {$|Groups_1|<|Groups_2|$} \label{alg:match_ids}
            \STATE $\mathbf{C}_1\gets \mathbf{ID}_1\mathbf{ID}^\top_1/|\mathbf{ID}_1[0]|,\mathbf{C}_2\gets \mathbf{ID}_2\mathbf{ID}^\top_2/|\mathbf{ID}_2[0]|$;
            \ELSE
            \STATE $\mathbf{C}_2\gets \mathbf{ID}_1\mathbf{ID}^\top_1/|\mathbf{ID}_1[0]|,\mathbf{C}_1\gets \mathbf{ID}_2\mathbf{ID}^\top_2/|\mathbf{ID}_2[0]|$;
            \ENDIF \label{alg:match_ide}
            \STATE Treat $\mathbf{C}_1$ as $\mathbf{V}$; Treat $\mathbf{C}_2$ as $\widetilde{\mathbf{V}}$;\label{alg:match_ihops} 
            \STATE $P\gets IHOP(\mathbf{V},\widetilde{\mathbf{V}},n_{iters},p_{free})$;\label{alg:match_ihope} 
            \STATE Extract and realign the columns and rows of $\mathbf{C}_1$ and $\mathbf{C}_2$ to $\mathbf{C}_1'$ and $\mathbf{C}_2'$ according to $P$; \label{alg:match_rs}
            \RIGHTCOMMENT {The group of $i$-th column and row of $\mathbf{C}_1'$ after realignment are matched to the group of the $i$-th column and row of $\mathbf{C}_2'$ after realignment;}
            \STATE $Cand\gets\emptyset$;
            \FOR{$i=1$ to $|\mathbf{C}_1'|$}
            \STATE $dis_i=||\mathbf{C}_1'[i]-\mathbf{C}_2'[i]||$;
            \STATE Append $(p,dis_i)$ to $Cand$;
            \RIGHTCOMMENT {$p$ is the corresponding pair of groups of $\mathbf{C}_1'[i]$ and $\mathbf{C}_2'[i]$;}
            \ENDFOR
            \STATE Sort $Cand$ according to $Cand.dis$;
            \STATE Remove $p_g\times|\mathbf{C}_1'|$ pairs with largest $dis$ from $P$;\label{alg:match_re}
            \RETURN Pairs of groups that are matched in $P$;
        \STATE \textbf{End Procedure}
    \end{algorithmic}
  \end{algorithm}
  \Description{Algorithm of match the groups from two observations, which are fully described in the text.}
\end{figure}

\section{Parameter Selection and Impact}
\label{app:parameter selection}

\begin{figure}
    \centering
    \subfloat[$\delta$, Enron, AP]
	{
 \label{fig:test_delta}
		\begin{minipage}{.46\linewidth}
			\centering
                \includegraphics[width=\linewidth]{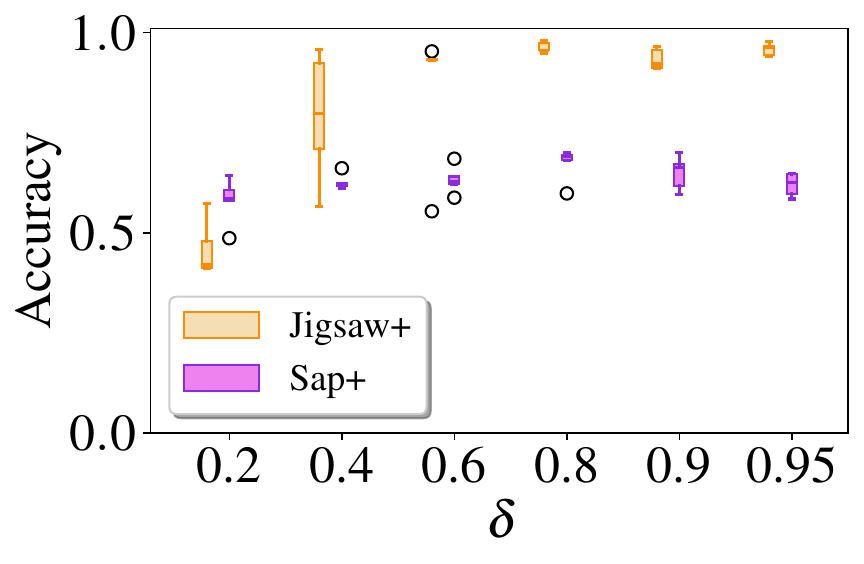}
		\end{minipage}
	}
    \subfloat[$maxlevel$, Enron, AP]
	{
 \label{fig:test_maxlevel}
		\begin{minipage}{.46\linewidth}
			\centering
			\includegraphics[width=\linewidth]{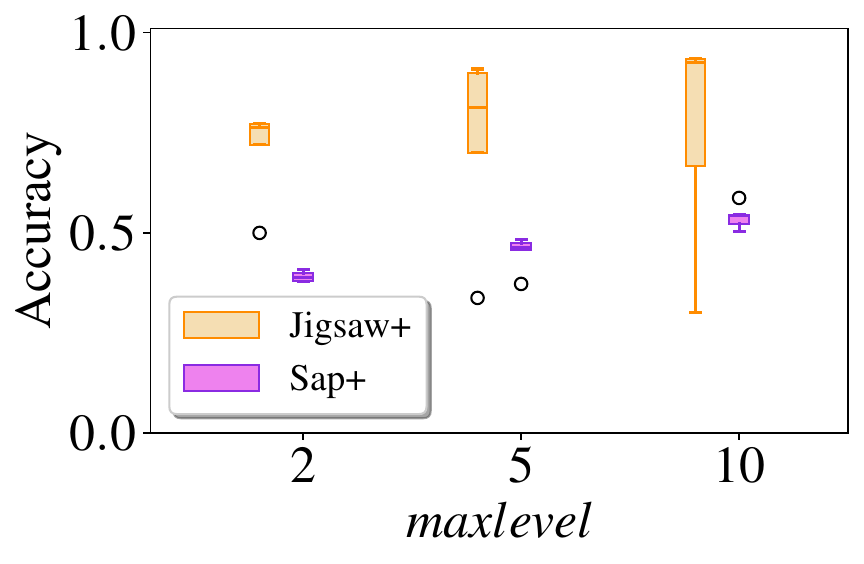}
		\end{minipage}
 	}
    
    \subfloat[$p_g$, Enron, AP]
	{
 \label{fig:test_p_g_AP}
		\begin{minipage}{.46\linewidth}
			\centering
                \includegraphics[width=\linewidth]{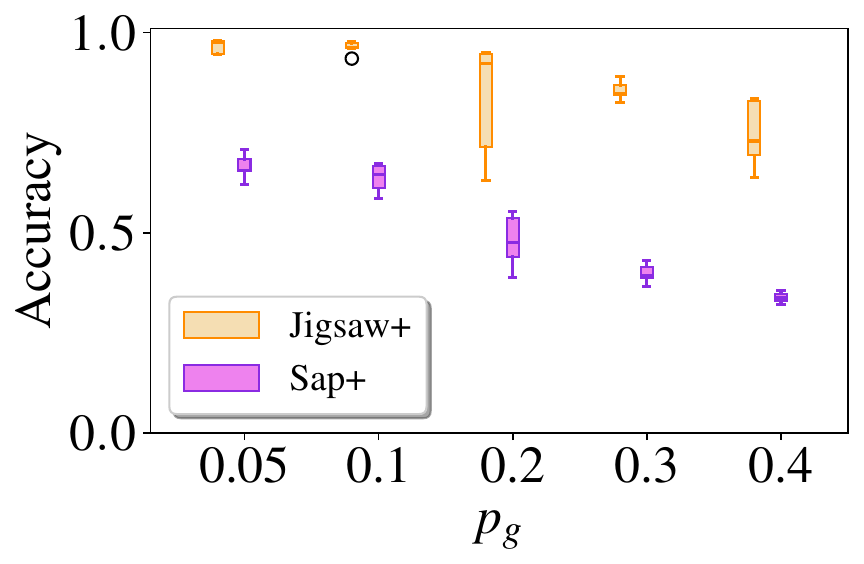}
		\end{minipage}
	}
    \subfloat[$p_g$, Enron, FVP]
	{
 \label{fig:test_p_g_FVP}
		\begin{minipage}{.46\linewidth}
			\centering
			\includegraphics[width=\linewidth]{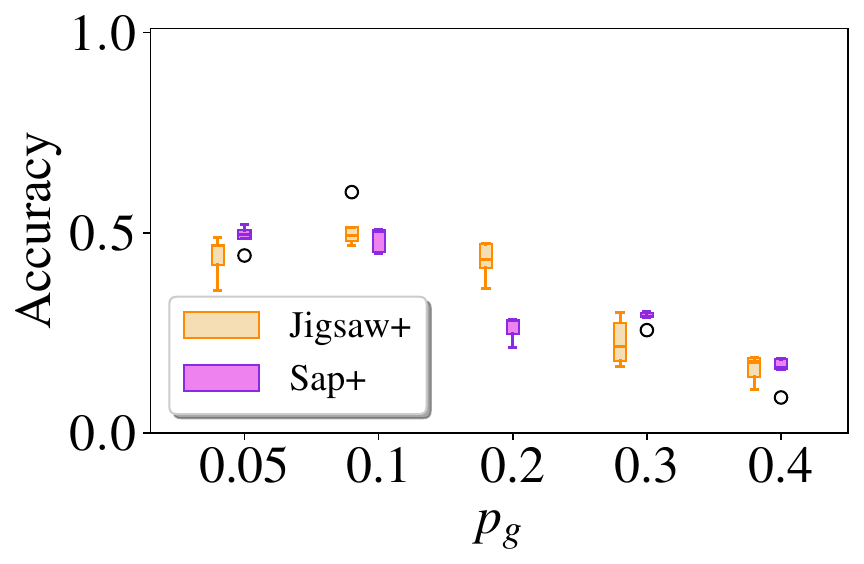}
		\end{minipage}
 	}

	\caption{The accuracy of Jigsaw+ and Sap+ with different hyperparameters, i.e., $\delta$, $maxlevel$, and $p_g$.}
	\label{fig:test_parameters}
    \Description{The figures present the impact of hyperparameters of Peekaboo for recovery accuracy, which are fully described in the text.}
\end{figure}
We provide the results of Jigsaw+ and Sap+ under different hyperparameter settings, i.e., $\delta$, $maxlevel$, and $p_g$. If the response similarity in Equations \ref{eq:rsp_fvp} and \ref{eq:rsp_ap} is no less than $\delta$, then the two queries are considered candidates for querying the same keyword. The results in Figure \ref{fig:test_delta} show that Jigsaw+ and Sap+ have the best accuracy of above $90\%$ and $0.6\%$, respectively, when $\delta$ is between 0.6 and 0.95. 
$maxlevel$ cancels the matching between two rounds if there are more than $maxlevel$ rounds between them. 
A small $maxlevel$ reduces runtime but may also lead to lower accuracy. 
We test Jigsaw+ and Sap+ with 16 rounds, and the results are shown in Figure \ref{fig:test_maxlevel}. 
The accuracy increases as $maxlevel$ increases, and Jigsaw+ and Sap+ reach about $80\%$ and $50\%$ accuracy,  respectively, when $maxlevel$ is no less than 5. 
$p_g$ is the ratio of removed matches in $\textsc{Match}$, in case some queries do not appear in both rounds of the input but still participate in matching, which causes errors. 
Other matching errors, such as those caused by noisy information, can also be filtered out by $p_g$. We evaluate $p_g$ with AP and FVP, and the results show that $p_g$ can be chosen between 0.05 and 0.2.

\section{Instantiation: IHOP+}
\label{app:IHOP+}

IHOP includes two coefficients for the costs of matching: $c$ for the quadratic terms and $d$ for the linear terms.
With the co-occurrence matrix information provided by Peekaboo, the $c$ and $d$ (in \cite{DBLP:conf/uss/OyaK22}, Equation 5) are set to
\begin{equation}
\begin{split}
    c_{i,i',j,j'}=-\sum\limits_{k=1}^\eta \mathbf{n}_D(\mathbf{C}^k_r[j][j']\log(\mathbf{C}^k_s[i][i'])\\
    -(1-\mathbf{C}^k_r[j][j'])\log(1-\mathbf{C}^k_s[i][i'])),
\end{split}
\end{equation}
\begin{equation}
\begin{split}  
\label{eq:d_1}
    d_{i,j}=-\sum\limits_{k=1}^\eta \mathbf{n}_D(\mathbf{C}^k_r[j][j])\log(\mathbf{C}^k_s[i][i])\\
    -(1-\mathbf{C}^k_r[j][j])\log(1-\mathbf{C}^k_s[i][i])),
\end{split}
\end{equation}
which summarizes the costs in different rounds. 
With the frequency information provided by Peekaboo, the $d$ (in \cite{DBLP:conf/uss/OyaK22}, Equation 6) is set to 
\begin{equation}
\begin{split}  
\label{eq:d_2}
    d_{i,j}=-\sum\limits_{k=1}^\tau \boldsymbol{\rho}[k](\mathbf{f}^k_r[j]\log(\mathbf{f}^k_s[i]),
\end{split}
\end{equation}
which represents the costs in each time slot.
As both the co-occurrence and frequency information are available, IHOP+ summarizes the Equation \ref{eq:d_1} and \ref{eq:d_2} as in IHOP. 
Following the attack strategy of IHOP, IHOP+ can recover the keywords of search queries in each group of $M$.

\begin{figure}
    \centering
		\begin{minipage}{.6\linewidth}
			\centering
                \includegraphics[width=\linewidth]{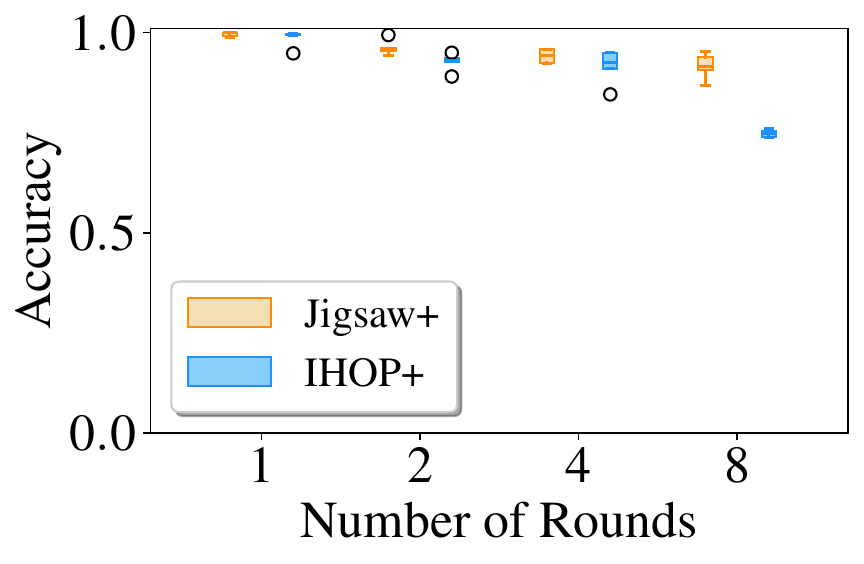}
		\end{minipage}

	\caption{The accuracy of Jigsaw+ and IHOP+ in Enron with AP leakage.}
	\label{fig:test_IHOP_plus}
    \Description{The figure presents the results of Jigsaw+ and IHOP+ for recovery accuracy in Enron, which is fully described in the text.}
\end{figure}

In \cite{DBLP:conf/uss/Nie00ZYL24}, it is reported that IHOP performs with accuracy similar to Jigsaw. 
We here compare IHOP+ with Jigsaw+ under the same settings as in Figure \ref{fig:test_comparison_enron}, Section \ref{sec:peekaboo_exp_2}, and present the results in Figure \ref{fig:test_IHOP_plus}. 
The accuracy of IHOP+ is comparable to that of Jigsaw+ and declines as the number of rounds increases.

\section{Peekaboo and Known-data Attacks} 
\label{app:peekaboo with known-data attack}
In Section \ref{sec:Peekaboo_query_rec}, Peekaboo calls the similar-data attacks to recover search queries. 
Previous known-data attacks can probably be modified and instantiated by Peekaboo to attack DSSE with an IOA. 
Based on the inferred SP of Peekaboo, the attacker can calculate the relationship between encrypted files in each round. 
For example, this relationship could be the number of keywords in both files. 
The attacker can further identify the ``search pattern'' of files and match the known historical files to the corresponding encrypted files. 
It is thus able to remove the updated encrypted files from the observed leakage, transforming the known-data attack in DSSE with an IOA into a traditional known-data attack. 
We note that current SOTA known-data attacks \cite{DBLP:conf/ndss/BlackstoneKM20,DBLP:conf/ccs/NingHPYL0D21} are able to provide nearly 100\% accuracy with only a small portion of the plaintexts of the encrypted files. 
If the client does not remove most of the attacker's known files from the encrypted database, the attacker can still recover search queries. 
Nevertheless, adapting known-data attacks for Peekaboo to target DSSE will not be the same as those used for similar-data attacks.

\section{Results of Jigsaw+ without Frequency}
\label{app:without_frequency}
\begin{figure}
    \centering
		\begin{minipage}{.6\linewidth}
			\centering
                \includegraphics[width=\linewidth]{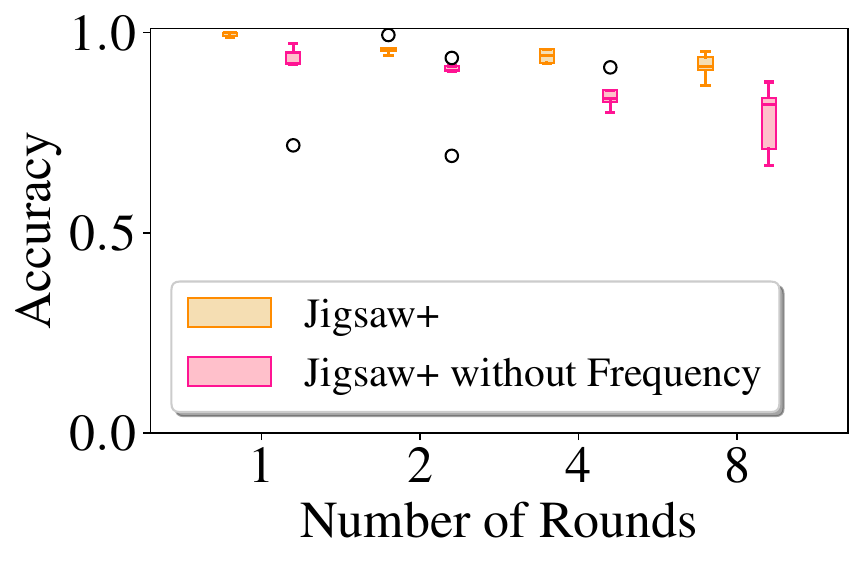}
		\end{minipage}

	\caption{The accuracy of Jigsaw+ and Jigsaw+ without query frequency information in Enron with AP leakage.}
	\label{fig:test_Jigsaw_plus_without_f}
    \Description{The figure presents the results of Jigsaw+ without query frequency for recovery accuracy in Enron, which is fully described in the text.}
\end{figure}

As Jigsaw can operate with or without auxiliary query frequency information \cite{DBLP:conf/uss/Nie00ZYL24}, we also evaluate Jigsaw+ without frequency information (denoted as Jigsaw+ without Frequency). 
We use the same experimental settings as those in Figure \ref{fig:test_comparison_enron}, Section \ref{sec:peekaboo_exp_2}. 
From Figure \ref{fig:test_Jigsaw_plus_without_f}, we see that Jigsaw+ without frequency achieves slightly lower accuracy compared to Jigsaw+, following the same trend: the accuracy slightly decreases  as the number of rounds increases.

\section{Results in Wikipedia}
\label{app:wiki}

\begin{figure*}[htp]
    \centering
            \begin{minipage}{\linewidth}
                \centering
                \begin{tikzpicture}
        \node[draw=gray!50, dashed, rectangle, rounded corners=0pt, thick, inner sep=-1pt] {  
\begin{tabular}{c@{\hskip 4pt}c@{\hskip 4pt}c@{\hskip 4pt}c}
    \begin{tikzpicture}
        \fill[darkorange] (0pt, 0pt) rectangle (6pt, 4pt);
    \end{tikzpicture}
    \begin{scriptsize}
        Jigsaw+
    \end{scriptsize}
    &
    \begin{tikzpicture}
        \fill[darkgreen] (0pt, 0pt) rectangle (6pt, 4pt);
    \end{tikzpicture}
    \begin{scriptsize}
        Jigsaw+ with SP
    \end{scriptsize}
    &
    \begin{tikzpicture}
        \fill[blueviolet] (0pt, 0pt) rectangle (6pt, 4pt);
    \end{tikzpicture} 
    \begin{scriptsize}
        Sap+
    \end{scriptsize}
    &
    \begin{tikzpicture}
        \fill[dodgerblue] (0pt, 0pt) rectangle (6pt, 4pt);
    \end{tikzpicture}  
    \begin{scriptsize}
        Sap+ with SP
    \end{scriptsize}

\end{tabular}
};
\end{tikzpicture}
            \end{minipage}
    \subfloat[Wikipedia, AP]
	{
 \label{fig:test_wiki_AP}
		\begin{minipage}{.23\linewidth}
			\centering
                \includegraphics[width=\linewidth]{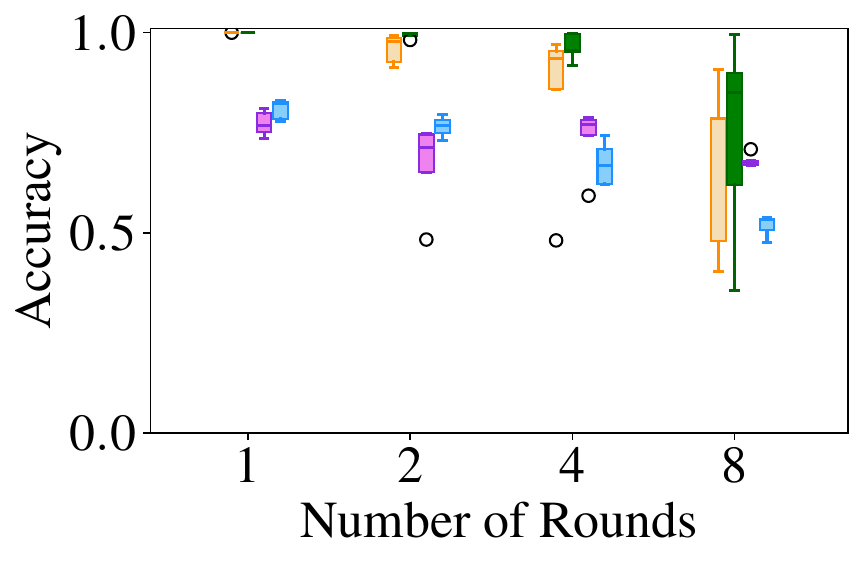}
		\end{minipage}
	}
    \subfloat[Wikipedia, FVP]
	{
 \label{fig:test_test_wiki_FVP}
		\begin{minipage}{.23\linewidth}
			\centering
			\includegraphics[width=\linewidth]{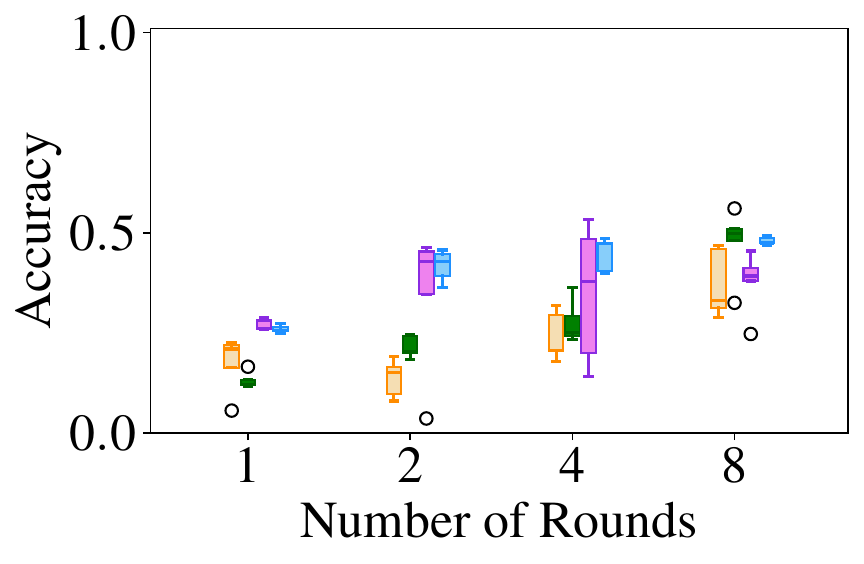}
		\end{minipage}
 	}
    \subfloat[Wikipedia, AP]
	{
 \label{fig:test_wiki_random_day}
		\begin{minipage}{.23\linewidth}
			\centering
                \includegraphics[width=\linewidth]{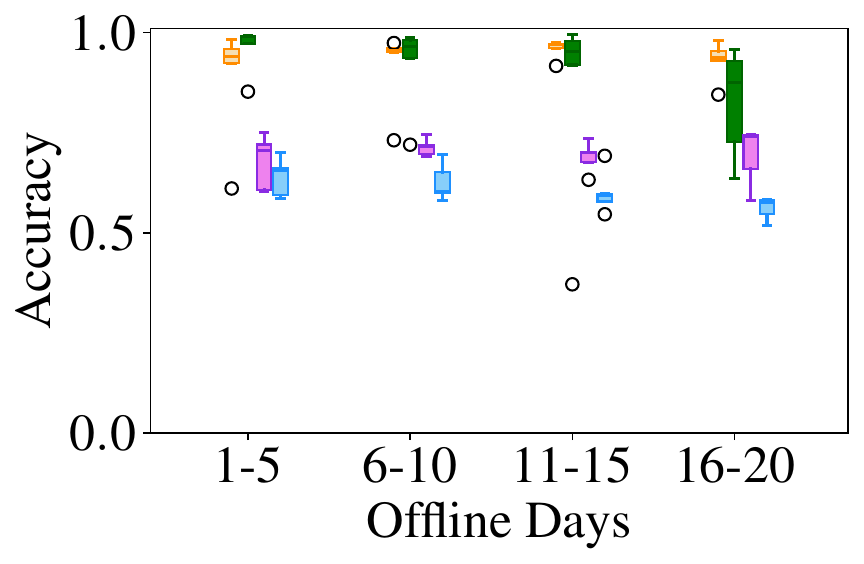}
		\end{minipage}
	}
    \subfloat[Wikipedia, AP]
	{
 \label{fig:test_wiki_query_number}
		\begin{minipage}{.23\linewidth}
			\centering
			\includegraphics[width=\linewidth]{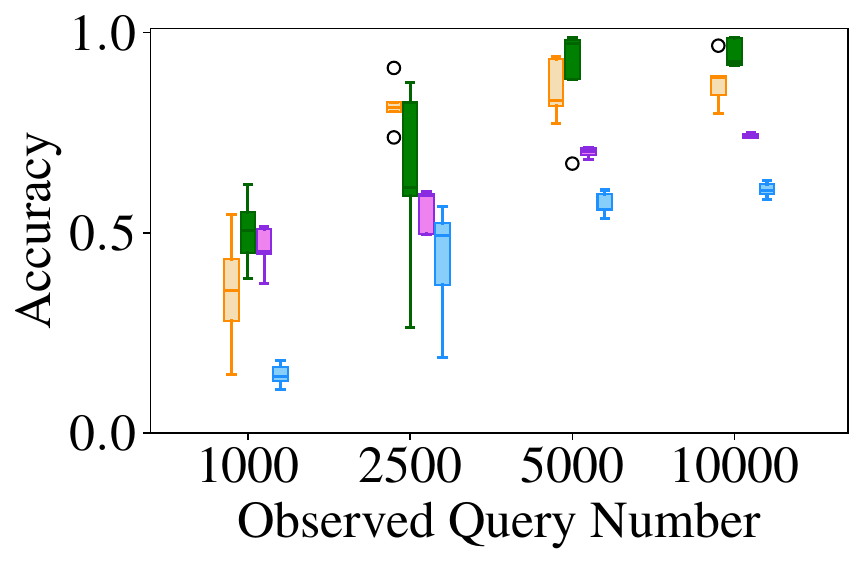}
		\end{minipage}
 	}

	\caption{
    The accuracy of Jigsaw+, Sap+, Jigsaw+ with SP, and Sap+ with SP in Wikipedia under varying conditions, i.e., number of rounds, leakage type (AP or FVP), number of offline days, and number of observed queries in each round.
    }
	\label{fig:test_comparison_wiki}
    \Description{The figures present the impact of different round numbers, leakage type (AP or FVP), number of offline days, and number of observed queries in each round for the recovery accuracy in Wikipedia, which are fully described in the text.}
\end{figure*}

\begin{figure*}
    \centering
    \begin{minipage}{\linewidth}
        \centering
        \begin{tikzpicture}
        \node[draw=gray!50, dashed, rectangle, rounded corners=0pt, thick, inner sep=-1pt] {  
\begin{tabular}{c@{\hskip 4pt}c@{\hskip 4pt}c@{\hskip 4pt}c}
    \begin{tikzpicture}
        \fill[darkorange] (0pt, 0pt) rectangle (6pt, 4pt);
    \end{tikzpicture}
    \begin{scriptsize}
        Jigsaw+
    \end{scriptsize}
    &
    \begin{tikzpicture}
        \fill[darkgreen] (0pt, 0pt) rectangle (6pt, 4pt);
    \end{tikzpicture}
    \begin{scriptsize}
        Jigsaw+ with SP
    \end{scriptsize}
    &
    \begin{tikzpicture}
        \fill[blueviolet] (0pt, 0pt) rectangle (6pt, 4pt);
    \end{tikzpicture} 
    \begin{scriptsize}
        Sap+
    \end{scriptsize}
    &
    \begin{tikzpicture}
        \fill[dodgerblue] (0pt, 0pt) rectangle (6pt, 4pt);
    \end{tikzpicture}  
    \begin{scriptsize}
        Sap+ with SP
    \end{scriptsize}

\end{tabular}
};
\end{tikzpicture}
    \end{minipage}
    \subfloat[Enron, AP]
	{
 \label{fig:test_query_number_enron_noupdate}
		\begin{minipage}{.23\linewidth}
			\centering
                \includegraphics[width=\linewidth]{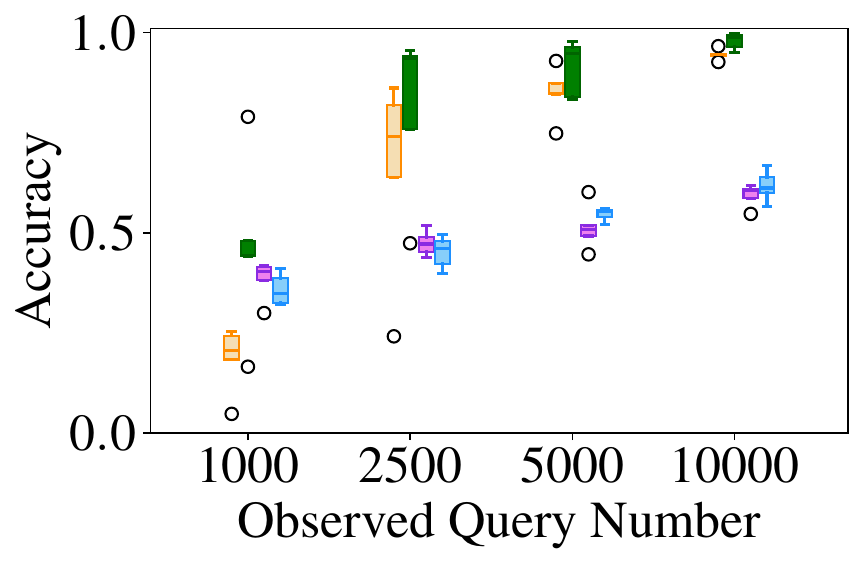}
		\end{minipage}
	}
    \subfloat[Lucene, AP]
	{
 \label{fig:test_query_number_lucene_noupdate}
		\begin{minipage}{.23\linewidth}
			\centering
			\includegraphics[width=\linewidth]{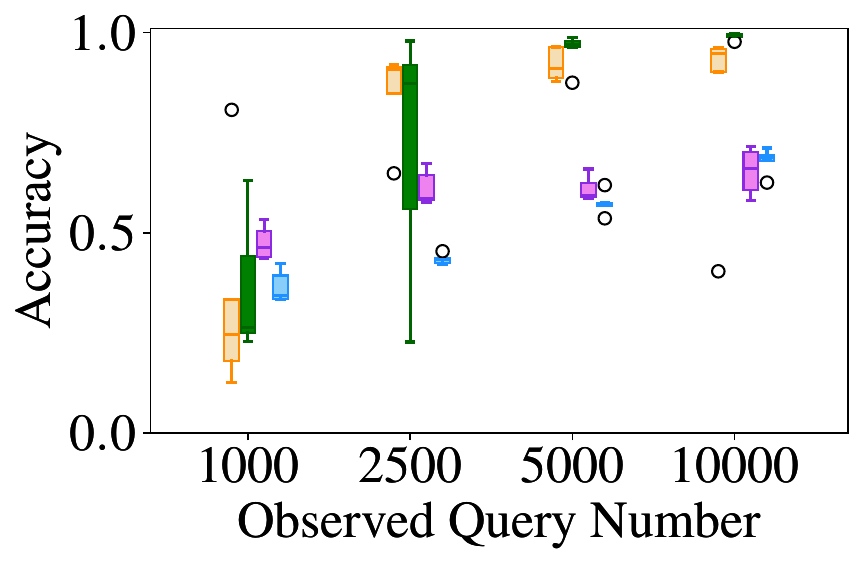}
		\end{minipage}
 	}
    \subfloat[Enron, FVP]
	{
 \label{fig:test_query_number_enron_fvp_noupdate}
		\begin{minipage}{.23\linewidth}
			\centering
                \includegraphics[width=\linewidth]{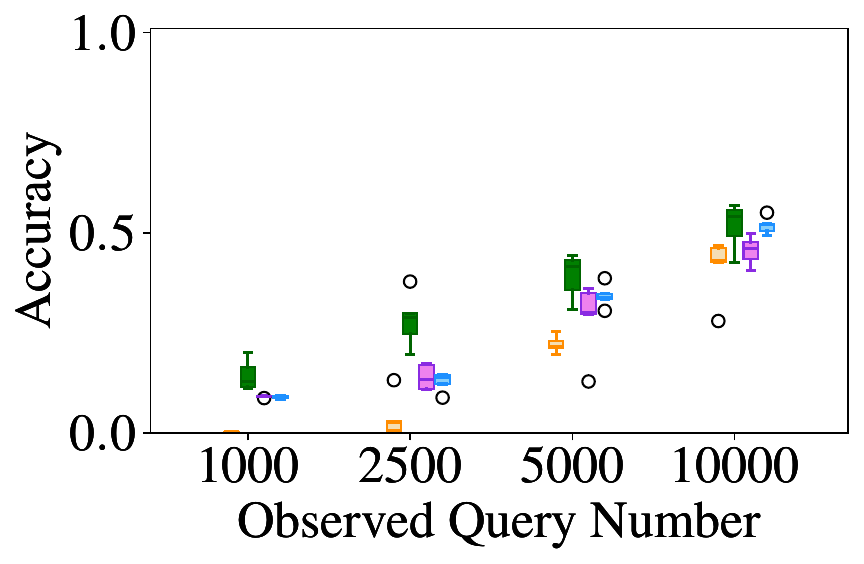}
		\end{minipage}
	}
    \subfloat[Lucene, FVP]
	{
 \label{fig:test_query_number_lucene_fvp_noupdate}
		\begin{minipage}{.23\linewidth}
			\centering
			\includegraphics[width=\linewidth]{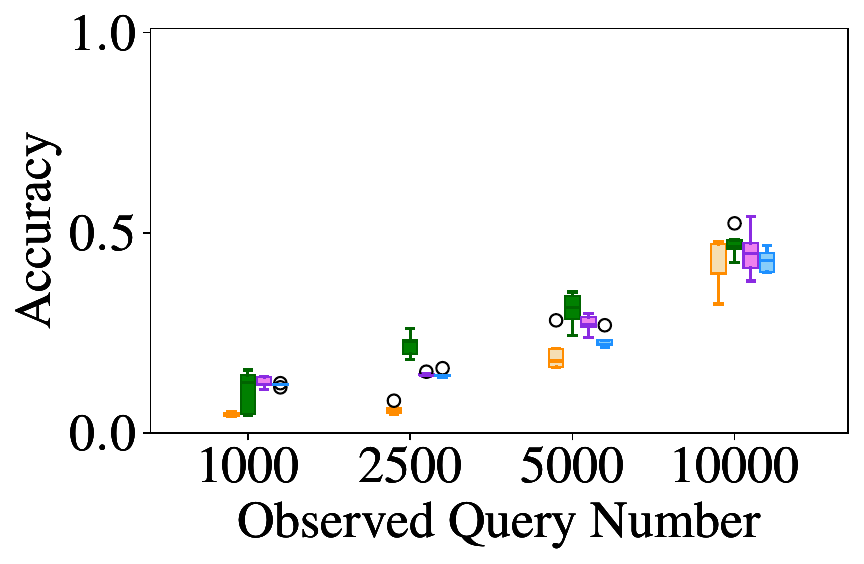}
		\end{minipage}
 	}

	\caption{The accuracy of Jigsaw+, Sap+, Jigsaw+ with SP, and Sap+ with SP in Enron and Lucene with different number of observed queries in each round with the AP or FVP leakage when the attacker only has a static database without updates.}
	\label{fig:test_query_number_noupdate}
    \Description{The figures present the impact of observed search queries for recovery accuracy when the attacker only has a static database without updates, which are fully described in the text.}
\end{figure*}

We evaluate the performance of Jigsaw+, Sap+, Jigsaw+ with SP, and Sap+ with SP on Wikipedia \cite{Wikipedia} to demonstrate that the ``+'' variants still remain effective on large-scale and non-email datasets.
We use the Wikipedia dataset from \cite{DBLP:conf/uss/Nie00ZYL24} and extract a subset of 200,000 files. Since the files in the Wikipedia dataset do not contain timestamps, we assign each file a timestamp randomly selected from one year. 
When using the FVP leakage, we set $p_g=0.1$.
Other experimental settings follow those described in Figure \ref{fig:test_comparison_enron}, \ref{fig:test_comparison_enron_fvp}, \ref{fig:test_random_enron}, and \ref{fig:test_query_number_enron}, Section \ref{sec:peekaboo_exp_2}.  
The results (Figure \ref{fig:test_comparison_wiki}) are generally consistent with those on Enron and Lucene.
This consistency is expected because the effectiveness of Peekaboo depends on the co-occurrence of queries and keywords in datasets, rather than data type. 
With the AP leakage (Figure \ref{fig:test_wiki_AP}), the accuracy of Jigsaw+ decreases as the number of rounds increases; however, the average accuracy remains above $80\%$.
Meanwhile, the accuracy of Sap+ fluctuates slightly around $75\%$.
Under the FVP leakage (Figure \ref{fig:test_test_wiki_FVP}), both Jigsaw+ and Sap+ yield the similar accuracy of approximately $40\%$. 
Figure \ref{fig:test_wiki_random_day} shows that the number of offline days between rounds has little impact on accuracy.
Finally, as shown in Figure \ref{fig:test_wiki_query_number}, the accuracy improves as the number of observed queries increases. 
Across all settings, Jigsaw+ and Sap+ perform similarly to Jigsaw+ with SP and Sap+ with SP, consistent with the results on Enron and Lucene.

\section{Peekaboo with Static Dataset}
\label{app:outdated data exp}
We provide the results while the Peekaboo attacker has only access to a static dataset without updates.
The experimental configuration is consistent with evaluating the impact of observed search queries (Figure \ref{fig:test_query_number}), except that the attacker's dataset has no updates. 
We illustrate the accuracy in Figure \ref{fig:test_query_number_noupdate}. Increasing the observed queries, we obtain better accuracy for Jigsaw+ and Sap+ and their with-SP versions. 
It is evident that without any updates, the accuracy only undergoes a slight decline and fluctuation, maintaining the performance as in Figure \ref{fig:test_query_number}.

\section{Ethical Consideration}
\label{app:ethical consideration}
This work introduces a more restricted yet practical attacker based on intermittent observation, aiming to inspire new studies on enhancing the security of DSSE. 
It delivers a positive impact on the development and deployment of DSSE. 
All experiments use publicly available datasets widely used in prior studies and the experiments are conducted for research purpose only, in an isolated virtual environment.   
Queries are randomly generated using public trend datasets, and leakage is simulated (instead of ``real'' leakage). 
The experiments are not 
deployed to any real-world applications. 
All tested attacks, defenses, and related libraries are open-source. 
To support mitigation on attacks, we also discuss potential countermeasures in Section \ref{sec:Against Countermeasures}.

\end{document}